%% file: main_with_corrections.tex
\documentclass[twocol]{ametsocV6.1}

\usepackage{booktabs}
\usepackage{arydshln}

\usepackage{subcaption}
\usepackage{rotating}  
\usepackage{soul}  
\usepackage[most]{tcolorbox}
\usepackage{listings}
\usepackage{xcolor}

\definecolor{hpheaderbg}{RGB}{233,205,170}
\definecolor{hpbodybg}{RGB}{248,247,243}
\definecolor{hpkey}{RGB}{27,127,80}
\definecolor{hpvalue}{RGB}{0,0,0}
\definecolor{hpcomment}{RGB}{52,108,142}
\definecolor{hpnumber}{RGB}{20,40,170}
\definecolor{hplinegutter}{RGB}{170,170,170}

\lstdefinestyle{hpconfig}{
    basicstyle=\ttfamily\footnotesize,
    keywordstyle=\bfseries\color{hpkey},
    commentstyle=\itshape\color{hpcomment},
    stringstyle=\color{hpvalue},
    numbers=left,
    numberstyle=\color{hplinegutter}\footnotesize,
    numbersep=8pt,
    showstringspaces=false,
    morekeywords={emb_dim,n_leads,clamp_max,clamp_min,mlp_hidden,channel_mult,
                  embedding_type,model_channels,N_grid_channels,attn_resolutions,
                  hr_mean_conditioning,
                  lr,seed,huber_delta,lambda_time,lambda_time_warmup_epochs,
                  total_batch_size,training_duration,batch_size_per_gpu,
                  grad_clip_threshold,split_fraction,
                  curriculum,enabled,leads_per_step,epochs_per_step,
                  fp_optimizations,dataloader_workers,
                  input_variables,output_variables,invariant_variables,
                  log_precip,crop_mult,interp_low_res, optimizer, loss,
                  lr_decay_rater, model_type, gridtype,
                  regression_checkpoint, lr_decay_rate},
    morecomment=[l]{\#},
    columns=fullflexible,
    keepspaces=true,
    breaklines=true,
}

\newtcolorbox{hpbox}[1]{%
    enhanced,
    colback=hpbodybg,
    colframe=hpheaderbg,
    boxrule=1.2pt,
    arc=2pt,
    left=2pt, right=2pt, top=2pt, bottom=2pt,
    fonttitle=\ttfamily\bfseries\color{black},
    coltitle=black,
    title={#1},
    colbacktitle=hpheaderbg,
    attach boxed title to top left={xshift=6pt, yshift=-2pt},
    boxed title style={
        colback=hpheaderbg, colframe=hpheaderbg,
        boxrule=0pt, arc=2pt, left=4pt, right=4pt
    },
    coltitle=black,
    overlay broken={\node[anchor=north east, inner sep=2pt] at (frame.north east) {};},
    before skip=8pt, after skip=8pt
}
\title{SwAIther-Precip: Lead-Time-Aware Bias Correction Enables Kilometer-Scale Downscaling of Global AI Precipitation Forecasts over Switzerland}

\authors{Dan Assouline\aff{a}\correspondingauthor{Dan Assouline, dassouline@ethz.ch},
Erwan Koch\aff{b},
Federico Amato\aff{a},
Filippo Quarenghi\aff{b,c},
Daniele Nerini\aff{d},
Thibaut Loiseau\aff{a},
Kyle van de Langemheen \aff{a},
and Tom Beucler\aff{b,c}.}

\affiliation{
\aff{a}{Swiss Data Science Center, ETH, Zürich, Switzerland}\\
\aff{b}{Expertise Center for Climate Extremes, University of Lausanne, Lausanne, Switzerland}\\
\aff{c}{Faculty of Geosciences and Environment, University of Lausanne, Lausanne, Switzerland}\\
\aff{d}{Federal Office of Meteorology and Climatology MeteoSwiss, Locarno-Monti, Switzerland}\\
}

\abstract{Skillful medium-range precipitation forecasting at kilometer scale remains challenging over complex terrain because precipitation arises from multiscale nonlinear processes that global models cannot resolve at affordable cost. Global AI weather models produce skillful medium-range forecasts, but their native 0.25° resolution limits direct use for local hazard applications. Statistical downscaling can bridge this gap, yet existing approaches often struggle with state-dependent, and especially lead-time-dependent, biases in global forecasts. We introduce SwAIther-Precip, a lead-time-aware downscaling framework that converts coarse-resolution deterministic AIFS forecasts into probabilistic km-scale precipitation fields over Switzerland. First, a U-Net conditioned on lead time via feature-wise linear modulation corrects systematic biases at coarse resolution. This targeted correction enables a cheaper super-resolution step conditioned only on corrected precipitation and topography, allowing direct training on observations rather than on the full atmospheric state. A CorrDiff-based diffusion model generates fine-scale spatial variability independently of lead time. Using AIFS forecasts and CombiPrecip radar–gauge observations, SwAIther-Precip reduces CRPS by 48\% relative to super-resolved raw AIFS. The generated fields reproduce observed spatial variability with spectral fidelity above 0.85 at large scales and 0.88 at small scales, corresponding to an effective resolution of $\approx$4 km on a 1 km grid for lead times up to 5 days. Training across lead times improves long-range performance, yielding a 13\% CRPS reduction at 6 days relative to single-lead models. These results, supported by a comparison with an end-to-end CorrDiff model, show that correcting lead-time-dependent biases before generative super-resolution is key to efficient km-scale probabilistic downscaling of global AI precipitation forecasts.
}

\begin{document}

%% Necessary!
\maketitle

% (To discuss; please add suggestions): \textbf{Potential Reviewer List (tentatively ranked)}: \href{https://joeloskarsson.github.io}{Joel Oskarsson (ETHZ)}, \href{https://www.physics.ox.ac.uk/our-people/antonio/publications}{Bobby Antonio (Oxford)}, \href{https://scholar.google.com/citations?hl=en&user=GylfPngAAAAJ&view_op=list_works&sortby=pubdate}{Fredrik Lindsten (Linkoping University)}, \href{https://scholar.google.es/citations?hl=en&user=kOD_Z30AAAAJ&view_op=list_works&sortby=pubdate}{Qidong Yang (MIT)}, \href{https://scholar.google.com/citations?user=Mjw-8yMAAAAJ&hl=en}{Xiaohui Zhong (Fudan University)}, \href{https://www.smhi.se/en/research/our-team/search-for-employees/tomas-landelius}{Tomas Landelius (SMHI)}, \href{https://research.google/people/108293/?&type=google}{Zhong Yi Wan (Google Research)}. 

%%%%%%%%%%%%%%%%%%%%%%%%%%%%%%%%%%%%%%%%%%%%%%%%%%%%%%%%%%%%%%%%%%%%%
% SIGNIFICANCE STATEMENT/CAPSULE SUMMARY
%%%
%%%%%%%%%%%%%%%%%%%%%%%%%%%%%%%%%%%%%%%%%%%%%%%%%%%%%%%%%%%%%%%%%%
%%%
%
% If you are including an optional significance statement for a journal article or a required capsule summary for BAMS 
% (see www.ametsoc.org/ams/index.cfm/publications/authors/journal-and-bams-authors/formatting-and-manuscript-components for details), 
% please apply the necessary command as shown below:
%
%Significance Statement (all journals except BAMS)
%
%	 Enter significance statement here, no more than 120 words. See \url{www.ametsoc.org/index.cfm/ams/publications/author-information/significance-statements/} for details.
%

\statement
AI weather models can now make skillful global forecasts up to two weeks ahead, but they are still too coarse for local precipitation hazards in mountainous regions such as Switzerland, which is a stringent testbed because terrain strongly shapes precipitation and high-quality observations allow rigorous evaluation. We show that correcting forecast errors in a lead-time-aware way before adding fine-scale detail makes these models much more useful locally. Using forecasts from a public global artificial intelligence weather model and Swiss radar–rain gauge observations, our method produces kilometer-scale precipitation forecasts, cuts forecast error in two relative to the raw model, and preserves realistic spatial patterns out to 6 days.

%% Capsule (BAMS only)
%%
%\capsule
%       Enter BAMS capsule here, no more than 30 words. See \url{www.ametsoc.org/index.cfm/ams/publications/author-information/formatting-and-manuscript-components/#capsule} for details.
%
%% * * If using twocol mode, you will need to use the commands "twocolsig" and "twocolcapsule" in place of "sig" and "capsule"
%%      to ensure that the text box correctly spans across both columns.
%

%%%%%%%%%%%%%%%%%%%%%%%%%%%%%%%%%%%%%%%%%%%%%%%%%%%%%%%%%%%%%%%%%%%%%
% MAIN BODY OF PAPER
%%%%%%%%%%%%%%%%%%%%%%%%%%%%%%%%%%%%%%%%%%%%%%%%%%%%%%%%%%%%%%%%%%%%%
%

%% In all cases, if there is only one entry of this type within
%% the higher level heading, use the star form: 
%%
% \section{Section title}
% \subsection*{subsection}
% text...
% \section{Section title}

%vs

% \section{Section title}
% \subsection{subsection one}
% text...
% \subsection{subsection two}
% \section{Section title}

%%%%%%%%%%%%%%%%%%%
\section{Introduction}
%%%%%%%%%%%%%%%%%%%

Accurate precipitation forecasting at high spatial resolution remains a central challenge in numerical weather prediction (NWP), particularly over regions with complex topography such as Switzerland. Precipitation fields exhibit strong spatial heterogeneity and nonlinear dynamics driven by orography, mesoscale organization, and local atmospheric conditions, which are often poorly resolved by global forecasting systems. In conventional NWP, the physically relevant quantity is the effective spatial resolution, i.e., the smallest wavelength at which the model retains realistic variance before the power spectrum becomes distorted \citep{skamarock2004evaluating}, which is often several grid lengths coarser than the nominal mesh \citep{klaver2020effective}. Reaching km-scale effective resolution means entering a non-hydrostatic regime tightly coupled with convection. In NWP, this forces a trade-off between horizontal resolution, ensemble size, and forecast range; AI can more readily target the spatiotemporal resolution. Deep generative radar nowcasting predicts high-resolution precipitation minutes ahead \citep{ravuri2021skilful}, while km-scale AI emulators extend this capability into the short range. StormCast \citep{pathak2024kilometerscaleconvectionallowingmodel} advances a kilometer-scale atmospheric state with a 1\,h autoregressive time step and shows competitive skill over roughly the first six forecast hours. HRRRCast \citep{abdi2026HRRRCast} instead covers the full 48\,h HRRR forecast schedule by training a single model on multiple lead-time increments and chaining predominantly 6\,h steps at inference, with shorter steps used to reach intermediate lead times, which reduces the number of autoregressive steps required at a given lead time. Beyond the short range, however, km-scale AI weather prediction at medium range (2\,days to 2\,weeks) remains rare, in part because forecast errors accumulate rapidly when a km-scale atmospheric state is rolled out over many short time steps. This constraint motivates refining a completed coarse forecast, rather than integrating a km-scale state forward in time.

\smallskip
\noindent In parallel, global AI weather prediction has advanced rapidly. Pangu-Weather \citep{bi2023accurate} and GraphCast \citep{lam2023GraphCast} established that neural networks trained on meteorological reanalyses can produce skillful medium-range deterministic global forecasts. These systems are now routinely evaluated with benchmarks such as WeatherBench~2 \citep{rasp2023WB2}. In this work we choose AIFS, the global AI weather prediction system of the European Centre for Medium-Range Weather Forecasts, as the coarse driver because its deterministic configuration  keeps post-processing inexpensive and provides the multivariate atmospheric context needed for precipitation correction \citep{lang2024aifs}. A probabilistic global AI model could be used, but propagating its uncertainty would require conditioning on, or separately post-processing, multiple possible coarse-resolution atmospheric states, increasing training and inference costs. Conditioning on the ensemble mean instead would avoid that cost but discard the very uncertainty that motivates using a probabilistic driver in the first place. Choosing AIFS is especially appropriate for precipitation because recent AIFS updates have both improved precipitation skill and expanded the set of output variables \citep{moldovan2025update}. Even so, current global AI forecasts remain coarse relative to the kilometer scale, typically around 25--30\,km, which limits their direct use for local and regional decision-making.

\smallskip
\noindent Bridging this spatial-resolution gap follows two strategies. The first learns or emulates regional atmospheric evolution directly. Machine-learning limited-area models (LAMs) have made this direction increasingly realistic \citep{adamov2025buildingmachinelearninglimited}: CRPS-LAM trains regional probabilistic forecasting with proper scoring rules rather than only with diffusion-based samplers \citep{larsson2025crpslamregionalensembleweather}, while other systems reach 3\,km and 1\,h resolution for non-precipitation surface variables \citep{xu2025artificial}. The design space of this first strategy extends beyond boundary-forced LAMs: stretched-grid models refine the mesh over the target region without explicit lateral boundaries \citep{nipen2025stretchedgrid}, and OneForecast combines global and regional prediction within a single multiscale framework \citep{gao2025oneforecast}. Hybrid AI--physics systems offer a lower-cost alternative; for example, Pangu-Weather can drive a regional WRF configuration targeting extreme precipitation \citep{xu2024improvement}, while \citet{sha2026ai} learn a stable AI regional emulator from reanalysis and climate-model forcings, and related work has sought to reduce the cost of dynamical downscaling through ML surrogates \citep{hobeichi2023downscalingGCMevapotranspiration}. Despite its physical appeal, this first strategy inherits the burdens of conventional regional modeling: lateral-boundary treatment, rollout stability, and the need to predict the full atmospheric state at high resolution. STCast, for example, explicitly addresses boundary mismatch through spatially aligned attention, underscoring the importance of such issues \citep{chen2025stcast}.

\smallskip
\noindent The second strategy, which we adopt here, bypasses many of these limitations through forecast refinement: a completed coarse forecast is downscaled post hoc to local scales without evolving the regional atmospheric state at kilometer resolution. CNN-based post-processing has been shown to improve precipitation forecasts from weather-model output \citep{badrinath2023precipCNN}, and multivariate approaches exploit the full set of NWP predictors \citep{rojascampos2023NNforprecipfromNWP}. Generative methods have progressed from GAN and VAE-GAN stochastic downscaling of imperfect IFS forecasts \citep{harris2022stochasticdownscalingprecipforecasts}, to joint bias correction and downscaling via CorrectorGAN \citep{price2022improvingprecip}, and to generative post-processing for regional rainfall over East Africa \citep{antonio2024postprocessingEastAfricaprecip}. CorrDiff introduced a coarse-regression-plus-residual-diffusion decomposition for km-scale atmospheric downscaling from coarse inputs \citep{mardani2025residual}, a successful framework since extended to forecasting, for example to downscale forecasts over China at 3\,km resolution for lead times up to 72\,h \citep{sun2026China3kmCorrDiff}.

\smallskip
\noindent This recent progress motivates extending forecast refinement to longer lead times, especially for km-scale precipitation. Arai et al.\ target medium-range precipitation super-resolution over Japan \citep{arai2025enhancingresolutionsprecipforecasts}, and Li et al.\ refine FuXi forecasts to 1\,km over China \citep{li2025one}; both employ deterministic neural networks. \citet{molinaro2026universaldiffusionbasedprobabilisticdownscaling} show that diffusion-based probabilistic downscaling can transfer across heterogeneous upstream models up to 90\,h for near-surface variables. Because precipitation is difficult to forecast even at 0.25$^{\circ}$ resolution, many learned refinement methods condition on the full atmospheric state rather than sharpening the model's own coarse precipitation field \citep{rojascampos2023NNforprecipfromNWP}; CorrDiff, for instance, excludes coarse precipitation from its inputs \citep{mardani2025residual}. By contrast, our setting is motivated by the fact that recent AIFS versions have improved precipitation skill enough for the coarse precipitation field itself to include predictive information (that a downscaling model can leverage) \citep{moldovan2025update}. This does not imply that the AIFS precipitation forecast is unbiased; rather, it already contains useful information about event timing and broad spatial organization. At the same time, AIFS still exhibits lead-time-dependent amplitude errors, wet-area biases, and spatial displacements. Applying generative refinement directly to such a drifting forecast risks amplifying existing errors rather than correcting them, which motivates a debias-then-generate strategy \citep{wan2023debiascoarselysampleconditionally}.

\smallskip
\noindent This motivates \textit{SwAIther-Precip}, a two-step Swiss-AI precipitation downscaling framework that post-processes coarse-resolution precipitation forecasts from a global neural weather model (here AIFS, though the pipeline could generalize to other global weather models) into high-resolution probabilistic precipitation fields over Switzerland. A key design choice is to introduce an explicit, lead-time-aware correction step at coarse resolution before stochastic super-resolution. A unified deterministic correction network uses the full AIFS context and the forecast lead time to produce a physically realistic, intermediate, coarse precipitation field, and a diffusion-based model then generatively super-resolves that corrected precipitation field only. In particular, we do not attempt to autoregressively integrate a km-scale atmospheric state forward in time given the medium-range lead times of interest. Our contributions are threefold:

\begin{enumerate}
    \item \textbf{A two-step downscaling decomposition.} We present a pipeline that explicitly separates lead-time-aware coarse-resolution bias correction from stochastic super-resolution, allowing each component to specialize: the deterministic first step corrects systematic spatial and intensity biases in the global model output, while the generative second step adds realistic fine-scale variability based on the observational training data. A key practical advantage is that the bias correction absorbs lead-time dependence at the resolved coarse scale, so the super-resolution model operates on only two channels (corrected precipitation and topography) and can be trained in a perfect-prognosis setting on observations alone, substantially reducing computational cost. A comparison with an end-to-end CorrDiff model at a 3-day lead time shows that single-step direct downscaling is clearly less accurate on most metrics. We further identify a mesoscale variance bottleneck (20--100\,km) introduced by the deterministic bias correction, pointing to an avenue for improvement.

    \item \textbf{Lead-time-aware downscaling via a lead-time-aware bias-correction step.} We introduce a downscaling framework whose output depends explicitly on forecast lead time. Lead-time awareness is injected at the bias-correction step (Step~1) through Feature-wise Linear Modulation (FiLM, \cite{perez2018film}) layers and a temporal-consistency regularization; the super-resolution step that follows (Step~2, CorrDiff) does not require lead-time conditioning. This multi-lead approach not only eliminates the need for separate single-lead models, but also outperforms them at longer lead times.

    \item \textbf{Km-scale probabilistic fields with realistic spatial structure.} By decoupling the two steps, we obtain probabilistic 1-km precipitation fields that retain realistic spatial structure across the full 6-hour to 6-day lead range, including the medium-range time frame. The ensemble forecasts notably reproduce the observed precipitation power spectrum down to an effective resolution of $\approx$4 km on a 1 km grid. This is achieved at substantially lower computational cost than a fully generative end-to-end model, since the diffusion component operates on a single variable rather than the full atmospheric state.
\end{enumerate}

\smallskip
\noindent The remainder of the article is organized as follows. Section \ref{sec:methods} describes the proposed methodology. Section \ref{sec:data} presents the datasets used for the study and for all experiments performed to test the methodology. Section \ref{sec:eval} defines the baseline models, evaluation metrics, and analysis tools used to assess the methodology across multiple variants. Section \ref{sec:results} presents and analyzes results of the conducted experiments, at both the bias-correction and full downscaling levels. Section~\ref{sec:discussion} summarizes the findings and outlines future challenges and directions.

\section{Methodology}
\label{sec:methods}
% ---------------------------------------------------------
%%%%%%%%%%%%%%%%%%%%%%%%%%%%%%%%%%%%%%%%%%%%%%%%%%%%%%%%%%%

% summarizing figure
\begin{figure*}[t]%
\includegraphics[width=1\textwidth]{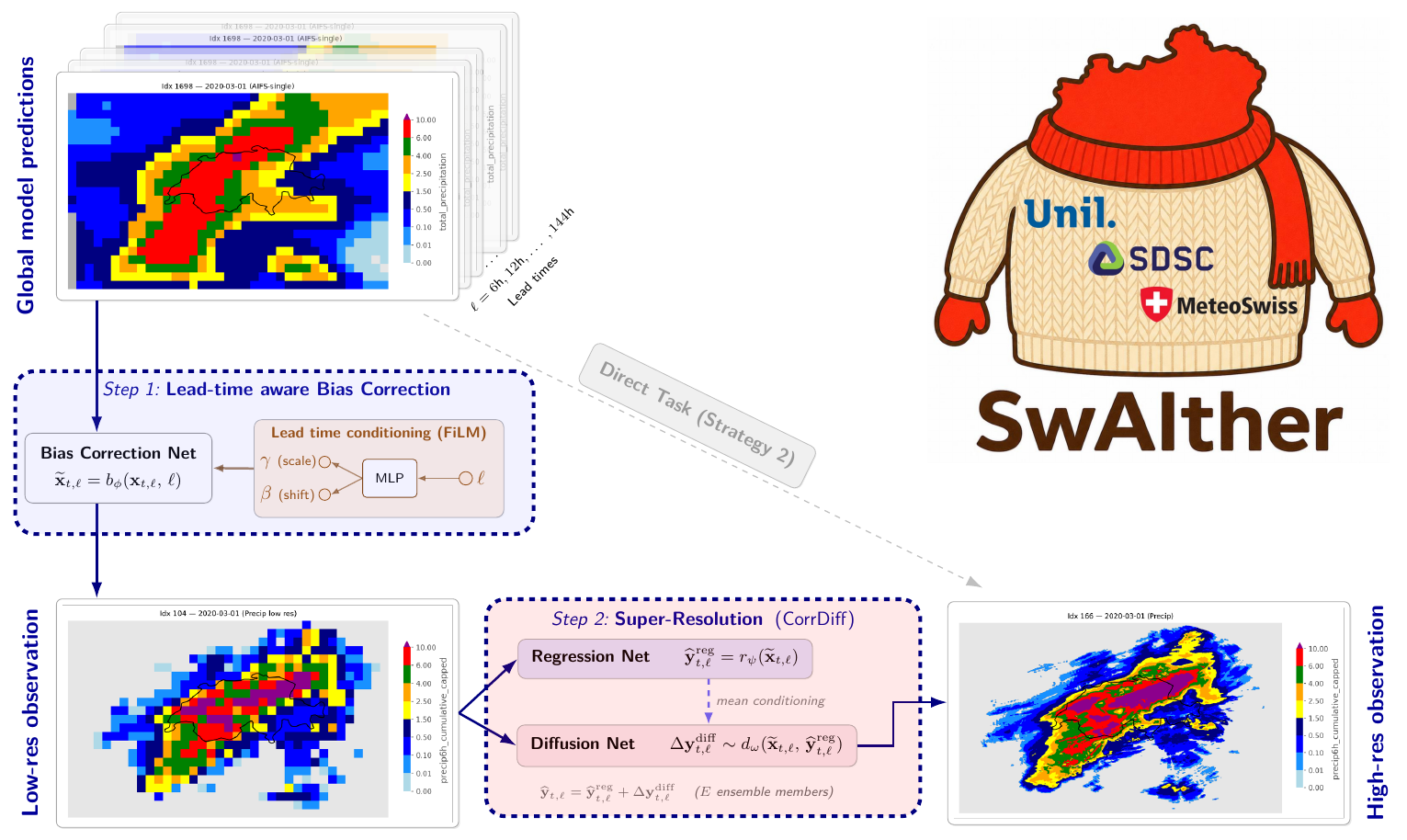}
{\caption{\textbf{The SwAIther-Precip strategy:} A first bias-correction step (Step~1) is followed by a super-resolution step (Step~2) which comprises a regression and a residual sub-network. During training, Step 1 and Step 2 are trained independently; during inference, a new global prediction performed by AIFS-single (with a provided lead time) is first bias-corrected, and then passed through the super-resolution model to generate diffusion samples at high resolution.}
\label{fig1}}
\end{figure*}

%%%%%%%%%%%%%%%%%%%%%%%%%%%%%%%%%%%%%%%%%
\subsection{Problem formulation: lead-time-aware downscaling}
%%%%%%%%%%%%%%%%%%%%%%%%%%%%%%%%%%%%%%%%%

Let $t \in \mathcal{T}$ denote the forecast valid time, i.e., the physical time at which the forecast is evaluated, and let $\ell \in \mathcal{L}$ denote the lead time, i.e., the elapsed time since forecast initialization. For each pair $(t,\ell)$, let $\mathbf{x}_{t,\ell} \in \mathbb{R}^{H \times W \times C_{\text{in}}}$ denote the coarse-resolution multichannel forecast on a grid of height $H$ and width $W$, where the $C_{\text{in}}$ channels include precipitation, dynamic atmospheric variables (e.g., wind and humidity), and static fields (e.g., topography). Let $\mathbf{y}_{t,\ell} \in \mathbb{R}^{P_H \times P_W \times C_{\text{out}}}$ denote the corresponding high-resolution reference field on a grid of height $P_H$ and width $P_W$, with $P_H \approx sH$ and $P_W \approx sW$ for spatial upscale factor $s$. In this study, precipitation is the only target variable, so $C_{\text{out}} = 1$, while the predictors are already multivariate ($C_{\text{in}} > 1$). The formulation itself does not restrict $C_{\text{out}}$, and further target variables could be accommodated by widening the output layer of both steps.
 
\smallskip 
\noindent The objective of lead-time-aware downscaling is to learn a parametrized mapping
\begin{equation}
\widehat{\mathbf{y}}_{t,\ell} = f_\theta(\mathbf{x}_{t,\ell}, \ell),
\qquad
f_\theta: \mathbb{R}^{H \times W \times C_{\text{in}}} \times \mathcal{L}
\rightarrow \mathbb{R}^{P_H \times P_W \times C_{\text{out}}},
\end{equation}
that transforms coarse-resolution forecasts into high-resolution precipitation fields while explicitly accounting for the forecast lead time. Unlike classical super-resolution, which aims solely at increasing spatial resolution, this mapping must also correct the systematic biases present in global model forecasts. Conditioning on $\ell$ is essential because forecast uncertainty, spatial displacement errors, and temporal decorrelation all increase with lead time, making the input--output relationship non-stationary across lead times. The use of one single-lead model per lead time, instead of a single lead-time-conditioned model, will be discussed throughout the study. 

\subsection{Two-step lead-time-aware downscaling strategy}

We propose to decompose the lead-time-aware downscaling problem into two complementary sub-tasks: (i) \textbf{Step 1: \emph{deterministic bias correction}} at coarse resolution, which maps the multi-channel global forecast $\mathbf{x}_{t,\ell} \in \mathbb{R}^{H \times W \times C_{\text{in}}}$ to an intermediate bias-corrected precipitation field $\widetilde{\mathbf{x}}_{t,\ell} \in \mathbb{R}^{H \times W \times C_{\text{out}}}$, and (ii) \textbf{Step~2: \emph{generative super-resolution}}, a CorrDiff-based generative model which refines low-resolution bias-corrected field $\widetilde{\mathbf{x}}_{t,\ell}$ into high-resolution precipitation fields. This decomposition is motivated by the magnitude and structure of the systematic errors in global model forecasts, particularly at longer lead times, and allows each component to specialize: the bias correction handles large-scale amplitude and distributional mismatches, while the super-resolution focuses on fine-scale spatial variability. The bias-correction network and the CorrDiff model operate on the AIFS forecast field
restricted to a rectangular bounding box around Switzerland (matching
the CombiPrecip domain, defined in the data section), rather than on the full global AIFS field.
 
\smallskip
\noindent Three training strategies can implement this decomposition. In \textbf{Strategy~1} (adopted in this work), the two steps are trained independently and applied sequentially. The bias correction model $b_\phi$ produces:
\begin{equation}
\widetilde{\mathbf{x}}_{t,\ell} = b_\phi(\mathbf{x}_{t,\ell}, \ell),
\end{equation}
and the super-resolution model $s_\psi$ is trained in a \textit{perfect super-resolution} setting using coarsened observations $\mathbf{y}_{t,\ell}^{\downarrow}$ as input:
\begin{equation}
\widehat{\mathbf{y}}_{t,\ell} \sim s_\psi(\mathbf{y}_{t,\ell}^{\downarrow}).
\end{equation}
At inference, $s_\psi$ is applied to the bias-corrected field $\widetilde{\mathbf{x}}_{t,\ell}$ instead. Reversing the bias-correction and super-resolution steps would instead require generating a fine-scale version of the uncorrected coarse forecast before correcting biases at high resolution, which would make the bias-correction step more challenging to train and computationally expensive. \textbf{Strategy~2} (used as ablation baseline) trains a single generative model to jointly perform bias correction and super-resolution, mapping $\mathbf{x}_{t,\ell}$ directly to high-resolution output. \textbf{Strategy~3} (left for future work) would train both steps end-to-end with joint loss terms, potentially mitigating error propagation from the sequential approach, as it would eliminate the training/inference distribution shift at the super-resolution stage.
 
\smallskip
\noindent We adopt Strategy~1 for three reasons. First, decoupling the two steps allows us to explicitly isolate and evaluate the bias correction component, which constitutes the main methodological contribution of this work. Second, preliminary experiments show that lead-time-dependent biases in global forecasts hinder the training of a single end-to-end model (Strategy~2), whereas an explicit correction provides a cleaner input distribution for the CorrDiff-based Step~2. Third, Strategy~1 offers computational advantages: Step~2 operates on bias-corrected precipitation and static fields only, avoiding the memory overhead of processing all atmospheric channels. For completeness, we compare Strategy~1 to Strategy~2 at a representative 3-day lead time in Section~\ref{sec:results_strategy2}.

%%%%%%%%%%%%%%%%%%%
\subsection{Step 1: Lead-time-aware bias correction}
\label{sec:bias_correction}
%%%%%%%%%%%%%%%%%%%

The first step of \textit{SwAIther-Precip} is a deterministic bias correction model operating at coarse resolution. It maps low-resolution multi-channel global model predictions to low-resolution observed precipitation fields, correcting systematic errors while preserving large-scale spatial structure. We maintain the same notations previously defined for the different quantities in the framework.

\subsubsection{Bias correction model architecture}

The model is implemented as a convolutional U-Net with attention at lower levels, following the regression architecture of CorrDiff \citep{mardani2025residual}. This choice aims to capture both large-scale spatial patterns and localized corrections while modeling long-range dependencies via attention, which are particularly relevant for precipitation over complex terrain.

\smallskip
\noindent\textbf{Lead-time conditioning via FiLM.}
To account for the strong dependence of forecast biases on lead time, we condition the model on $\ell$ via Feature-wise Linear Modulation \citep[FiLM;][]{perez2018film}, mirroring standard time-step conditioning in diffusion models. The discrete lead time index is mapped to a continuous embedding $\mathbf{e}_\ell$, from which a small multilayer perceptron $\varphi$ produces channel-wise scale and shift parameters $(\boldsymbol{\gamma}_\ell, \boldsymbol{\beta}_\ell) = \varphi(\mathbf{e}_\ell)$ applied to the input forecast as
\begin{equation}
\bar{\mathbf{x}}_{t,\ell}
=
\mathrm{FiLM}(\mathbf{x}_{t,\ell} \mid \ell)
=
\mathbf{x}_{t,\ell} \odot (1 + \boldsymbol{\gamma}_\ell)
+
\boldsymbol{\beta}_\ell,
\label{eq:film}
\end{equation}
where $\odot$ denotes channel-wise multiplication. 
In practice, FiLM conditioning proved more effective than appending lead time as an additional input channel during early testing. It appears to match better lead time's scalar, spatially-uniform nature.

\smallskip
\noindent\textbf{Residual formulation.}
To anchor predictions to the original forecast and encourage the network to learn systematic corrections rather than reconstruct the full field, we adopt a residual formulation. Denoting by $\bar{p}_0 = \bar{\mathbf{x}}_{t,\ell}^{(\mathrm{prec})}$ the precipitation channel of the FiLM-modulated forecast, the final bias-corrected output is:
\begin{equation}
\widetilde{\mathbf{x}}_{t,\ell}
=
%\mathrm{clip}_{[0,\,20]}\!\left(
\bar{p}_0
+
\gamma_\mathrm{res} \cdot \mathrm{UNet}_\phi\!\left(\bar{\mathbf{x}}_{t,\ell}\right)
%\right),
\label{eq:biascorr}
\end{equation}
where $\gamma_\mathrm{res}$ is a learned scalar initialized at zero. Anchoring on $\bar{p}_0$ rather than on the raw forecast precipitation $\mathbf{x}_{t,\ell}^{(\mathrm{prec})}$ provides a useful inductive bias: when the U-Net residual is uninformative, the model reduces to a lead-time-aware affine recalibration of the raw forecast rather than to an identity copy, generalizing the standard residual setup \citep{he2016deep} to a lead-time-aware baseline. Output values are clipped to $[0, 20]$~mm/6\,h to enforce non-negativity and provide a numerical safeguard during training (cap value discussed in the Supplementary Material, Section 3).

\smallskip
\noindent The detailed architecture of the U-Net and the full Step 1 pipeline are presented in Appendix~\ref{app:archi}.

\subsubsection{Regularization for cross-lead-time consistency}
\label{subsec:regularization-cross-lead-time-consistency}

Multiple forecasts with different lead times may correspond to the same valid time $t$, in which case the bias-corrected field should be approximately invariant to $\ell$ for this given $t$. We enforce this via a regularization term penalizing discrepancies across lead times:
\begin{equation}
\mathcal{L}_{\text{time}}
=
\sum_{t \in \mathcal{T}}
\frac{1}{|\mathcal{L}_t|^2}
\sum_{\ell, \ell' \in \mathcal{S}_t}
\left\|
\widetilde{\mathbf{x}}_{t,\ell}
-
\widetilde{\mathbf{x}}_{t,\ell'}
\right\|_2^2,
\end{equation}
where $\mathcal{S}_t$ is the set of available lead times for a valid time $t$. This regularization term encourages consistency across lead times while still allowing the model to adapt to systematic differences in forecast skill as a function of $\ell$. It acts as a form of temporal smoothing and reduces spurious lead-time-dependent artifacts in the bias-corrected fields. In practice, samples sharing a valid time (from forecasts at different lead times) share an integer time identifier, and the consistency loss is computed within each mini-batch by grouping samples by this identifier. The train/test split is applied at the valid-time level so that same-valid-time samples are never mixed across splits, and the batch size (1024) is significantly larger than the number of lead times, ensuring strong pairwise coverage per batch.

\subsubsection{Bias-correction training objective}

The bias correction model is trained using a combination of a reconstruction loss at coarse resolution and the temporal regularization term introduced above. For each training sample $(t,\ell)$, we define a Huber loss between the bias-corrected output and the corresponding low-resolution observed precipitation:

\begin{equation}
\mathcal{L}_{\text{H}}(t,\ell)
=
\frac{1}{N}\sum_{i}
\left\{
\begin{array}{@{}l@{}l@{}}
\frac{1}{2}\left(\widetilde{x}_{t,\ell}^{(i)} - y_{t,\ell}^{\downarrow,(i)}\right)^2,
  & \text{if } \left|\widetilde{x}_{t,\ell}^{(i)} - y_{t,\ell}^{\downarrow,(i)}\right| \leq \delta, \\[6pt]
\delta \left(\left|\widetilde{x}_{t,\ell}^{(i)} - y_{t,\ell}^{\downarrow,(i)}\right| - \frac{1}{2}\delta\right),
  & \text{otherwise},
\end{array}
\right.
\end{equation}

where $\delta$ is the Huber threshold, $N = H \times W \times C_{\text{out}}$ is the total number of elements, and $\mathbf{y}_{t,\ell}^{\downarrow} \in \mathbb{R}^{H \times W \times C_{\text{out}}}$ denotes the observed precipitation field coarsened to the AIFS grid. Unless specified, we choose $\delta = 1$. The Huber loss acts as mean squared error for small residuals and transitions to a mean absolute error for large residuals, making the training more robust to outliers.
As opposed to an MSE loss, it limits the influence of a few high-intensity pixels on the gradient, given the residual heavy-tailed structure of the precipitation distribution after log-transformation.
The total training loss for the bias correction U-Net is then given by
\begin{equation}
\mathcal{L}_{\text{bias}}
=
\mathbb{E}_{(t,\ell) \sim \mathcal{D}}
\left[
\mathcal{L}_{\text{H}}(t,\ell)
\right]
+
\lambda \, \mathcal{L}_{\text{time}},
\end{equation}
where $\lambda$ is a hyperparameter controlling the strength of the cross-lead-time regularization.

\subsubsection{Curriculum learning and multi-stage fine-tuning}
\label{sec:multistep_finetuning}
 
Training a single model across all lead times is challenging because error characteristics vary strongly between short-range and medium-range forecasts. To improve convergence across lead times, we explore two progressive training strategies , both based on the idea of gradually exposing the model to longer (and more difficult) lead times.
 
\smallskip
\noindent \emph{Curriculum learning} \citep{bengio2009curriculum} operates within a single training phase: batches are filtered to include only a subset of lead times that expands at fixed epoch intervals (e.g., lead times 6--48\,h for the first 4 epochs, then 48--96\,h, etc.).
 
\smallskip
\noindent \emph{Multi-stage fine-tuning} \citep{siddiqui2024exploring} organizes training into multiple stages with separate hyperparameters. We adopted a three-stage scheme. Stage~1 trains on short-range lead times only (first third of lead times) with $\lambda = 0$, establishing a strong spatial representation. Stage~2 fine-tunes from the best Stage~1 checkpoint on medium-to-long-range lead times, with $\lambda = 0.3$, to adapt to the distinct bias patterns at longer lead times. Stage~3 fine-tunes on all lead times jointly with a smaller learning rate and $\lambda = 1$, harmonizing behavior across lead times while preserving the lead-time-specific corrections learned in earlier stages.

%%%%%%%%%%%%%%%%%%%
\subsection{Step 2: Generative super-resolution}
\label{sec:super_resolution}
%%%%%%%%%%%%%%%%%%%

The second step of \textit{SwAIther-Precip} focuses on spatial super-resolution using the CorrDiff framework \citep{mardani2025residual}, which combines a deterministic regression model with a conditional diffusion model. Starting from a low-resolution precipitation field that has already been bias-corrected, the goal is to reconstruct realistic high-resolution precipitation patterns consistent with observed fine-scale structure. Crucially, this step is not explicitly conditioned on forecast lead time: the upstream bias correction handles lead-time-dependent errors resolved at the coarse scale, while the super-resolution model learns the conditional distribution of fine-scale precipitation fields from observations. This super-resolution step can therefore correct residual fine-scale distributional biases consistent with the corrected coarse state, but not lead-time-dependent fine-scale errors that are not encoded in the coarse state.

\subsubsection{Super-resolution model architecture}
 
The architecture of the super-resolution layer is the same as defined by Corrdiff \citep{mardani2025residual}. The regression net $r_\psi$ is a UNet which upsamples the low-resolution input (precipitation and topographic elevation only) to produce an initial high-resolution estimate $\widehat{\mathbf{y}}^{\text{reg}}_{t,\ell} = r_\psi(\mathbf{y}_{t,\ell}^{\downarrow})$. A conditional diffusion model $d_\omega$ then generates stochastic residual corrections to restore fine-scale variability that the deterministic regression cannot capture:
\begin{equation}
\widehat{\mathbf{y}}_{t,\ell}
=
\widehat{\mathbf{y}}^{\text{reg}}_{t,\ell}
+
\Delta \mathbf{y}^{\text{diff}}_{t,\ell},
\qquad
\Delta \mathbf{y}^{\text{diff}}_{t,\ell}
\sim
d_\omega(\widetilde{\mathbf{x}}_{t,\ell}, \widehat{\mathbf{y}}^{\text{reg}}_{t,\ell}).
\end{equation}
Both models are trained independently following the standard CorrDiff protocol.

\subsubsection{Training and inference configurations}

During training, the super-resolution model operates in a \emph{perfect super-resolution} setting: low-resolution inputs are obtained by coarsening high-resolution observations, isolating the spatial refinement task from upstream forecast biases. At inference time, the model is instead applied to the bias-corrected fields $\widetilde{\mathbf{x}}_{t,\ell}$ produced by Step~1. This separation ensures that the super-resolution model never compensates for global model biases.

\subsubsection{Generative inference pipeline}
\label{sec:generation}

Since the super-resolution model is not lead-time-aware, inference is performed independently per lead time: for each lead time, the regression and diffusion networks are applied sequentially to the bias-corrected inputs. For each sample, $E$ ensemble members are generated using independent random seeds, producing stochastic realizations that share large-scale structure but differ in fine-scale detail. We note that sharing diffusion seeds across lead times has been proposed to encourage temporal coherence between lead times \citep{andrae2024continuous}; however, in our setting, independent seeds yielded better results, likely because the bias correction step already provides lead-time-specific conditioning that makes artificial seed-based coherence unnecessary. The disadvantage of our present framework is the lack of temporal consistency between individual ensemble members: the super-resolution samples are independent when conditioned on the bias-correction outputs. Model outputs, generated in log-transformed and min--max-normalized space, are transformed back to physical units (mm/6h) and clipped to non-negative values.

%%%%%%%%%%%%%%%%%%%
\subsection{Summary of the end-to-end inference pipeline}
\label{sec:end_to_end_mapping}
%%%%%%%%%%%%%%%%%%%

Figure~\ref{fig1} summarizes the full inference pipeline. Given a coarse-resolution forecast $\mathbf{x}_{t,\ell}$, the pipeline applies the following sequential transformations:
  
\begin{align}
\widetilde{\mathbf{x}}_{t,\ell}
&=
b_\phi(\mathbf{x}_{t,\ell}, \ell), \\
\widehat{\mathbf{y}}_{t,\ell}
&=
r_\psi(\widetilde{\mathbf{x}}_{t,\ell})
+
\Delta \mathbf{y}^{\text{diff}}_{t,\ell},
\qquad
\Delta \mathbf{y}^{\text{diff}}_{t,\ell}
\sim
d_\omega(\widetilde{\mathbf{x}}_{t,\ell}, \widehat{\mathbf{y}}^{\text{reg}}_{t,\ell}).
\end{align}
 
where $b_\phi$ is the lead-time-conditioned bias correction U-Net, $r_\psi$ the regression U-Net, and $d_\omega$ the conditional diffusion model. This decomposition explicitly separates lead-time-dependent bias correction from lead-time-agnostic spatial refinement.

%%%%%%%%%%%%%%%%%%%%%%%%%%%%%%%%%%%%%%%%%%%%%%%%%%%%%%%%%%%
% ---------------------------------------------------------
\section{Data}
\label{sec:data}
% ---------------------------------------------------------
%%%%%%%%%%%%%%%%%%%%%%%%%%%%%%%%%%%%%%%%%%%%%%%%%%%%%%%%%%%

\begin{figure*}[t]
%\hspace{-1cm}
\includegraphics[width=1\textwidth]{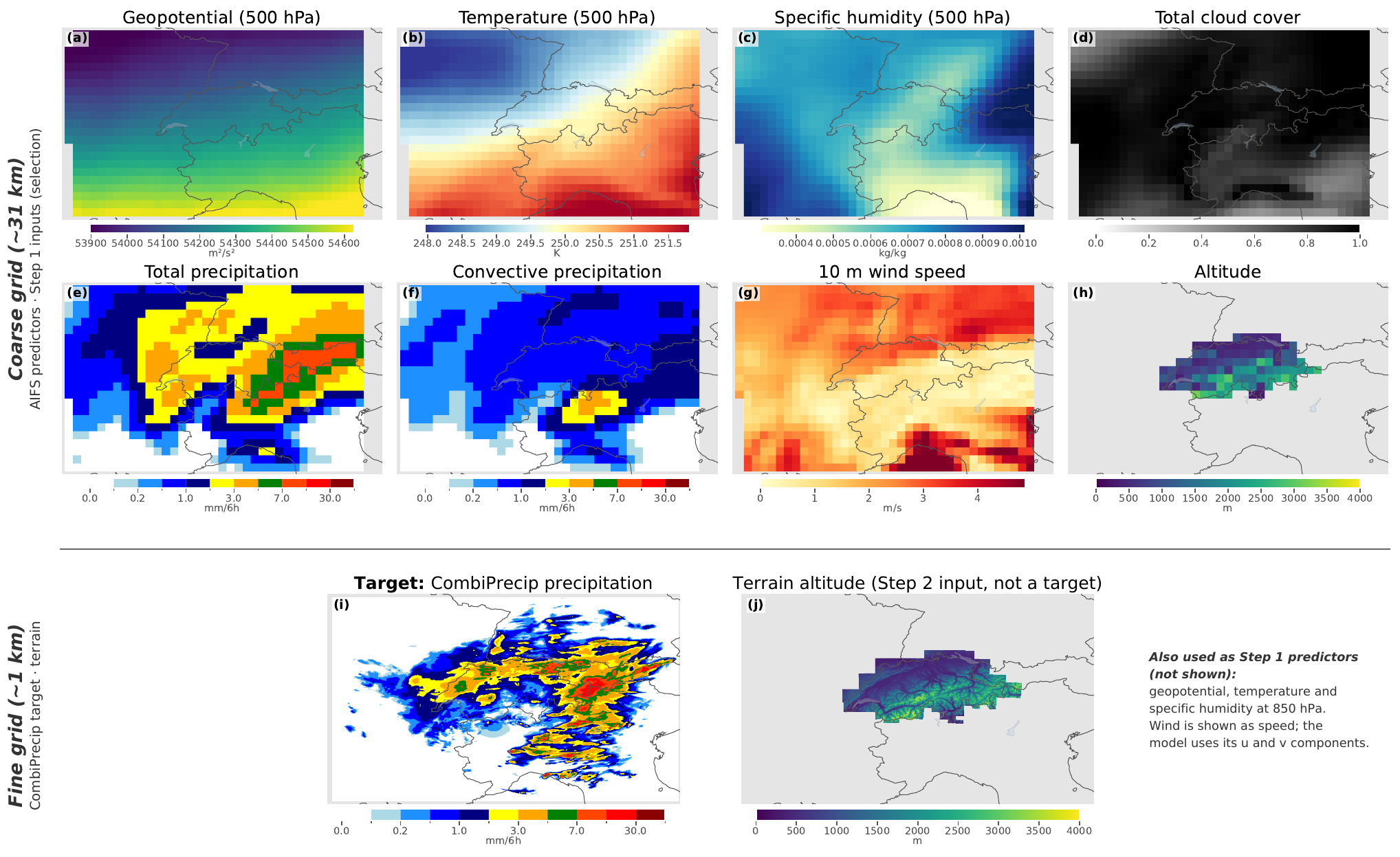}
\caption{\textbf{Resolution gap between the coarse predictors and the fine-scale target, for one example valid time (selection of predictors).} (a--h)~Coarse grid ($\sim$31\,km): a selection of the AIFS forecast fields used as predictors for the bias correction (Step~1): geopotential, temperature and specific humidity at 500\,hPa, total cloud cover, total and convective precipitation, 10\,m wind speed (the model uses the zonal and meridional wind
components), and altitude. Geopotential, temperature and specific humidity at 850\,hPa are also used as predictors but are not shown. (i)~Fine grid ($\sim$1\,km): the CombiPrecip 6\,h accumulated precipitation, the only target variable. (j)~Fine grid ($\sim$1\,km): terrain altitude from the SwissTopo DHM25 digital elevation model, used as an additional input of the super-resolution step (Step~2) and not a target. The resolution gap between the coarse AIFS grid ($23\times37$ points) and the CombiPrecip target ($640\times710$ points) corresponds to a $\sim$30$\times$ downscaling factor, illustrating the challenge of recovering fine-scale precipitation structure from global model output.}
\label{fig:data_overview}
\end{figure*}

\subsection{Target precipitation data}

The target dataset for the downscaling task is the MeteoSwiss CombiPrecip product \citep{sideris2014combiprecip}. This dataset combines measurements from a national weather radar system and rain gauge observations distributed across Switzerland. The radar–gauge combination provides the best estimate of ground-level sub-daily precipitation distribution currently available for Switzerland (as claimed by MeteoSwiss, https://www.meteoswiss.admin.ch/). The original CombiPrecip data are expressed in units of $\mathrm{mm/h}$.

\smallskip
\noindent The dataset provides precipitation fields at an approximate spatial resolution of $1\,\mathrm{km}$ and a temporal resolution of $10$ minutes, and is available continuously from 2005 onward. The dataset considers a spatial windows covering Switzerland and its immediate surroundings, corresponding to a grid of $640 \times 710$ pixels. %In this study, we slightly tighten up the window to a size of 608 $\times$ 640 so that the window does not include too many nan pixels.

\smallskip
\noindent To ensure temporal consistency with the global neural weather model forecasts, all CombiPrecip fields are aggregated into $6$-hour accumulated precipitation amounts (mm/6h) by summing the corresponding high-frequency observations. This temporal aggregation aligns the observational targets with the lead times considered in this work.

\smallskip
\noindent For the bias correction task, the same CombiPrecip observations are aggregated to the coarse AIFS grid by mass-conservative block-mean: for each coarse cell, we average all high-resolution pixels whose centers fall inside its footprint. This coarsened version of the observations serves as the low-resolution target for the deterministic bias correction model. The aggregation is performed such that the resulting grid corresponds approximately to a $31\,\mathrm{km}$ resolution, consistent with typical global forecasting systems.

\subsection{Input forecast data}

\subsubsection{AIFS global forecast inputs for bias correction}

The input data for Step~1 (bias correction) stage are forecasts from ECMWF's Artificial Intelligence Forecasting System \citep[AIFS;][]{lang2024aifs}, specifically the \texttt{AIFS-single-1.0} configuration (the deterministic version of the system), which offers improved precipitation skill over earlier versions through physics-informed constraints \citep{moldovan2025update}. Forecasts were generated using the official implementation\footnote{\url{https://huggingface.co/ecmwf/aifs-single-1.0}} with IFS analysis initial conditions, covering the period 2019--2023 at lead times $\ell \in \{6\text{h}, 12\text{h}, \dots, 144\text{h}\}$.
% We use AIFS-Single, the deterministic version of ECMWF's AI Forecasting System, whose most recent release introduces bounded activations for precipitation and shows substantial improvements in precipitation skill relative to the previous AIFS version 
% The available ensemble configuration was not used, as its deterministic precipitation scores are lower than those of the single-member model. 

\smallskip
\noindent From the large set of available AIFS outputs, we retain a subset of physically relevant variables summarized in Figure~\ref{fig:data_overview} (with geopotential, temperature and specific humidity at 850\,hPa additionally included). The full list of AIFS outputs can be found in the Supplementary Material in Section 1, together with the selected subset used as inputs for the SwAIther-Precip framework. This subset captures the key physical drivers of precipitation over complex terrain: precipitation fields provide the direct correction signal, near-surface winds inform orographic forcing, cloud cover indicates the condensation regime, and upper-air variables at 500 and 850\,hPa characterize large-scale steering flow, atmospheric stability, and moisture availability - two levels established as among the most informative for statistical precipitation downscaling \citep{hessami2008automated, wilby2002sdsm}. The 850\,hPa level is particularly relevant over Switzerland, where it intersects the Alpine crest and captures low-level moisture transport. Retaining a compact input set also keeps dimensionality manageable and limits overfitting to redundant channels.

\smallskip
\noindent All input fields are interpolated to a common coarse spatial grid covering Switzerland, with a spatial resolution of approximately $31\,\mathrm{km}$. Forecast fields are temporally aligned with the corresponding CombiPrecip observations at each valid time and lead time. 

\subsubsection{Inputs to the super-resolution model}

\noindent Both sub-networks of the Step~2, i.e. the regression and residual diffusion ones, operate on a restricted set of input channels. Specifically, the low-resolution inputs consist only of precipitation and static topographic information. We experimented with including all additional atmospheric variables predicted by the global neural model and used in the bias correction model, but found no consistent performance improvements at the spatial scales considered  (see Section~\ref{sec:results}\ref{sec:results_super_resolution}\ref{sec:step2_channels}). Consequently, these variables are omitted from Step~2. The output of the super-resolution model is the high-resolution precipitation field.

\subsection{Weekly Train-Test split}
\label{sec:train_test}

\noindent To preserve a realistic seasonal distribution of weather regimes in both splits and avoid the sensitivity to inter-annual distribution shifts that affects year-based partitions, we adopt a rolling weekly Train-Test split strategy inspired by Google's MetNet training scheme \citep{sonderby2020metnet, espeholt2022deep}, which employed a similar interleaved temporal splitting approach for precipitation forecasting. For each calendar week, the first six days are assigned to training and the seventh to testing; all lead times sharing a given valid time belong to the same split. Within each day, only the 06--12\,UTC and 18--00\,UTC accumulation windows are used, which introduces a minimum six-hour gap to the held-out windows (00--06\,UTC and 12--18\,UTC), reducing residual temporal autocorrelation while preserving the unused windows as an independent testbed for temporal generalization.

\smallskip
\noindent We do not reserve a separate validation set for hyperparameter tuning. Given the limited temporal coverage of the dataset (2019--2023), we prioritize maximizing both training and test data volume. Hyperparameters (learning rate, number of epochs, $\lambda$, curriculum schedule, see Supplementary Material, Section 3) were set based on standard practices in the literature and brief preliminary experiments; no systematic optimization (e.g., grid search or Bayesian tuning) was performed against test set metrics. The comparison of multiple training configurations (Section~\ref{sec:results}) should therefore be interpreted as an architectural and methodological study rather than a hyperparameter search.

\smallskip
\noindent Additional experiments validating these design choices - including autocorrelation analysis, evaluation on held-out time windows, and comparison with a yearly split - are presented in the Supplementary Material (Section 4), confirming that (i) temporal autocorrelation is not a significant confound, (ii) inter-year performance differences reflect distribution shift rather than leakage, and (iii) held-out time windows reflect intrinsic difficulty rather than a generalization gap.

%%%%%%%%%%%%
\subsection{Data preprocessing and postprocessing}
\label{sec:preprocessing}

\subsubsection{Preprocessing of AIFS precipitation forecasts}

The precipitation outputs produced by the global neural weather model (AIFS in our case) are first converted to a common physical unit and temporal resolution. All precipitation forecasts are expressed as accumulated precipitation in millimeters over a $6$-hour interval (mm/6h), consistent with the evaluation lead times considered in this study. Since negative precipitation values are physically meaningless and occasionally arise due to numerical artifacts in neural model outputs, all negative values are clipped to zero. This operation is applied consistently across all lead times and forecast instances.

\subsubsection{Preprocessing of CombiPrecip precipitation observations}

The MeteoSwiss Combi precipitation product is available at a higher temporal resolution than the global model forecasts. To ensure temporal alignment, the observations are aggregated into $6$-hour accumulated precipitation fields (mm/6h) by summing the corresponding hourly values.

\smallskip 
\noindent To avoid training the model on uninformative samples, we remove \emph{dry samples}, defined as precipitation fields in which more than $99.5\%$ of grid points exhibit zero precipitation. Such samples provide little learning signal for either bias correction or super-resolution and disproportionately dominate the dataset due to the sparsity of precipitation events. Filtering is observation-based only; AIFS false-alarm pixels remain abundant in the retained samples and contribute to the training loss.

\smallskip 
\noindent In addition, we mitigate the impact of extreme outliers, which are likely attributable to measurement or processing errors. For each precipitation field, we fit a Gamma distribution, a commonly used parametric model for precipitation, to the non-zero pixel values. Let $F_{\Gamma}$ denote the cumulative distribution function of the fitted Gamma distribution. We compute the $99.5^{\text{th}}$ percentile threshold
\[
\tau = F_{\Gamma}^{-1}(0.995),
\]
and all values exceeding this 99.5th-percentile threshold are capped (clipped) to the threshold value rather than removed, preserving the ordinal structure of the tail.
This procedure preserves the overall structure of intense precipitation events while preventing a small number of extreme values from dominating the learning process.

\subsubsection{Shared normalization and transformation}

Following the above operations, both input and target precipitation fields are transformed using a logarithmic variance-stabilizing transformation: $x \leftarrow \log(1 + x)$, which reduces skewness and mitigates the influence of large precipitation values. Note that clipping and the log transform play complementary roles: log-transform compresses the heavy tail, while clipping removes rare unrepresentative outliers that would still dominate after log-transformation.

\smallskip 
\noindent Finally, precipitation fields are normalized using min--max scaling computed over the training dataset, where $x_{\min}$ and $x_{\max}$ are determined exclusively from training samples and applied consistently during test and inference. All other AIFS input variables and the DHM25 altitude are min-max normalized in the same way, but without the log transform, since their distributions are not heavy-tailed. This normalization ensures numerical stability during optimization and consistent scaling across datasets. Percentile-based min-max scaling is preferred over z-normalization as it preserves the physical non-negativity of precipitation and is robust to its heavy-tailed distribution.

\subsubsection{Postprocessing of downscaled forecasts}
\label{subsec:postprocessing}

Throughout this study, a wet/dry threshold of 0.1\,mm/6h is used, as 0.1mm corresponds to the minimum reporting resolution of WMO-standard rain gauges (WMO, 2018, No.~8, Ch.~6), regardless of the accumulation time. The high-resolution diffusion samples undergo two postprocessing operations: (1)~values are masked to regions where the bias-corrected low-resolution field indicates precipitation, enforcing spatial consistency between the two steps of the pipeline; (2)~the 0.1\,mm/6h threshold is applied to suppress residual drizzle artifacts. The same threshold is applied to observations for consistent evaluation. At Step~1 no thresholding is applied, as the untouched field is passed directly as input to super-resolution.

%%%%%%%%%%%%%%%%%%%%%%%%%%%%%%%%%%%%%%%%%%%%%%%%%%%%%%%%%%%
% ---------------------------------------------------------
\section{Evaluation framework}
\label{sec:eval}
% ---------------------------------------------------------
%%%%%%%%%%%%%%%%%%%%%%%%%%%%%%%%%%%%%%%%%%%%%%%%%%%%%%%%%%%

%%%%%%%%%%%%%%%%%%%%%%%
\subsection{Baseline models}
\label{subsec:baselines}
%%%%%%%%%%%%%%%%%%%%%%%

To quantify the benefits of the proposed two-step approach, we compare against classical post-processing baselines and architectural alternatives, summarized in Table~\ref{tab:baselines}. All baselines are evaluated at coarse resolution for the bias correction task; for the full downscaling pipeline, only a selected subset is evaluated since performance is dominated by the upstream bias correction quality. Let $\mathbf{x}^{(\text{prec})}_{t,\ell} \in \mathbb{R}^{H \times W}$ denote the raw coarse-resolution precipitation forecast and $\mathbf{y}^{\downarrow}_{t,\ell} \in \mathbb{R}^{H \times W}$ the corresponding observational target.

\begin{table}[t]
\centering
\caption{Baseline models for bias correction evaluation}
\label{tab:baselines}
\resizebox{\columnwidth}{!}{%
\begin{tabular}{ll}
\toprule
Model & Description \\
\midrule
B0 (raw AIFS) & Unprocessed AIFS precipitation prediction \\
B1 & Global affine rescaling (one gain and offset for the whole domain and all leads) \\
B2 & Affine rescaling fitted per lead time \\
B3 & Per-pixel, per-lead affine rescaling on precipitation only \\
B4 & Per-pixel multivariate linear regression (ridge) on all predictors \\
B5 & Per-pixel per-lead quantile mapping \\
B7 & Global multivariate ridge regression \\
\midrule
ViT + FiLM & Vision Transformer with lead-time conditioning \\
CorrDiff reg. & Regression-only direct downscaling (Strategy~2 proxy) \\
Single-lead UNets & Independent UNets at 6\,h, 3\,days and 6\,days, without lead-time conditioning \\
\bottomrule
\end{tabular}%
}
\end{table}

\begin{table*}[t]
\centering
\caption{Model configurations evaluated in this study. $\lambda$ is the strength of the cross-lead-time consistency term (Section 2c(2)); Curri. denotes curriculum learning and Multi-step FT the multi-stage fine-tuning of Section 2c(4). Note that for the Multi FT, the value of $\lambda$ changes in each step of the fine tuning, hence the multiple values. Please refer to the supplementary material for the full list of steps and $\lambda$ values. A \emph{multi-lead} model is conditioned on lead time via FiLM and trained jointly on all 24 lead times; a \emph{single-lead} model is trained on one lead time only, without lead-time conditioning.}
\label{tab:configs}
\resizebox{\textwidth}{!}{%
\begin{tabular}{llcccl}
\toprule
Name (text and Figs.~3--5) & $\lambda$ & Curri. & Multi-step FT & Lead-time cond. & Purpose \\
\midrule
UNet$_{\lambda=0}$ & 0   & no  & no  & FiLM, 24 leads & Reference multi-lead model. Baseline for the temporal-consistency term \\
UNet$_{\lambda=0}$, Curri & 0   & yes & no  & FiLM, 24 leads & Isolates curriculum learning from the temporal-consistency term \\
UNet$_{\lambda=2}$ & 2   & no  & no  & FiLM, 24 leads & Isolates the cross-lead-time consistency regularization \\
UNet$_{\lambda=2}$, Curri  & 2   & yes & no  & FiLM, 24 leads & Tests whether curriculum and temporal regularization are complementary \\
UNet Multi FT  & 0, 0.3, 1 & no  & yes & FiLM, 24 leads & Tests progressive multi-stage fine-tuning across lead-time ranges \\
ViT & 0   & yes & no  & FiLM, 24 leads & Architectural alternative to the U-Net, protocol of UNet$_{\lambda=0}$, Curri \\
\midrule
UNet single (6\,h / 3\,d / 6\,d) & --- & no & no & none & Single-lead models; quantify the benefit of joint multi-lead training \\
UNet Strat2 proxy (3\,d) & --- & --- & --- & none & CorrDiff regression from AIFS directly to 1\,km: Strategy~2 proxy. \\
End-to-end CorrDiff (3\,d) & --- & --- & --- & none & Full CorrDiff from AIFS directly to 1\,km: Test full Strategy 2. \\
\bottomrule
\end{tabular}%
}
\end{table*}

\smallskip 
\noindent The deep learning baselines, which require more details, are as follows: 
 
\smallskip
\noindent \textit{Vision Transformer (ViT) with FiLM conditioning.} An architectural alternative to the U-Net: the FiLM-conditioned input is split into non-overlapping patches and processed by a Transformer encoder with multi-head self-attention, and decoded via transposed convolutions to reconstruct the coarse-resolution precipitation field. We include a ViT variant to test whether its global receptive field would better capture long-range spatial dependencies than the local U-Net convolutions. During training, ViTs indeed exhibited a smaller train/validation gap, motivating their inclusion as a candidate architecture. The ViT variant was trained under the same recipe as the \textit{UNet, $\lambda=0$, Curri} configuration to isolate the architectural change from other hyperparameter differences.

\smallskip
\noindent \textit{CorrDiff regression-only (Strategy~2 proxy).} The regression component of CorrDiff trained directly from coarse forecasts to high-resolution precipitation without diffusion refinement, testing whether a single-step deterministic mapping can substitute for the two-step decomposition.

\smallskip
\noindent \textit{End-to-end CorrDiff (full Strategy~2, single lead).}
The CorrDiff regression-only model above is a proxy for Strategy~2 that omits the
diffusion stage. To evaluate Strategy~2 directly, we also trained a complete CorrDiff
model end-to-end, from the AIFS predictors (11 variables and altitude, bilinearly
interpolated to the 1~km grid) to CombiPrecip, without an explicit bias-correction step.
Its regression stage is the regression-only model above. Because conditioning the
diffusion stage on 12 channels over the full 1~km domain is computationally prohibitive,
its diffusion stage is trained on random $448 \times 448$ patches with global
low-resolution conditioning, following the patch-based training of CorrDiff for large
domains. Owing to this cost, the model is trained at a single lead time (3~days), with the
same architecture size, optimization settings and batch size (2) as our Step~2 diffusion
model, and for the same wall-clock time (about 30~h on one NVIDIA GH200 GPU). We compare
it with the 3-day single-lead specialist of Strategy~1 (single-lead bias-correction U-Net
followed by Step~2), generating 12-member ensembles with the same stochastic sampler for
both models. As Strategy~2 produces no bias-corrected low-resolution field, its
low-resolution mask in post-processing (Section~\ref{subsec:postprocessing}) is computed
from the raw AIFS precipitation, as for the AIFS raw baseline.
 
\smallskip
\noindent \textit{Single-lead U-Net models.} Separate U-Net models trained independently at 6\,h, 3\,days, and 6\,days, sharing the same architecture as the multi-lead U-Net but without lead-time conditioning, to assess the benefit of joint multi-lead training.

%%%%%%%%%%%%%%%%%%%%%%%
\subsection{Forecast verification metrics}
%%%%%%%%%%%%%%%%%%%%%%%

\begin{table}[t]
\centering
\caption{Summary of evaluation metrics}
\label{tab:metrics_summary}
\resizebox{0.5\textwidth}{!}{%
\begin{tabular}{llccl}
\toprule
Category & Metric & Range & Optimal & Purpose \\
\midrule
Point-wise & MSE & $[0, \infty)$ & 0 & Error magnitude \\
 & MSE\textsubscript{wet} & $[0, \infty)$ & 0 & Error for rain events \\
\midrule
Categorical & CSI & $[0, 1]$ & 1 & Overall detection skill \\
\midrule
Spatial & FSS & $[0, 1]$ & 1 & Scale-dependent skill \\
 & SAL-S & $[-2, 2]$ & 0 & Structure error \\
 & SAL-A & $[-2, 2]$ & 0 & Amplitude error \\
 & SAL-L & $[0, 2]$ & 0 & Location error \\
\midrule
Ensemble & CRPS & $[0, \infty)$ & 0 & Overall probabilistic skill \\
 & avFSS & $[0, 1]$ & 1 & Scale-dependent skill \\
 & eS & $[-2, 2]$ & 0 & Median Structure error \\
 & eA & $[-2, 2]$ & 0 & Median Amplitude error \\
 & eL & $[0, 2]$ & 0 & Median Location error \\
\bottomrule
\end{tabular}%
}
\end{table}

We evaluate bias correction and downscaling performance using metrics spanning four complementary aspects of forecast quality. Full definitions of metrics and implementation details are provided in the Supplementary Material (Section 6). The same 0.1\,mm/6h threshold is used to define wet pixels for all event-based and spatial metrics.

\smallskip
\noindent \textbf{Pointwise and categorical metrics.} We report MSE restricted to wet pixels (MSE\textsubscript{wet}, with a threshold 0.1\,mm/6h, as mentioned in Section \ref{sec:data}\ref{sec:preprocessing}) to focus on precipitation events rather than the dominant dry fraction, and the Critical Success Index (CSI) for binary event detection skill.

\smallskip
\noindent \textbf{Spatial verification.} The Fractions Skill Score \citep[FSS;][]{roberts2008scale} evaluates spatial agreement at multiple neighborhood sizes ($n = 1, 5, 17$, corresponding to pixel-level, $\sim$25\,km, and $\sim$85\,km scales). For ensemble forecasts we use the member-averaged FSS \citep[avFSS;][]{necker2024fss}. The Structure--Amplitude--Location diagnostic \citep[SAL;][]{wernli2008sal} decomposes forecast error into shape, intensity, and displacement components; for ensembles we report the median across members \citep[eSAL;][]{radanovics2018verification}.

\smallskip
\noindent \noindent \textbf{Probabilistic verification.} The Continuous Ranked Probability Score \citep[CRPS;][]{hersbach2000decomposition} provides a strictly proper measure of ensemble skill. Ensemble calibration is assessed via randomized Probability Integral Transform histograms \citep{gneiting2007strictly}, with departure from uniformity quantified by KL divergence.

\smallskip
\noindent \textbf{Spectral evaluation.} Following \cite{sinclair2005empirical, pulkkinen2019pysteps}, we compute radially-averaged power spectral densities (PSD) of co-masked log-precipitation fields and report band-averaged spectral ratios (predicted/observed PSD) at large (100--600\,km), meso (20--100\,km), and small (2--20\,km) scales. The effective resolution is defined as the wavelength where the spectral ratio drops below 0.5 \citep{klaver2020effective}.

%%%%%%%%%%%%%%%%%%%%%%%
\subsection{Uncertainty estimates}
%%%%%%%%%%%%%%%%%%%%%%%

We quantify the sampling uncertainty of the aggregated bias-correction metrics with a
moving-block bootstrap over the test times (1000 replicates), resampling all lead
times of a valid time together and using blocks of four consecutive test times (about two weeks) to account for the persistence of weather regimes. Since the same resamples are applied to all models, differences between models are assessed with paired 95\% confidence intervals; a difference is called significant when its interval excludes zero. Details and full results are given in Supplementary Section S7.

%%%%%%%%%%%%%%%%%%%%%%%
\subsection{Verification protocol and Implementation details}
%%%%%%%%%%%%%%%%%%%%%%%

\noindent All metrics are computed separately for each lead time, allowing us to assess how model performance degrades with increasing lead time. We evaluate 24 lead times ranging from 6 hours to 6 days (144 hours) at 6-hour intervals. Additionally, we report metrics averaged across all lead times to provide a summary measure of overall model performance. 

\noindent For all probabilistic metrics, we generate $M = 12$ ensemble members from the diffusion model for each forecast initialization. Metrics are computed at each lead time separately and additionally averaged across all lead times. Table~\ref{tab:metrics_summary} summarizes all evaluation metrics used in this study, their optimal values, and their primary purpose.

%%%%%%%%%%%%%%%%%%%%%%%%%%%%%%%%%%%%%%%%%%%%%%%%%%%%%%%%%%%
% ---------------------------------------------------------
\section{Results}
\label{sec:results}
% ---------------------------------------------------------
%%%%%%%%%%%%%%%%%%%%%%%%%%%%%%%%%%%%%%%%%%%%%%%%%%%%%%%%%%%

\begin{figure*}[t]
%\hspace{-1cm}
\includegraphics[width=0.9\textwidth]{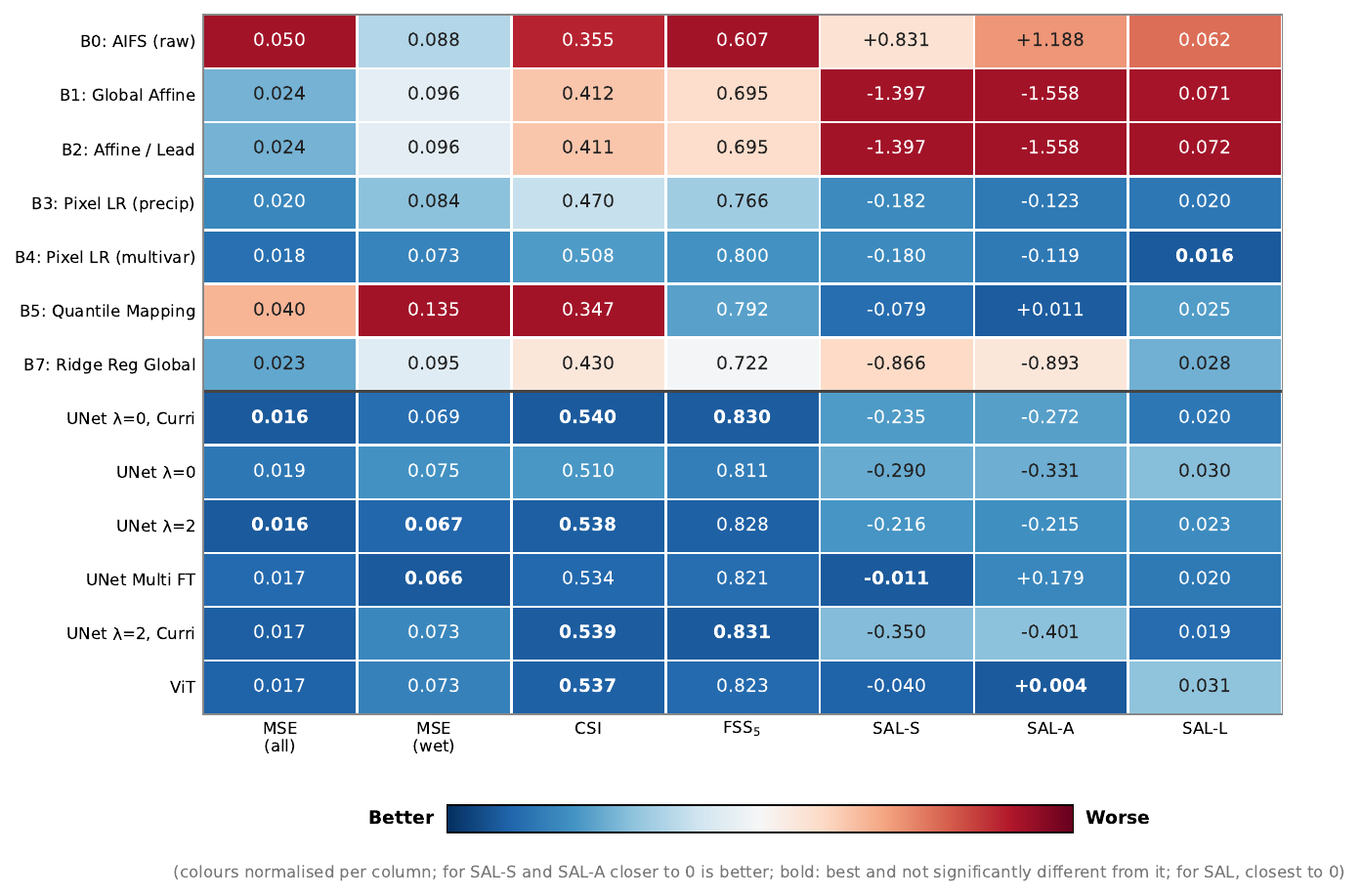}
\caption{\textbf{Bias Correction (Step 1): General model comparison, for deterministic metrics, averaged across all lead times} Colors are normalized independently per column. For SAL-S and SAL-A, values closer to zero indicate better performance. See Figure \ref{tab:step1_combined} for the lead-time-resolved view of key metrics. Bold: best value of each column and values not significantly different from it (paired moving-block bootstrap, 95\%; Supplementary Section S7).
}
\label{tab:metrics_compact}
\end{figure*}

This section reports results for (i) the coarse-resolution bias correction task and (ii) the full downscaling pipeline. All scores follow the evaluation protocol described in Sections \ref{sec:data}\ref{sec:train_test} and \ref{sec:data}\ref{sec:preprocessing}, and are computed per lead time unless stated otherwise.

\smallskip
\noindent We evaluate several training configurations of the lead-time-aware bias correction U-Net, varying the temporal regularization strength $\lambda$ (Section~\ref{sec:bias_correction}), the use of curriculum learning, and whether multi-stage fine-tuning is applied (Section~\ref{sec:multistep_finetuning}). The complete list of variants is presented in Table \ref{tab:configs}. Model names encode these choices : e.g., UNet$_{\lambda=2}$, Curri denotes $\lambda=2$ with curriculum learning and no multi-stage fine-tuning; UNet Multi FT denotes the multi-stage fine-tuned UNet model (Section~\ref{sec:multistep_finetuning}). We also include a ViT baseline trained with the same protocol as UNet$_{\lambda=0}$, Curri. All multi-lead models are conditioned on lead time via FiLM and trained jointly across all 24 lead times (6\,h to 144\,h). Single-lead U-Nets trained independently at 6\,h, 3\,d, and 6\,d, as well as a Strategy~2 baseline (Section~\ref{sec:results_strategy2}), are included for comparison.
% The configurations compared below, their abbreviated names and the reason each was tested are defined in Section~\ref{sec:eval}\ref{subsec:baselines} (Tables~\ref{tab:baselines} and~\ref{tab:configs}).

%%%%%%%%
\subsection{Results for lead-time-aware bias correction (Step~1)}
\label{sec:results_bias_correction}

\begin{sidewaysfigure*}
\centering
\includegraphics[width=1\textheight]{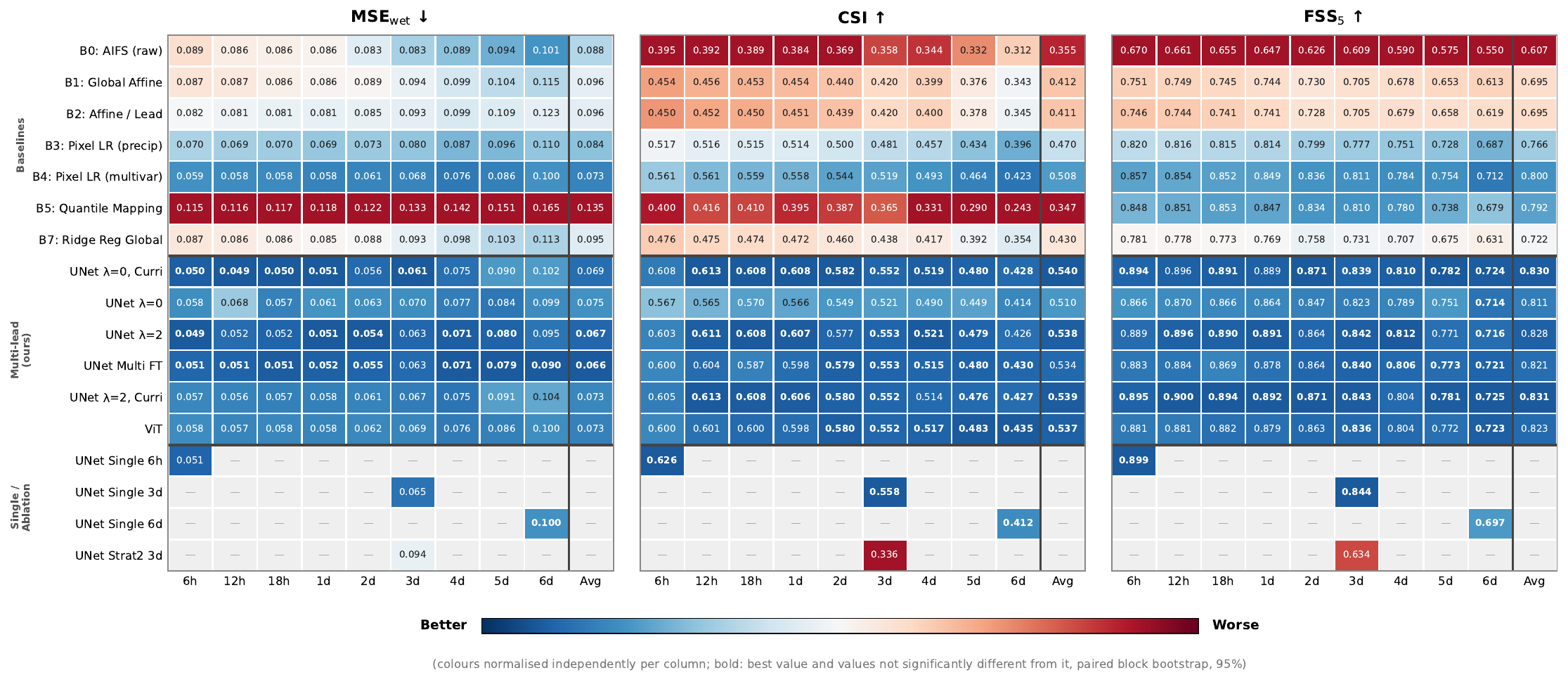}
\caption{\textbf{Bias correction (Step~1): MSE (wet pixels), CSI, and FSS$_5$ by lead time.} Models are grouped into classical baselines (top), multi-lead learned models (middle), and single-lead / ablation models (bottom) (see Section 5a3 for discussion). Colors are normalized independently per column to highlight relative model ranking within each lead time and metric. Bold: best value of each column and values not significantly different from it (paired moving-block bootstrap, 95\%; Supplementary Section S7).
}
\label{tab:step1_combined}
\end{sidewaysfigure*}

\subsubsection{Performance averaged across lead times}

Figure \ref{tab:metrics_compact} summarizes all metrics averaged across lead times. Without post-processing, raw AIFS (B0) yields average MSE\_wet of 0.088. Global affine corrections (B1--B2) do not improve MSE$_{wet}$ and severely distort precipitation structure (SAL-S\,$\approx$\,$-1.40$, SAL-A\,$\approx$\,$-1.56$). Among classical baselines, the strongest is B4 (per-pixel multivariate regression), achieving MSE\_wet of 0.073 - an 18\% reduction over B0 - highlighting the importance of both local calibration and additional atmospheric predictors. All learned models outperform B4 in CSI and FSS$_5$, and the best ones also in MSE: the lowest average MSE\_wet is reached by UNet Multi FT (0.066, a 26\% reduction over B0 and 10\% over B4), statistically indistinguishable from UNet$_{\lambda=2}$ (0.067). Combining curriculum learning with temporal regularization (UNet$_{\lambda=2}$, Curri) does not improve over UNet$_{\lambda=2}$ alone (0.073 vs.\ 0.067), suggesting partial redundancy between the two mechanisms. The ViT achieves 0.073, on par with B4 and UNet$_{\lambda=2}$, Curri, and ahead of UNet$_{\lambda=0}$ (0.075), but behind the best UNet variants.

% Figure \ref{tab:metrics_compact} summarizes all metrics averaged across lead times. Without post-processing, raw AIFS (B0) yields average MSE\_wet of 0.088. Global affine corrections (B1--B2) do not improve MSE and severely distort precipitation structure (SAL-S\,$\approx$\,$-1.38$, SAL-A\,$\approx$\,$-1.54$). Among classical baselines, the strongest is B4 (per-pixel multivariate regression), achieving MSE\_wet of 0.073 - a 17\% reduction over B0 - highlighting the importance of both local calibration and additional atmospheric predictors. All learned models substantially outperform B4: the best variant, UNet Multi FT, reaches average MSE\_wet of 0.065 (26\% reduction over B4). Combining curriculum learning with temporal regularization (UNet$_{\lambda=2}$, Curri) does not improve over UNet$_{\lambda=2}$ alone (0.072 vs.\ 0.066), suggesting partial redundancy between the two mechanisms. The ViT achieves 0.071, comparable to UNet$_{\lambda=0}$ but behind the best UNet variants.

 \smallskip
\noindent For event detection and spatial skill, the pattern is consistent. UNet$_{\lambda=0}$, Curri achieves the highest average CSI (0.540), while UNet$_{\lambda=2}$, Curri leads in FSS$_5$ (0.831). All learned models exceed 0.81 in FSS$_5$, compared to 0.800 for B4 and 0.607 for raw AIFS. Interestingly, CSI rankings differ slightly from MSE: UNet$_{\lambda=0}$, Curri leads in CSI despite not being the best in MSE, suggesting that curriculum learning may benefit event detection more than squared error reduction, although the differences among the best variants are small and largely within the sampling uncertainty.

% \smallskip
% \noindent For event detection and spatial skill, the pattern is consistent. UNet$_{\lambda=0}$, Curri achieves the highest average CSI (0.541), while UNet$_{\lambda=2}$, Curri leads in FSS$_5$ (0.832). All learned models exceed 0.81 in FSS$_5$, compared to 0.800 for B4 and 0.607 for raw AIFS. Interestingly, CSI rankings differ slightly from MSE: UNet$_{\lambda=0}$, Curri leads in CSI despite not being the best in MSE, suggesting that curriculum learning may benefit event detection more than squared error reduction.

The SAL diagnostic (Figure~\ref{tab:metrics_compact}) confirms these trends: raw AIFS shows strong positive structure and amplitude biases (overly broad and intense fields), which spatially varying baselines (B3--B4) and all learned models reduce to near-neutral values. UNet Multi FT stands out with a near-zero structure and the largest positive amplitude bias (SAL-A\,$=$\,+0.18), foreshadowing the bias-variance trade-off analyzed in Section~\ref{sec:discussion}. Location errors remain small across all learned models (SAL-L\,$\leq$\,0.031).

% The SAL diagnostic (Figure~\ref{tab:metrics_compact}) confirms these trends: raw AIFS shows strong positive structure and amplitude biases (overly broad and intense fields), which spatially varying baselines (B3--B4) and all learned models reduce to near-neutral values. UNet Multi FT stands out with a near-zero structure and the only positive amplitude bias, foreshadowing the bias-variance trade-off analyzed in Section~\ref{sec:discussion}. Location errors remain small across all learned models (SAL-L\,$\leq$\,0.027).
 
\subsubsection{Lead-time dependence of skill}

%  \begin{figure*}[t!]
% \includegraphics[width=1\textwidth]{Results_Plot_BiasCorrection_ByLead_MSE_wet.png}
% {\caption{\textbf{Bias correction (Step~1): MSE over wet pixels ($>0.1$\,mm/6h) as a function of forecast lead time.} Results are shown for multi-lead UNet and ViT variants, all baselines, and single-lead models evaluated at their respective target lead times. Confidence intervals and paired differences between models are given in Supplementary Figure S14.}
% \label{BC_plot_WetMSEPerLead}}
% \end{figure*}

\begin{figure*}[t!]
\includegraphics[width=1\textwidth]{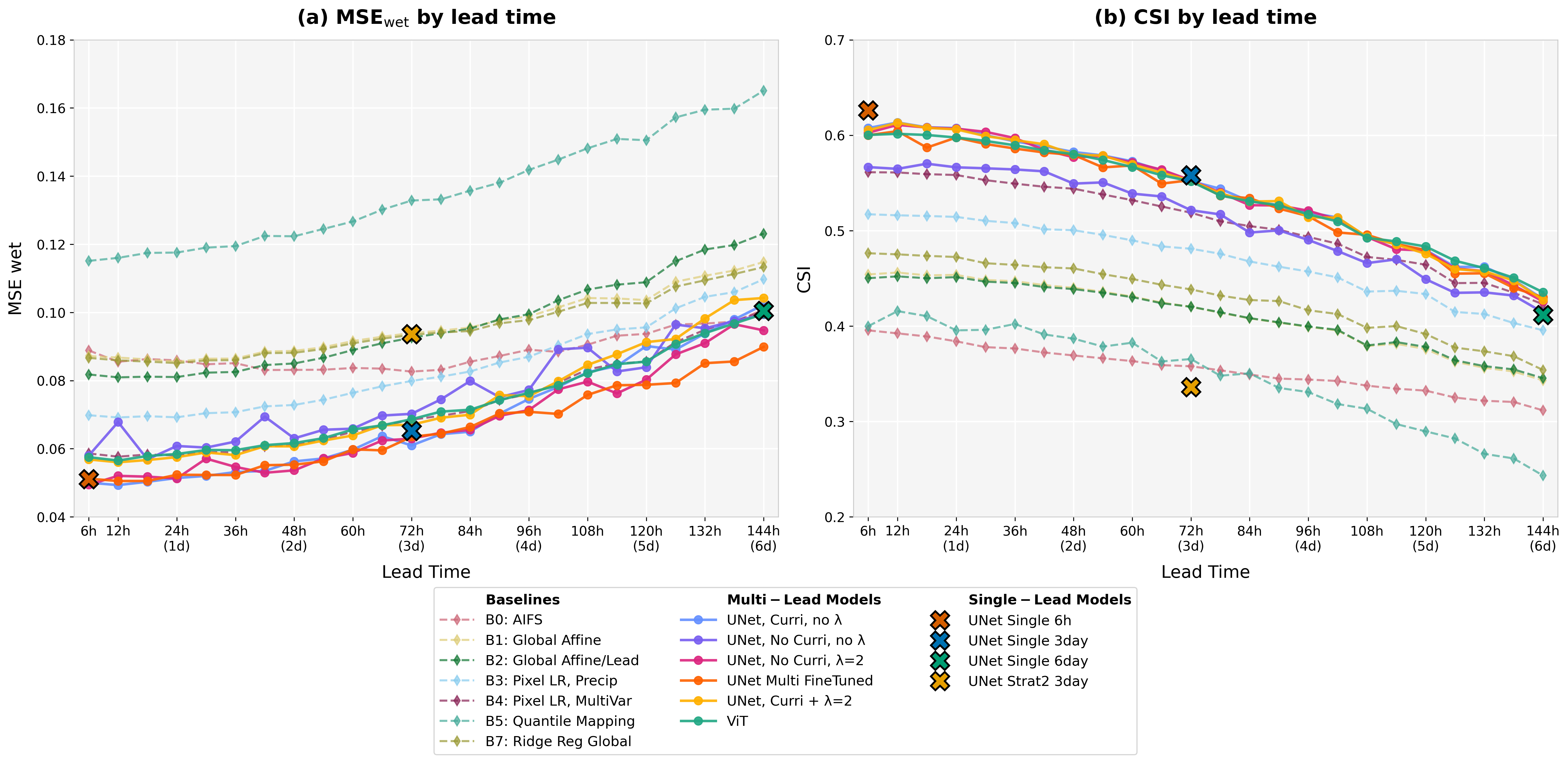}
{\caption{\textbf{Bias correction (Step~1) as a function of forecast lead time: (a) MSE over wet pixels ($>0.1$\,mm/6h) and (b) critical success index (CSI, threshold
0.1\,mm/6h).} Results are shown for multi-lead UNet and ViT variants, all baselines, and single-lead models evaluated at their respective target lead times (crosses). Confidence intervals and paired differences between models are given in Supplementary Figure S14.}
\label{BC_plot_WetMSEPerLead}}
\end{figure*}

Figure~\ref{BC_plot_WetMSEPerLead} provides a view of bias correction performance across all lead times, for MSE\_wet (a) and CSI (b). The best deep-learning models are significantly better than every baseline on all four metrics (e.g.\ UNet$_{\lambda=2}$), and the temporal-consistency term yields a significant improvement over the model trained without it (UNet$_{\lambda=2}$ vs.\ UNet$_{\lambda=0}$). In contrast, the differences among the best multi-lead variants are small and mostly within the sampling uncertainty (Supplementary Section S7). Figure~\ref{tab:step1_combined} further details MSE\_wet, CSI, and FSS$_5$ at representative lead times. All models degrade with increasing lead time, as expected.

% Figure~\ref{BC_plot_WetMSEPerLead} provides a view of bias correction performance across all lead times, showing that all deep-learning models are significantly better than every baseline on all four metrics, and the temporal-consistency term yields a significant improvement over the model trained without it (UNet$_{\lambda=2}$ vs.\ UNet$_{\lambda=0}$). In contrast, the differences among the best multi-lead variants are small and mostly within the sampling uncertainty (Supplementary Section S7). Figure~\ref{tab:step1_combined} further details MSE\_wet, CSI, and FSS$_5$ at representative lead times. All models degrade with increasing lead time, as expected.

Among the UNet configurations, no single variant dominates across all 
lead times and metrics. At short-to-medium lead times (6\,h to 2\,days), 
several configurations  (UNet$_{\lambda=0}$ (Curri), UNet$_{\lambda=2}$, and UNet Multi FT) perform within noise of one  another on MSE\_wet, often with near-ties (e.g., 0.051--0.052 at 1\,d). UNet$_{\lambda=0}$ (Curri) is strongest on CSI in this regime, while UNet$_{\lambda=2}$ (Curri) leads on FSS$_5$ at most lead times. At longer lead times (4--6\,days), the picture is 
clearer: UNet Multi FT consistently achieves the lowest MSE\_wet 
(0.090 at 6\,d vs.\ 0.095 for UNet$_{\lambda=2}$), suggesting that 
the multi-stage fine-tuning strategy particularly benefits long-range 
error correction, where per-lead signal is weakest and the model 
benefits most from statistical strength borrowed from shorter leads.

% Among the UNet configurations, no single variant dominates across all 
% lead times and metrics. At short-to-medium lead times (6\,h to 2\,days), 
% several configurations  (UNet$_{\lambda=0}$ (Curri), 
% UNet$_{\lambda=2}$, and UNet Multi FT) perform within noise of one 
% another on MSE\_wet, often with multiple ties (e.g., three-way tie at 
% 0.050 at 1\,d). UNet$_{\lambda=0}$ (Curri) is strongest on CSI in 
% this regime, while UNet$_{\lambda=2}$ (Curri) leads on FSS$_5$ at 
% most lead times. At longer lead times (4--6\,days), the picture is 
% clearer: UNet Multi FT consistently achieves the lowest MSE\_wet 
% (0.085 at 6\,d vs.\ 0.090 for UNet$_{\lambda=2}$), suggesting that 
% the multi-stage fine-tuning strategy particularly benefits long-range 
% error correction, where per-lead signal is weakest and the model 
% benefits most from statistical strength borrowed from shorter leads.

The ViT configuration is broadly competitive, particularly at longer 
lead times, where it reaches the highest CSI at 5\,days (0.483) and 6\,days (0.435), although by small margins. This suggests that the ViT architecture's global attention may offer some 
advantage in the long-lead regime, though the best UNet configurations remain preferable overall when averaged across metrics.

% The ViT configuration is broadly competitive, particularly at longer 
% lead times (e.g., CSI 0.492 at 5\,days, matching the best UNet, and 
% 0.432 at 6\,days, tied with UNet$_{\lambda=0}$ Curriculum). This 
% suggests that the ViT architecture's global attention may offer some 
% advantage in the long-lead regime, though the best UNet configurations remain preferable overall when averaged across metrics.

% Figure~\ref{BC_plot_WetMSEPerLead} provides a comprehensive view of bias correction performance across all lead times, clearly illustrating the consistent advantage of learned models over classical baselines at every horizon. Figure \ref{tab:step1_combined} further details MSE\_wet, CSI, and FSS$_5$ at representative lead times. All models degrade with increasing lead time, but Unet models maintain a consistent advantage over baselines across all lead times (Figures~\ref{BC_plot_WetMSEPerLead} achieves the lowest MSE at the short lead times). UNet$_{\lambda=2}$ achieves the lowest MSE at short leads (0.048 at 6\,h), while UNet Multi FT is strongest at extended ranges (0.085 at 6\,d vs.\ 0.090 for UNet$_{\lambda=2}$), suggesting that multi-stage fine-tuning improves long-range error accumulation. 
% The ViT is competitive at the longest lead times (CSI of 0.432 at 6\,d), suggesting robustness to increased input uncertainty at extended ranges. 
 
\subsubsection{Multi-lead versus single-lead training}

To assess whether explicit lead-time conditioning is preferable to training one model per lead time, we compare multi-lead models to single-lead UNets trained independently at 6\,h, 3\,d, and 6\,d (Figure~\ref{tab:step1_combined}). The advantage of multi-lead training grows with lead time. At 6\,h, the single-lead model achieves significantly higher CSI (0.626 vs.\ 0.608) and FSS$_5$ (0.899 vs.\ 0.895) than the best multi-lead models, while its MSE\_wet is slightly higher (0.051 vs.\ 0.049). At 3 days, performance is comparable across approaches. At 6 days, however, the multi-lead models are ahead: the single-lead UNet reaches only 0.101 MSE\_wet and 0.412 CSI, compared to 0.090 and 0.430 for UNet Multi FT, although the CSI and FSS$_5$ differences are within the sampling uncertainty. This pattern suggests that single-lead training provides marginal benefits at short lead times, but that the broader training distribution of multi-lead models provides regularization that is increasingly beneficial at longer lead times. Operationally, a single model that remains competitive with or surpasses dedicated single lead models is also far more practical than maintaining a separate model per lead time.

% To assess whether explicit lead-time conditioning is preferable to training one model per lead time, we compare multi-lead models to single-lead UNets trained independently at 6\,h, 3\,d, and 6\,d (Figure~\ref{tab:step1_combined}). The advantage of multi-lead training grows with lead time. At 6\,h, the single-lead model achieves slightly higher CSI (0.636 vs.\ 0.613) and FSS$_5$ (0.904 vs.\ 0.898) while matching the best multi-lead model in MSE (0.049). At 3 days, performance is comparable across approaches. At 6 days, however, the multi-lead models are clearly superior: the single-lead UNet reaches only 0.100 MSE and 0.408 CSI, compared to 0.085 and 0.429 for UNet Multi FT. This pattern suggests that single-lead training provides marginal benefits at short lead times, but that the broader training distribution of multi-lead models provides regularization that is increasingly beneficial at longer lead times. Operationally, a single model that remains competitive with or surpasses dedicated single lead models is also far more practical than maintaining a separate model per lead time.

\subsubsection{False alarms and missed precipitation}
\label{sec:miss_fa}

To assess whether bias correction removes spurious precipitation and recovers precipitation
that AIFS misses, we classified each grid cell of the evaluation set on the bias-correction
grid (0.25$^\circ$) with a 0.1\,mm/6\,h threshold. False alarms (AIFS wet, observation dry)
are by far the dominant AIFS error, with about 18 times more cells than misses (observation
wet, AIFS dry): AIFS spreads light precipitation over much larger areas than observed
(Table~\ref{tab:miss_fa}).

\smallskip
\noindent Both bias-correction models remove about 65\% of the AIFS false alarms and over
80\% of the falsely predicted precipitation amount, while retaining about 90\% of the
correctly forecast precipitation. Isolated spurious precipitation, more than two grid cells
away from any observed rain, is removed most reliably (78\%), whereas false alarms adjacent
to observed rain, which mostly reflect errors in the extent of precipitation systems, are
harder to correct (44\%). Suppression decreases from about 70\% at lead times up to 3 days
to 61\% beyond. The effect is clearest in nearly dry situations, in which at most 5\% of the
domain is observed wet (one third of the evaluation samples, 1.1\% observed wet area on
average): AIFS predicts precipitation over 19\% of the domain, against 5\% after bias
correction with UNet$_{\lambda=2}$, and 81\% of its false alarms are removed.

\smallskip
\noindent Recovering missed precipitation is considerably harder, since the model must
create a precipitation signal absent from the AIFS precipitation field. UNet$_{\lambda=2}$
recovers 19\% of the missed cells and the Multi-FT model 42\%. Recovery is much more frequent next to AIFS precipitation (26\% and 52\%), where it amounts to correcting the extent or position of a precipitation system, than for isolated misses with no AIFS precipitation within two grid cells (9\% and 30\%), and the recovered amounts remain small (6\% and 15\% of the missed precipitation). The added precipitation nonetheless carries a real signal: 25\% (UNet$_{\lambda=2}$) and 20\% (Multi FT) of the cells made wet where AIFS was dry are observed wet, against about 4\% if precipitation were added at random. Multi FT recovers more misses at the cost of more new false alarms (6.5\% vs.\ 2.2\% of the cells where both AIFS and the observation are dry), consistent with its wetter output.

\begin{table}[ht]
\centering
\caption{False alarms and missed precipitation on the evaluation set (0.25$^\circ$ grid, threshold 0.1\,mm/6\,h, all lead times). Isolated: more than two grid cells away from observed precipitation (false alarms) or from AIFS precipitation (misses). Precision: share of the cells made wet by the model where AIFS was dry that are observed wet. 95\% bootstrap confidence intervals of the overall rates are within $\pm 1$ percentage point.}
\label{tab:miss_fa}
\begin{tabular}{lcc}
\toprule
 & UNet$_{\lambda=2}$ & Multi FT \\
\midrule
\multicolumn{3}{l}{\textit{AIFS false alarms (1\,837\,340 cells)}} \\
Suppressed (all)                    & 65.0\% & 64.1\% \\
\quad isolated / near observed rain & 78.4\% / 44.0\% & 77.2\% / 43.2\% \\
\quad AIFS $\geq 1$\,mm             & 48.1\% & 52.1\% \\
False-alarm amount removed          & 83.1\% & 79.5\% \\
Correct precipitation retained      & 89.6\% & 90.4\% \\
\midrule
\multicolumn{3}{l}{\textit{AIFS misses (99\,640 cells)}} \\
Recovered (all)                     & 18.9\% & 42.2\% \\
\quad near AIFS rain / isolated     & 25.9\% / 9.4\% & 51.7\% / 29.5\% \\
\quad observed $\geq 1$\,mm         & 27.2\% & 56.6\% \\
Missed amount recovered             & 5.5\%  & 14.5\% \\
Precision of added precipitation    & 25.4\% & 20.0\% \\
New false alarms (both dry)         & 2.2\%  & 6.5\% \\
\bottomrule
\end{tabular}
\end{table}

\subsubsection{Qualitative examples of bias-corrected precipitation}

\begin{figure*}[t]%
%\hspace{-1cm}
\includegraphics[width=1\textwidth]{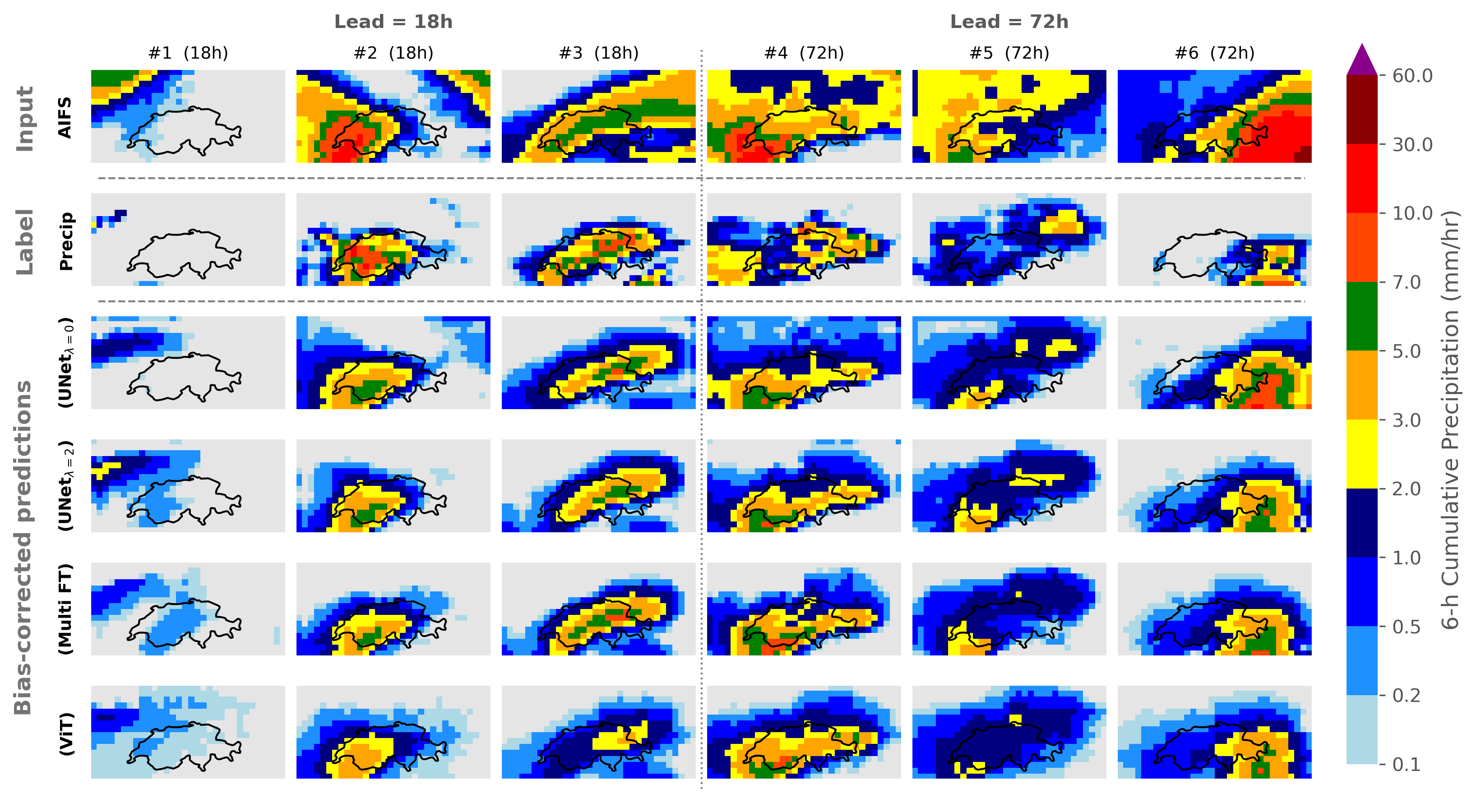}
\caption{\textbf{Qualitative bias correction results at 18h (left) and 72h (right) lead time, for random time steps in the test set}. Rows show, from top to bottom: raw AIFS input, ground-truth precipitation, and predictions from four of our multi-lead models. 
%All models sharpen the over-extended AIFS fields toward the observed spatial structure, with visibly smoother outputs at longer lead times.
}
\label{fig:BC_viz_compare_lead2and11}
\end{figure*}

Figure~\ref{fig:BC_viz_compare_lead2and11} presents bias-corrected fields from four multi-lead models at 18\,h and 72\,h lead times, for examples of time steps in the test dataset. 
% Note that the examples were randomly sampled among test samples presenting interesting patterns (for example not fully dry), yet not based on performance of our models. 
At 18\,h, the behavior is case-dependent. In examples \#2 and \#3, where AIFS produces a broad, over-extended precipitation band, all models contract the field toward the observed structure and recover tighter spatial gradients. UNet$_{\lambda=2}$ and Multi~FT are the sharpest, consistent with their leading MSE\_wet scores. Example \#1 behaves differently, as the observations are essentially dry over Switzerland apart from a small patch at the north-western edge. AIFS forecasts a weak band, and all models retain a broad area of light precipitation rather than suppressing it. This shows on one hand an effective correction of placement and intensity, on the other weaker suppression of spurious light rain. Such an asymmetry is a genuine limitation of the deterministic bias correction, and is quantified in the false-alarm evaluation of Section~\ref{sec:results}\ref{sec:results_bias_correction}\ref{sec:miss_fa}. At 72\,h, predictions are visibly smoother but all models substantially reduce the large AIFS biases. Multi~FT maintains relatively sharp features at this longer lead time, consistent with its positive amplitude bias (SAL-A $= +0.18$). Across both lead times, differences between neural models remain subtle compared to the large shared improvement over the AIFS and classical baselines, consistent with the tight metric spread among learned models (CSI: 0.510-0.540; Figure~\ref{tab:metrics_compact}).

%%%%%%%%%%%%
\subsection{Results for the full downscaling pipeline (Steps~1 and ~2)}
\label{sec:results_super_resolution}

\begin{figure*}[t]
\centering
\begin{subfigure}[t]{0.46\textwidth}
    \centering
    \includegraphics[width=1\textwidth]{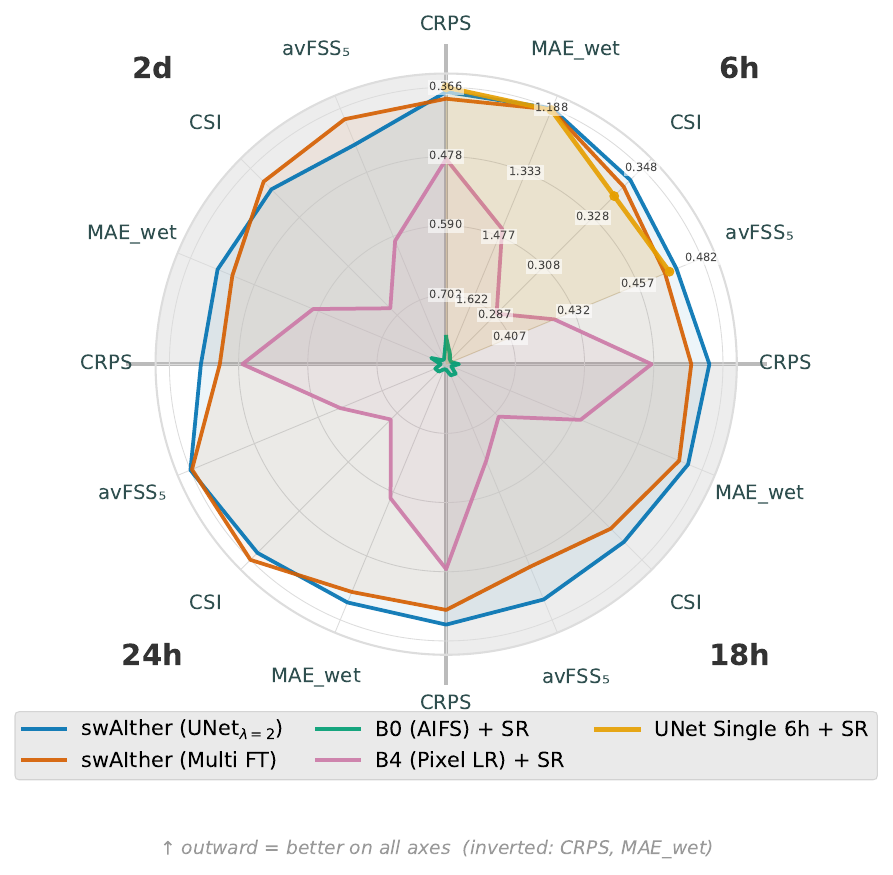}
    \caption{Short-range (6h -- 2 days)}
    \label{fig:spider_low}
\end{subfigure}
\hfill
\begin{subfigure}[t]{0.53\textwidth}
    \centering
    \includegraphics[width=1\textwidth]{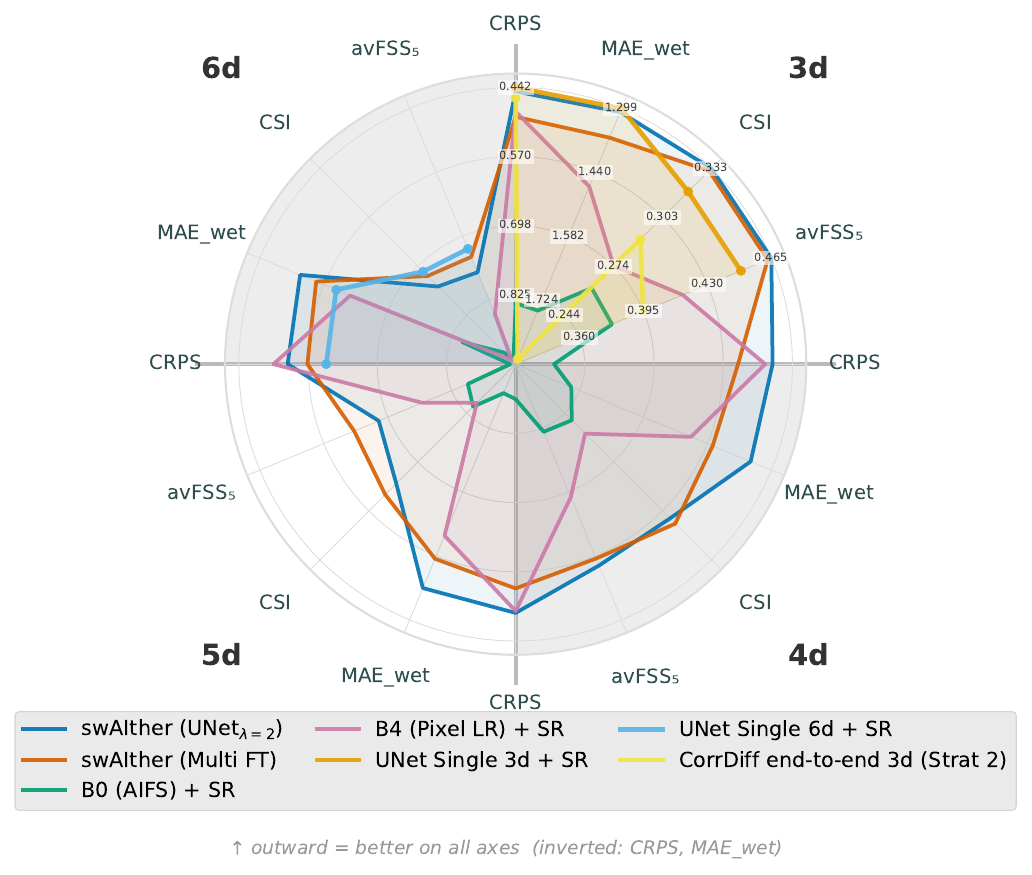}
    \caption{Extended-range (3 days -- 6 days)}
    \label{fig:spider_high}
\end{subfigure}

\vspace{0.3cm}

\begin{subfigure}[t]{0.75\textwidth}
    %\hspace{-1cm}
    \includegraphics[width=\textwidth]{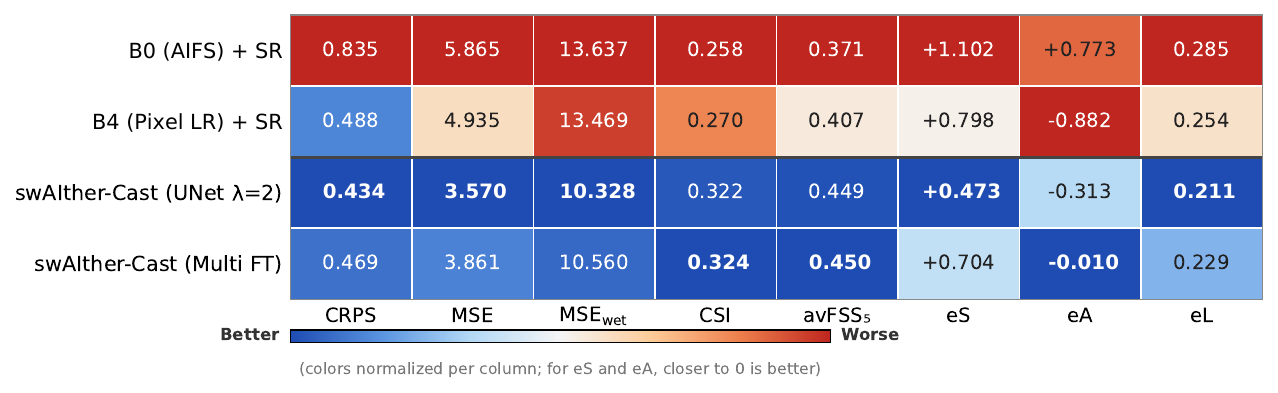}
    \caption{Average metrics across all lead times}
    \label{fig:ds_heatmap}
\end{subfigure}

\caption{\textbf{Full downscaling pipeline evaluation.}
(a,b)~Multi-metric radar charts across lead times. Each quadrant corresponds to a lead time, with four metrics per quadrant: CRPS, MAE (wet pixels), CSI, and avFSS$_5$. Within each quadrant, metrics are arranged clockwise in this order starting from the quadrant's left boundary, so that each spoke belongs unambiguously to the quadrant on its right. Outward displacement indicates better performance; single-lead UNet models appear as filled wedges at their respective lead times, including the end-to-end CorrDiff model trained on a 3 day lead time to evaluate Strategy 2 (see Section ~\ref{sec:results}\ref{sec:results_super_resolution}\ref{sec:results_single_diff} and Section~\ref{sec:results}\ref{sec:results_super_resolution}\ref{sec:results_strategy2} for discussion).  (c) Average metrics across all lead times (colors normalized per column; for eS and eA, closer to zero is better). 
%Both SwAIther-Cast configurations substantially outperform baselines across all metrics and horizons, with UNet$_{\lambda=2}$ achieving the best probabilistic skill (CRPS\,$=$\,0.434) and UNet Multi~FT the best amplitude calibration ($eA = -0.010$).
}
\label{fig:ds_evaluation}
\end{figure*}

We now evaluate the full two-step downscaling pipeline, where the bias-corrected low-resolution precipitation is refined using CorrDiff-based generative super-resolution. Due to the computational cost of diffusion sampling - generating 12 ensemble members per time step across all lead times - we restrict the super-resolution evaluation to two bias correction models and two baselines. 

\smallskip
\noindent We select UNet$_{\lambda=2}$ and UNet Multi FT, which achieve the lowest average wet-pixel MSE in the bias correction evaluation (0.067 and 0.066 respectively) and represent two distinct training strategies. We detail the full list of hyperparameters for these final models in the Supplementary Material, Section 3. As baselines, we include B0 (raw AIFS) and B4 (per-pixel multivariate linear regression). We generate $M=12$ ensemble members for each forecast initialization. The evaluation is performed on approximately one quarter of the test times, while keeping the same lead-time coverage and evaluation protocol (we consider every 4 time steps in the test set, to cover all seasons through the year). Since the same super-resolution model is used in all pipelines, performance differences reflect only the upstream bias correction. To avoid confusion however, we refer to the baseline pipelines as \texttt{B0 (AIFS) $+$ SR} and \texttt{B4 (Pixel LR) $+$ SR}, where \texttt{SR} is short for super resolution, being the perfect super-resolution Corrdiff model trained separately and applied here at inference time. The number of ensemble members (12) was chosen as a compute/storage trade-off that is sufficient for stable probabilistic verification and consistent with recent generative-downscaling literature.

\subsubsection{Performance averaged across lead times}

The heatmap shown in Figure \ref{fig:ds_evaluation}c summarizes the average performance across all lead times. UNet$_{\lambda=2}$ achieves the best CRPS (0.434), MSE (3.570), and wet-pixel MSE (10.328). UNet Multi FT achieves the highest CSI (0.324) and avFSS$_5$ (0.450), and notably near-zero amplitude bias ($eA = -0.010$), a substantial improvement over B0 ($eA = 0.773$, over-prediction) and B4 ($eA = -0.882$, under-prediction).
 
\smallskip
\noindent The B0 baseline (AIFS $+$ SR) unsurprisingly performs worst across all metrics (CRPS 0.835), demonstrating that applying super-resolution to uncorrected AIFS output amplifies rather than compensates for biases. B4 improves upon B0 but remains substantially behind the UNet-based pipelines. The structure component $eS$ remains positive for all models (0.473--1.102), reflecting a common tendency of generative models to produce slightly smoother precipitation objects than observed.

\subsubsection{Lead-time dependence of skill}

Figures~\ref{fig:ds_evaluation}a and~\ref{fig:ds_evaluation}b show multi-metric comparisons across lead times (for per-lead values and additional avFSS / Ensemble SAL diagnostics, please see the Supplementary Material, Section 2). Both SwAIther-Precip configurations dominate all metrics at every lead, with UNet$_{\lambda=2}$ achieving the best CRPS throughout (0.373 at 6\,h to 0.532 at 6\,d, a 51\% CRPS reduction over B0 at short range). Performance is remarkably stable from 6\,h through 2\,d and degrades gradually beyond, with all models converging at 6\,d as predictability decreases. UNet Multi FT trails slightly in CRPS at extended ranges but maintains competitive CSI and avFSS$_5$, and provides better amplitude calibration overall.
 
 \begin{figure*}[t]
%hspace{-1cm}
\includegraphics[width=1\textwidth]{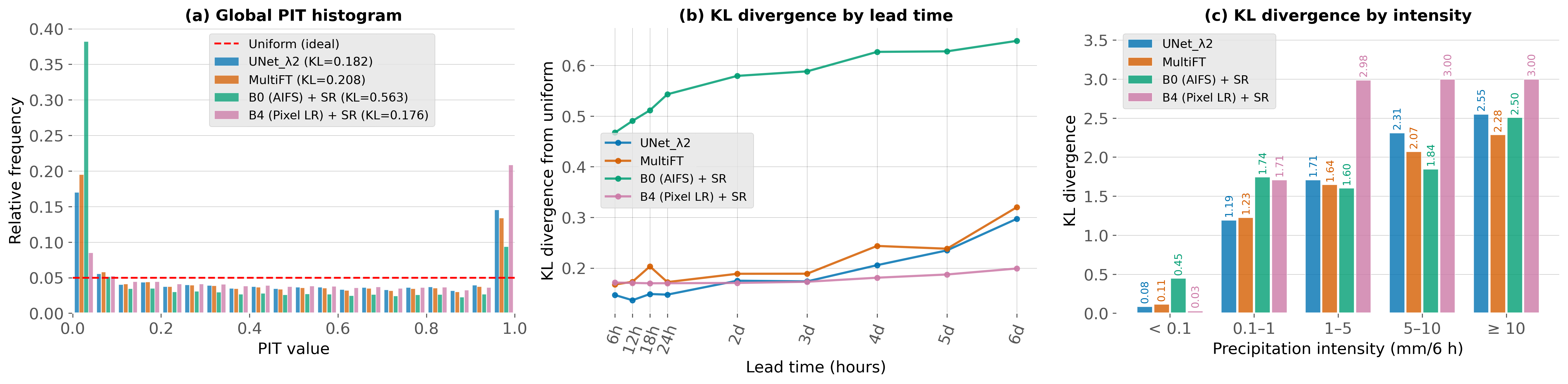}
\caption{\textbf{PIT calibration diagnostics for both SwAIther-Precip variants and the considered downscaling baselines.}
(a)~Global PIT histograms with KL divergence from a uniform distribution. 
(b)~KL divergence as a function of lead time. 
(c)~KL divergence stratified by precipitation intensity. In (c), intensity classes are defined by the observed precipitation; since conditioning on the
observation shifts PIT values towards one even for a calibrated forecast, KL values in the highest classes are best compared across models. The maximum possible value with 20 bins is $\ln 20 \approx 3.0$.}
\label{fig:pit_comparison}
\end{figure*}

 \smallskip 
 \noindent Figure~\ref{fig:pit_comparison} summarises PIT diagnostics aggregated over ${\sim}248 \times 10^{6}$ pixel--time samples. Both neural models are substantially better calibrated (KL\,$=$\,0.182 and 0.208) than AIFS (0.563) and competitive with B4 (0.176). Calibration remains stable up to 3\,days before increasing moderately (panel~b), while AIFS degrades rapidly beyond 24\,h. 
 Stratification by intensity (panel~c) reveals that intense precipitation is the most challenging regime, with KL increasing with intensity for all models, partly because conditioning on the observed intensity penalizes even a calibrated forecast \citep{lerch2017forecaster}. Among the SwAIther models, UNet$_{\lambda=2}$ is better calibrated below 1\,mm/6\,h and Multi FT above, including for $\geq$10\,mm/6\,h (KL 2.28 vs.\ 2.55), while B4 reaches the maximum possible value at all intensities above 1\,mm/6\,h. The super-resolved raw AIFS remains better calibrated between 1 and 10\,mm/6\,h, reflecting its positive amplitude bias, which indicates that the pipeline improves the overall structure and lower-intensity skill but remains challenged for intense precipitation.
 % Stratification by intensity (panel~c) reveals that heavy precipitation ($\geq 1$\,mm) is the most challenging regime (KL\,$\approx$\,1.7--3.0); UNet$_{\lambda=2}$ achieves the best calibration among bias-corrected models across all intensity bins, although the raw AIFS forecast remains better calibrated in the highest-intensity ($\geq 1$\,mm) bin, which indicates that the pipeline improves overall shape and lower-intensity skill but is still challenged at extremes.

\subsubsection{Multi-lead training versus single-lead training}
\label{sec:results_single_diff}

% Single-lead UNet models appear as filled wedges in Figure~\ref{fig:ds_evaluation}. At 6h, the single-lead closely matches SwAIther-Precip (CRPS 0.366 vs. 0.373). At 3d, it falls slightly behind in CSI and avFSS$_5$. At 6d, the single-lead model degrades substantially (CRPS 0.604 vs.\ 0.532 for UNet$_{\lambda=2}$, MSE\_wet 14.506 vs.\ 11.889), confirming that multi-lead training provides a regularization benefit that is increasingly important at longer lead times, where it benefits most from statistical strength borrowed from the shorter-lead examples and from the temporal-consistency term, which ties long-lead corrections to those of shorter-lead forecasts verifying at the same instant.

Single-lead UNet models (postprocessed with the SR model as all full pipelines compared ) appear as filled wedges in Figure~\ref{fig:ds_evaluation}, enabling direct visual comparison  with the multi-lead swAIther variants at their respective training horizons. The pattern varies by both lead time and metric. At 6\,h (panel a), the single-lead wedge nearly overlaps the multi-lead envelope across all four metrics, indicating that single- and multi-lead training perform comparably at this short horizon. At 3\,d (panel b), the single-lead wedge remains close to the multi-lead envelope on CRPS and MAE\_wet but is visibly inside on CSI and avFSS$_5$, indicating a modest disadvantage on the categorical and spatial metrics. At 6\,d, the single-lead wedge is clearly inside the multi-lead envelope on all four metrics, with the largest gaps on CRPS and MAE\_wet. This progression (near-overlap at 6\,h, mixed at 3\,d, substantial gap at 6\,d) is consistent with the regularization benefit of multi-lead training becoming increasingly important as per-lead training signal becomes sparser at long horizons. The multi-lead model benefits here from statistical strength borrowed from the shorter-lead examples and the single-lead specialists models regress more strongly towards the conditional mean.

\subsubsection{Two-step decomposition versus direct downscaling}
\label{sec:results_strategy2}

To test whether the explicit bias-correction step is beneficial, we compare the two-step
approach with a full end-to-end CorrDiff model trained to map the AIFS fields directly to
1\,km precipitation (Strategy~2, Section \ref{sec:eval}\ref{subsec:baselines}), bypassing the intermediate correction. Because this model is expensive to train, the comparison is made at a single lead time (72\,h), against the 72\,h single-lead specialist of Strategy~1 trained with the same budget.

\smallskip
\noindent The results favor the two-step approach (Table~\ref{tab:strategy2}). Strategy~1 is significantly better on MAE$_\mathrm{wet}$ (1.30 vs.\ 1.87\,mm), CSI (0.319 vs.\ 0.290)
and avFSS$_5$ (0.448 vs.\ 0.395), while both strategies reach a comparable CRPS (0.442
vs.\ 0.462\,mm; difference not significant at the 95\% level). The end-to-end model
over-predicts both rainfall amount and spatial extent, with an ensemble-mean bias of
$+0.51$\,mm (vs.\ $-0.18$\,mm) and a false-alarm ratio of 0.67 (vs.\ 0.60). Its larger
ensemble spread partly compensates for this in the CRPS (Section~\ref{sec:discussion}).
The same conclusions hold without the low-resolution mask in post-processing.

\smallskip
\noindent The regression stage of this model, evaluated alone as a lighter proxy for
Strategy~2, points the same way: at 72\,h it reaches an MSE$_{wet}$ of 0.094 (vs.\ 0.061-0.064
for the best bias-correction models), a CSI of 0.336 (vs.\ 0.552–0.553) and an FSS$_5$
of 0.634 (vs.\ 0.839--0.843), with both structural distortion (SAL-S\,$=$\,$-0.421$)
and amplitude over-prediction (SAL-A\,$=$\,0.594). Together, these results suggest that
simultaneously correcting biases, bridging the resolution gap and learning
convective-scale structure is considerably harder than decomposing the problem into
specialized steps.

\smallskip
\noindent Beyond performance, the two-step decomposition is also much cheaper. Strategy~2
requires a separate model for every lead time, each conditioned on 12 high-resolution
input channels and trained on spatial patches, whereas our Step~2 is a single
lead-independent model with two input channels, trained on observations directly. This
restriction does not cost skill: adding atmospheric predictors to Step~2 does not improve it
(Section~\ref{sec:step2_channels}).

\begin{table*}[ht]
\centering
\caption{Comparison of the two-step approach (Strategy~1: single-lead bias-correction
U-Net followed by the Step~2 super-resolution model) with an end-to-end CorrDiff model
(Strategy~2: regression and patch-based diffusion trained directly from the multivariate
AIFS state to 1~km CombiPrecip), at the 72\,h lead time. Mean over the 117 validation
times, 12 ensemble members. $\Delta$: Strategy~2 minus Strategy~1, with 95\% paired
bootstrap confidence intervals over times. Best values in bold.}
\label{tab:strategy2}
\resizebox{1.4\columnwidth}{!}{%
\begin{tabular}{lcccc}
\toprule
 & CRPS $\downarrow$ & MAE$_\mathrm{wet}$ $\downarrow$ & CSI $\uparrow$ & avFSS$_5$ $\uparrow$ \\
\midrule
Strategy 1 (two-step)    & \textbf{0.442} & \textbf{1.299} & \textbf{0.319} & \textbf{0.448} \\
Strategy 2 (end-to-end)  & 0.462 & 1.866 & 0.290 & 0.395 \\
\midrule
$\Delta$ (S2 $-$ S1)     & $+0.020$ & $+0.567$ & $-0.029$ & $-0.057$ \\
95\% CI                  & $[-0.010,\ 0.047]$ & $[0.394,\ 0.774]$ & $[-0.043,\ -0.015]$ & $[-0.076,\ -0.041]$ \\
\bottomrule
\end{tabular}
}
\end{table*}

\subsubsection{Ablation: atmospheric predictors in the super-resolution step}
\label{sec:step2_channels}

Our Step~2 receives only the coarse-resolution precipitation and altitude. Whether additional
atmospheric predictors should be included is a common design question, with a direct impact on
data volume. To test it, we trained a second Step~2 regression that additionally receives ten
AIFS atmospheric fields: convective precipitation, total cloud cover, the 10\,m wind components,
and specific humidity, temperature and geopotential at 500 and 850\,hPa. All else is equal
(architecture, hyperparameters, training budget and data split). As no observations of these
variables are available on our grid (and we expect the reader to not have easy access to it either), we used the $+6$\,h AIFS forecasts valid at the same time, which are the most accurate atmospheric fields available.

\smallskip
\noindent In the perfect super-resolution setting (coarsened CombiPrecip as input, 117 test times), the additional channels do not improve Step~2
(Table~\ref{tab:step2_channels}). The detection and spatial structure of precipitation are
unchanged: CSI and avFSS$_5$ differ by less than 0.1\%, which is not significant. Errors on
precipitation amounts increase slightly but consistently: MAE, MAE$_\mathrm{wet}$, RMSE and
MAE$_\mathrm{heavy}$ are all 3--5\% higher, significantly so at the 95\% level (paired bootstrap
over times). The coarse precipitation field thus appears to carry most of the information needed
at the 1\,km scale, while the atmospheric forecasts mainly add noise.

\smallskip
\noindent The additional channels also come at a substantial data cost. They multiply the input size per sample by six, and once interpolated to the 1\,km grid, as is standard for CorrDiff, they increase the preloaded training inputs from 2.6 to about 28\,GB. This ablation covers the
regression stage of Step~2, which carries the deterministic mapping; the diffusion stage was not
retrained.

\begin{table*}[ht]
\centering
\caption{Effect of adding atmospheric predictors to the Step~2 regression, in the perfect super-resolution setting (coarsened CombiPrecip as input), over 117 test times. ``Precip.'': coarse precipitation and altitude (our Step~2). ``+ Atmos.'': with ten additional AIFS $+6$\,h atmospheric fields. $\Delta$: relative difference; CI: 95\% paired bootstrap confidence interval of the absolute difference over times. Bold: significantly better. Data: input size per sample and training inputs preloaded at 1\,km (float16).}
\label{tab:step2_channels}
\resizebox{\textwidth}{!}{%
\begin{tabular}{lcccccccc}
\toprule
 & MAE $\downarrow$ & MAE$_\mathrm{wet}$ $\downarrow$ & RMSE $\downarrow$ & MAE$_\mathrm{heavy}$ $\downarrow$
 & CSI $\uparrow$ & avFSS$_5$ $\uparrow$ & Input (MB) & Inputs at 1\,km (GB) \\
\midrule
Precip.  & \textbf{0.135} & \textbf{0.507} & \textbf{0.425} & \textbf{1.002} & 0.650 & 0.804 & 3.1  & 2.6  \\
+ Atmos. & 0.140 & 0.526 & 0.448 & 1.046 & 0.649 & 0.804 & 18.7 & 28.1 \\
\midrule
$\Delta$ & $+3.4\%$ & $+3.7\%$ & $+5.4\%$ & $+4.3\%$ & $-0.1\%$ & $0.0\%$ & $\times 6$ & $\times 11$ \\
CI & $[0.002,\,0.008]$ & $[0.006,\,0.036]$ & $[0.012,\,0.035]$ & $[0.015,\,0.085]$
   & $[-0.003,\,0.002]$ & $[-0.003,\,0.003]$ & & \\
\bottomrule
\end{tabular}
}
\end{table*}

%%%%%%%%%%%%%%%%%%%%%%%
\subsubsection{Spectral fidelity of the downscaled forecasts}
\label{sec:results_spectral}
%%%%%%%%%%%%%%%%%%%%%%%

\begin{figure*}[t]
%\hspace{-1cm}
\includegraphics[width=1\textwidth]{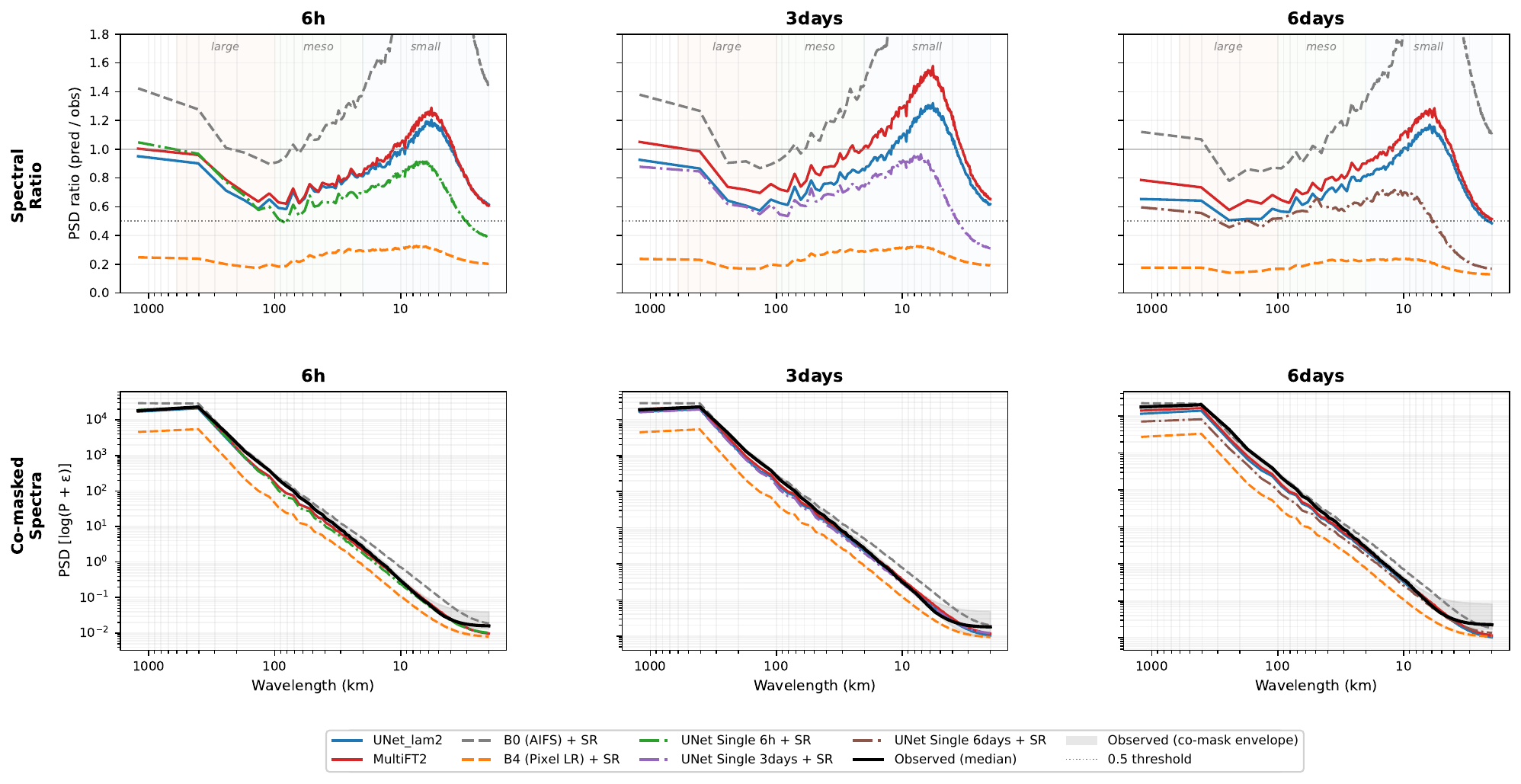}
\caption{\textbf{Spectral analysis of the downscaling pipeline across lead times (6h, 3days, 6days).} Top row: spectral ratio (predicted/observed PSD) computed on co-masked log-precipitation fields. A ratio of 1 indicates perfect spectral fidelity; the dotted line marks the 0.5 effective-resolution threshold. Shaded bands denote the large-scale (100--600km), mesoscale (20--100km), and small-scale (2--20km) wavelength ranges. Bottom row: radially averaged power spectral densities of log-transformed precipitation, computed within the common wet area where both prediction and truth exceed 0.1\,mm/6h (co-masking). The black line shows the median observed spectrum across models, with the gray envelope indicating the range due to model-dependent co-mask differences. 
%UNet$_{\lambda=2}$ maintains a spectral ratio close to 1 at large and mesoscales, with a gradual roll-off at small scales. The single-horizon models trained at 6h, 3days, and 6days are shown at their respective lead times for comparison.
}
\label{fig:spectral_composite}
\end{figure*}

\begin{figure*}[t]
%\hspace{-1cm}
\includegraphics[width=1\textwidth]{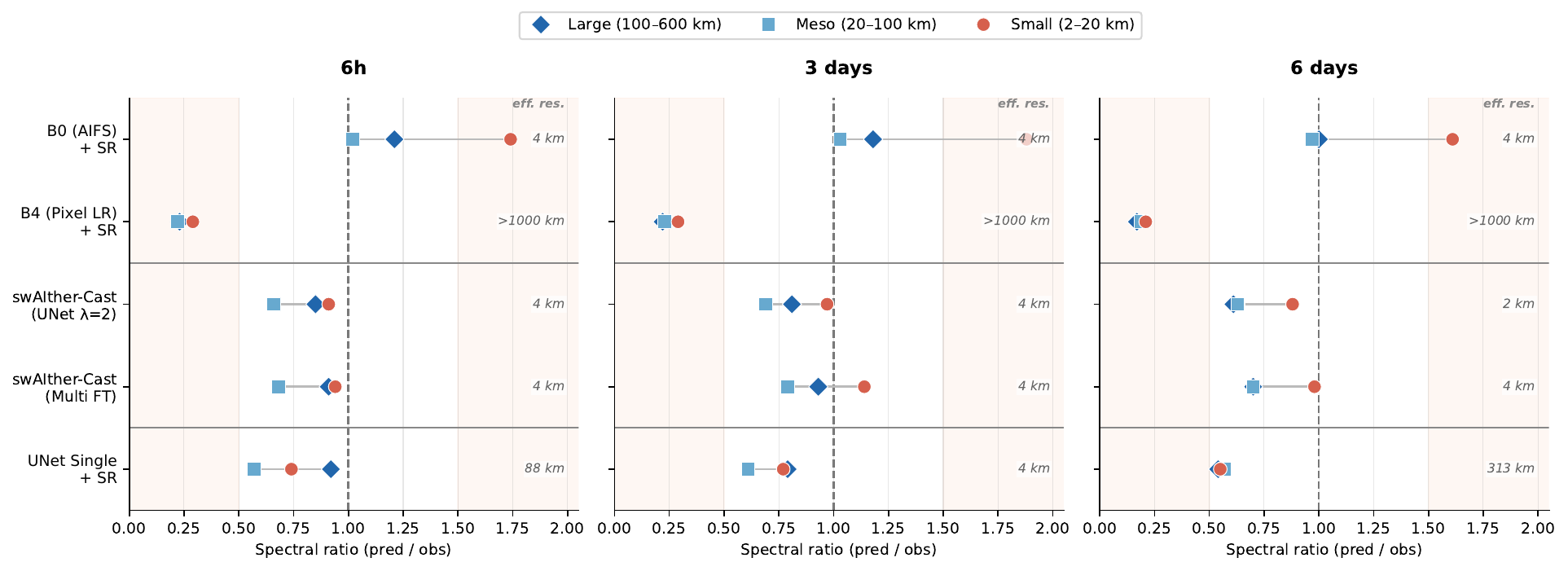}
\caption{\textbf{Band-averaged spectral ratios (predicted/observed PSD) at three lead times.} Each dot represents the ratio averaged over a wavelength band: large scale (100--600km, diamond), mesoscale (20--100km, square), and small scale (2--20km, circle). The dashed line marks perfect spectral fidelity (ratio$=$1). Effective resolution (km) is shown on the right for each lead time. The SwAIther-Precip models maintain ratios close to 1 across scales, with a consistent mesoscale deficit. B0 (AIFS) + SR systematically exceeds 1 (over-dispersion), while B4 + SR collapses below 0.3 (severe under-dispersion).}
\label{fig:spectral_dotplot}
\end{figure*}

Figures~\ref{fig:spectral_composite} and~\ref{fig:spectral_dotplot} present the spectral diagnostics across three lead times. Both SwAIther-Precip models achieve spectral ratios close to unity at large scales (0.85--0.93) and small scales (0.88--0.98), substantially outperforming the baselines: B0 $+$ SR systematically exceeds 1 (reaching 1.7–1.9 at small scales; see the note below), while B4 $+$ SR collapses below 0.3 at all scales, producing fields with less than a third of the observed variance. Both SwAIther-Precip models maintain stable characteristics from 6\,h to 3\,d, with moderate large-scale degradation at 6\,d, and achieve an effective resolution of $\sim$4\,km on the 1\,km grid, consistent with state-of-the-art NWP models \citep{klaver2020effective, skamarock2004evaluating}. Single-lead models perform comparably or worse spectrally (UNet Single 6\,d drops to 0.55 at small scales), further supporting the multi-lead regularization benefit.

\smallskip
\noindent The most consistent spectral deficiency is a variance deficit at mesoscales (20--100\,km), with band ratios of 0.63--0.79, visible in Figure~\ref{fig:spectral_dotplot} as the mesoscale marker pulling left of the large- and small-scale ones. This deficit, and its origin in the deterministic bias correction, is discussed in Section \ref{sec:discussion}\ref{sec:limitations}. 

% \smallskip
% \noindent Note that elevated small-scale power in the raw AIFS spectrum should not be interpreted as physical resolution of small-scale precipitation: it most probably reflects interpolation artifacts from regridding AIFS's native N320 reduced Gaussian grid to the 1,km evaluation grid, and sharp precipitation-edge features from AIFS's bounded (ReLU) output activations \citep{moldovan2025update}.

\smallskip
\noindent Note that the AIFS + SR curve corresponds to the super-resolution of the uncorrected AIFS precipitation, not to AIFS itself. Its elevated small-scale power does not indicate a resolution of small-scale precipitation: it reflects the fine-scale texture generated by the super-resolution model over the widespread light precipitation of
uncorrected AIFS, amplified by the logarithmic transform used in the spectral analysis.
%%%%%%%%%%%%%%%%%%%%%%%
\subsubsection{Ensemble characteristics and sample quality}
\label{sec:results_ensemble}
%%%%%%%%%%%%%%%%%%%%%%%

\begin{figure*}
%\hspace{-1cm}
\includegraphics[width=1\textwidth]{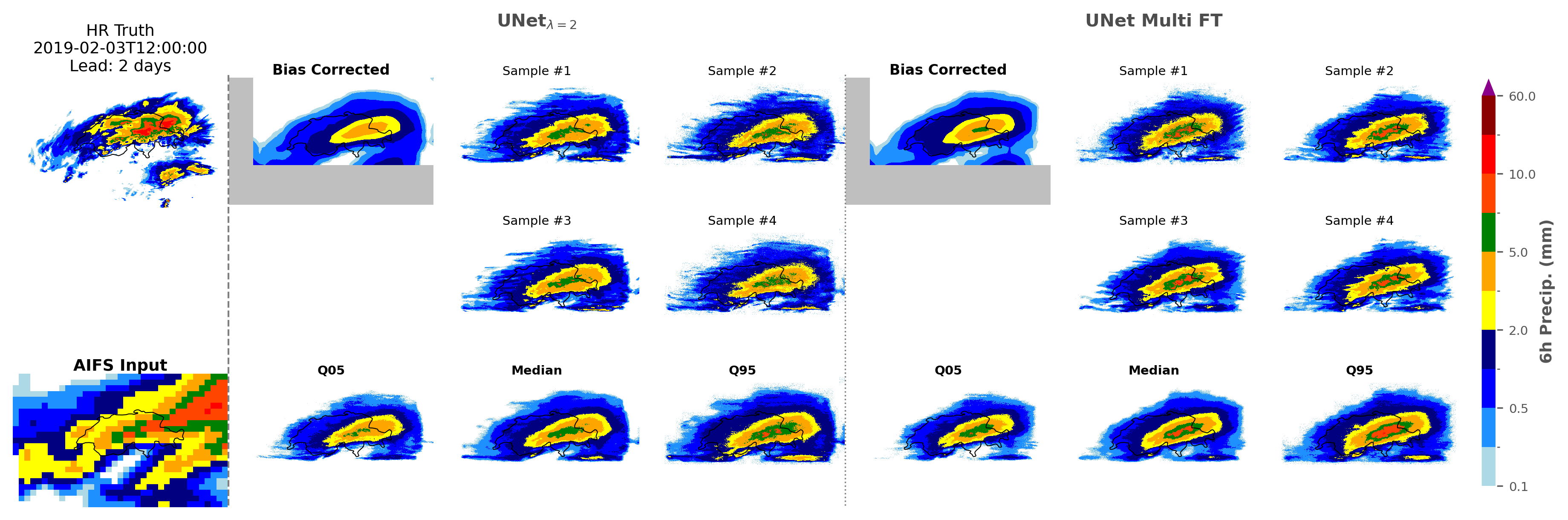}
\caption{\textbf{High-resolution precipitation samples generated by the conditional diffusion model (Step~2) for our two SwAIther-Precip variants (using UNet$_{\lambda=2}$ and Multi~FT bias-correction backbones) at 2-day lead time, for a random time step in the test set}. Left column: high-resolution ground truth and raw AIFS input. Per model: the deterministic bias-corrected field used as conditioning input, four ensemble members (out of the existing 12 members), and summary statistics (Q05, median, Q95) computed over all 12 members. Individual samples exhibit realistic fine-scale variability while the ensemble median recovers the broad spatial structure of the ground truth.}
\label{fig:DIFF_viz_2daysLead_ex2}
\end{figure*}

Figure~\ref{fig:DIFF_viz_2daysLead_ex2} presents ensemble members for a strong winter precipitation event (2019-02-03, 2-day lead time). For both pipelines, ensemble members share a consistent large-scale structure while exhibiting meaningful diversity in fine-scale details - edges of the precipitation region, isolated features, and the shape of the high-intensity core vary across members, reflecting genuine forecast uncertainty at convective scales. The quantile summaries (Q05, median, Q95) reveal a plausible ensemble spread, although the ensembles remain under-dispersive overall (Section~\ref{sec:discussion}\ref{sec:limitations}).

%reveal well-calibrated ensemble spread, though both pipelines seem to slightly underestimate peak intensities compared to the truth, an attenuation inherited from the deterministic bias correction.
 
\smallskip
\noindent UNet Multi FT produces noticeably higher peak intensities than UNet$_{\lambda=2}$, consistent with its near-zero amplitude bias ($eA = -0.010$ vs.\ $-0.313$) and higher mesoscale spectral ratio (0.79 vs.\ 0.69 at 3\,days). The multi-stage fine-tuning partially mitigates the variance compression inherent in the deterministic bias correction, yielding more intense fields at the cost of slightly higher MSE (3.861 vs.\ 3.570).
 
\begin{sidewaysfigure*} 
\centering
\includegraphics[width=1\textheight]{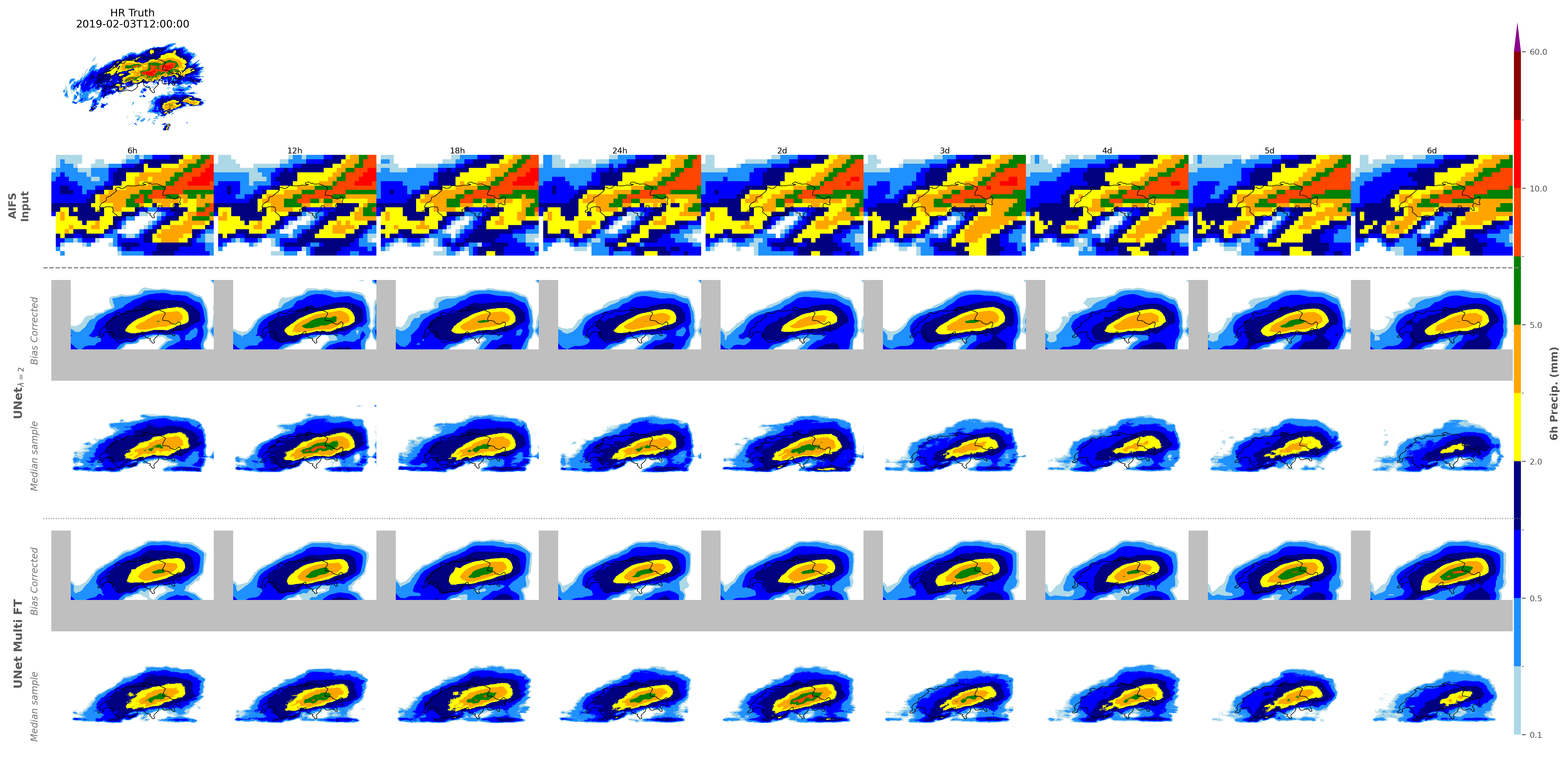}
\caption{\textbf{High-resolution precipitation median samples across all considered lead times (6h to 6d), for our two SwAIther-Precip variants (using UNet$_{\lambda=2}$ and Multi~FT bias-correction backbones), for a random time step in the test set.} Top left: high-resolution ground truth. First row: raw AIFS input at each lead time. Per model: the deterministic bias-corrected field and the ensemble 
  median of the conditional diffusion model (Step~2).}
\label{fig:all_leads_overview}
\end{sidewaysfigure*}

\smallskip
\noindent Figure~\ref{fig:all_leads_overview} shows ensemble-median predictions across all lead times. Both pipelines maintain remarkably stable spatial structure from 6h through 3\,d, with gradual degradation at 4--6\,d reflecting increasing synoptic-scale uncertainty. Despite progressive smoothing of the bias-corrected input at longer lead times, the diffusion model continues to generate realistic fine-scale texture at all lead times, consistent with the stable small-scale spectral ratios reported above. Additional ensemble-median predictions across all lead times for a particularly heavy precipitation event from November 12$^{\text{th}}$ to November 15$^{\text{th}}$, 2023 can be observed in the Supplementary Material, Section 5.

%%%%%%%%%%%%%%%%%%%%%%%%%%%%%%%%%%%%%%%%%%%%%%%%%%%%%%%%%%%
% ---------------------------------------------------------
\section{Discussion and conclusions}
\label{sec:discussion}
% ---------------------------------------------------------
%%%%%%%%%%%%%%%%%%%%%%%%%%%%%%%%%%%%%%%%%%%%%%%%%%%%%%%%%%%

\subsection{Summary of findings}

Two configurations emerge as strongest overall: UNet$_{\lambda=2}$ for deterministic and spatial-structure scores, and UNet Multi FineTuned for long-lead robustness and intensity calibration. Both are strong candidates for downstream use, with the choice depending on which aspect of forecast skill is prioritized. 
Basing ourselves on these two versions, the results presented in Sections \ref{sec:results}\ref{sec:results_bias_correction} - \ref{sec:results}\ref{sec:results_super_resolution} support the three claims of this work.

% \smallskip
% \noindent \textit{The two-step decomposition is effective.} The full pipeline achieves a 48\% CRPS reduction over raw AIFS (UNet$_{\lambda=2}$) and 11\% over the strongest classical baseline, with well-calibrated ensemble spread (UNet Multi FT, which achieves 
% near-zero amplitude error eA $= -0.010$) and skill across all metrics (Figure~\ref{fig:ds_evaluation}). The Strategy~2 ablation study (Section~\ref{sec:results}\ref{sec:results_super_resolution}\ref{sec:results_strategy2}) shows that it clearly degrades performance on most metrics when bias correction and super-resolution are collapsed into a single step, suggesting that the resolution gap and systematic biases in AIFS precipitation benefit from explicit intermediate correction.

\smallskip
\noindent \textit{The two-step decomposition is effective.} The full pipeline achieves a 48\% CRPS reduction over raw AIFS (UNet$_{\lambda=2}$) and 11\% over the strongest classical baseline, with near-zero amplitude error (UNet Multi FT, eA $= -0.010$) and skill across all metrics (Figure~\ref{fig:ds_evaluation}). Compared with an end-to-end CorrDiff model at a 3-day lead time (Strategy~2; Section~\ref{sec:results}\ref{sec:results_super_resolution}\ref{sec:results_strategy2}), the two-step approach is clearly better on most metrics, suggesting that the resolution gap and systematic biases in AIFS precipitation benefit from explicit intermediate correction.

\smallskip
\noindent \textit{Lead-time awareness is both necessary and efficient.} A single lead-time-conditioned multi-lead model is competitive with single-lead models at short lead times and and outperforms them at longer lead times (13\% CRPS improvement at 6\,days with UNet Multi FT; Figure~\ref{fig:ds_evaluation}). Spectral quality remains stable across lead times, with effective resolution at $\sim$4\,km from 6\,h through 6\,days (Figure~\ref{fig:spectral_dotplot}). This is operationally attractive, as a single model replaces a suite of single-lead models.

\smallskip
\noindent \textit{The end-to-end pipeline produces km-scale, medium-range forecasts with realistic spatial structure.} Spectral ratios remain close to unity at both large and small scales across all lead times (UNet$_{\lambda=2}$; Figure~\ref{fig:spectral_dotplot}), and the $\sim$4\,km effective resolution on a 1\,km grid is consistent with state-of-the-art NWP models \citep{klaver2020effective, skamarock2004evaluating}. Ensemble members exhibit realistic fine-scale texture with meaningful inter-member diversity (Figure~\ref{fig:DIFF_viz_2daysLead_ex2}).

\subsection{Limitations and future work}
\label{sec:limitations}

Several limitations of the current framework suggest directions for future work.

\smallskip
\noindent The main limitation is a recurring trade-off between spatial bias correction and variance preservation. The deterministic bias correction excels at correcting spatial placement and extent of precipitation, reducing AIFS's wet fraction from $\sim$58\% to a realistic $\sim$28\%, but tends to underestimate peak intensities (negative SAL-A values; Figure~\ref{tab:metrics_compact}) and exhibits a persistent mesoscale variance deficit (band ratios of 0.63--0.79; Figure~\ref{fig:spectral_dotplot}). This is a direct consequence of the Huber loss, which regresses toward the conditional mean at scales where precipitation is genuinely stochastic. The CorrDiff super-resolution partially compensates at small scales (2--20\,km) but cannot recover mesoscale variance lost in the intermediate representation. UNet Multi FT---fine-tuned with a multi-stage loss---partially mitigates this effect, achieving the best amplitude calibration ($eA = -0.010$ at the super-resolution level) and the highest mesoscale spectral ratios (0.79 at 3 days), and visibly capturing peak intensities better than UNet$_{\lambda=2}$ in qualitative examples (Figures~\ref{fig:BC_viz_compare_lead2and11} and~\ref{fig:DIFF_viz_2daysLead_ex2}). The intensity attenuation is particularly concerning for extreme events, which are precisely the cases where high-resolution probabilistic forecasts are most valuable. Promising mitigation directions include spectral regularization terms in the bias correction loss, stochastic bias correction (e.g., conditional diffusion or VAE) that can represent the multi-modal nature of mesoscale precipitation, multi-stage or spectral-aware losses building on the Multi FT result, and training the CorrDiff super-resolution to target a broader spectral range. Future work should also specifically evaluate the pipeline's performance on heavy precipitation events and explore loss functions that preserve tail behavior.

\smallskip
\noindent Another scope limitation is the lack of temporal consistency: the super-resolution samples are independent when conditioned on the bias-correction outputs. Extending the diffusion model to generate temporally consistent trajectories is an important direction for applications requiring consistent precipitation time-series or accumulations. A promising extension would be to try a correlated-noise scheme during the diffusion based generation (e.g., partial correlation between successive lead times), getting inspiration from the fully shared diffusion seeds across lead times, but in a more balanced way. This could preserve most of the temporal consistency benefits while reducing the static artifacts observed when using identical seeds.

\smallskip
\noindent CombiPrecip is available at 10-min resolution. This would in principle allow for a finer output frequency than the 6\,h accumulations used here. However, the limiting factor is not the target, as AIFS-Single produces 6-hourly steps. An higher-frequency product would require temporal disaggregation within the accumulation window in addition to spatial refinement. This would depend on the temporally coherent sampling discussed above, and on reexamining verification at accumulation periods for which the radar-gauge product is noisier.

% \smallskip
% \noindent Our comparison with single-step direct downscaling used only the regression component of CorrDiff as a proxy, since running the complete diffusion model on all input channels was computationally prohibitive. A complete end-to-end pipeline for the ``direct task'' (Strategy 2) would provide a more definitive comparison.

% \smallskip
% \noindent The comparison with the Strategy 2, meaning end-to-end CorrDiff model, is restricted to the 3-day lead time and to a single training run per model. Both diffusion models were trained with the same hyperparameters, chosen for our Step~2 model and not tuned separately for the end-to-end model, and with the same compute budget (about 30~h on one GH200 GPU). Dedicated tuning and could therefore change the results quantitatively.
\smallskip
\noindent The comparison with the end-to-end CorrDiff model (Strategy~2) is restricted to the 3-day lead time and to a single training run per model. Both diffusion models were trained with the same hyperparameters, chosen for our Step~2 model and not tuned separately for the end-to-end model, and with the same compute budget (about 30\,h on one GH200 GPU); dedicated tuning could therefore change the results quantitatively.

\smallskip
\noindent  The Strategy 2 end-to-end CorrDiff model has a substantially larger ensemble spread than the two-step approach (spread--skill ratio 0.62 against 0.21). This explains why its CRPS remains close to that of the two-step approach despite a less accurate ensemble mean. It also points to a limitation of our design: since Step~1 is deterministic and Step~2 is conditioned on its output, most of the forecast uncertainty is fixed before the diffusion stage, which leads to under-dispersive ensembles.
% If the PIT analysis already shows this, add: (see also Section~\ref{sec:pit}).
Propagating the uncertainty of Step~1, for instance with a probabilistic bias-correction model, is a promising direction for future work.

\smallskip
\noindent We did not perform a systematic attribution of the bias-correction model to its individual predictors; feature ablation or gradient-based attribution methods would help identify which atmospheric fields contribute most to the correction, and could guide a more compact choice of predictors.

\smallskip
\noindent As our framework is trained over Switzerland using AIFS and CombiPrecip, performance may depend on domain characteristics (orographic complexity, convective regime) and the bias structure of the driving model. If retrained elsewhere, we expect our weather downscaling framework to remain applicable, with less reliance on topography in flatter regions and additional static inputs, such as a land-sea mask, in coastal or island regions. Evaluation on other domains and with other global models would help assess spatial generalization \citep{harder2026benchmarking}.

\acknowledgments
TB acknowledge support from the Swiss National Science Foundation (SNSF) under Grant No. 10001754 (``RobustSR'' project). DA, FA, TL, KV, EK, and TB acknowledge support from the Swiss Data Science Center (SDSC) End-User Innovation Project under grant CI24-04: NWF4CH - Democratizing Neural Weather Forecasting for Switzerland : An Open Platform Approach. We thank Max Defez, Shivanshi Asthana, Milton Gomez, David Leutwyler, Mary McGlohon, Petar Stamenkovic for advice that helped develop SwAIther-Precip. 

%  Keep acknowledgments (note correct spelling: no ``e'' between the ``g'' and
% ``m'') as brief as possible. In general, acknowledge only direct help in
%  writing or research. Financial support (e.g., grant numbers) for the work done, 
%  for an author, or for the laboratory where the work was performed must be 
%  acknowledged here rather than as footnotes to the title or to an author's name.
%  Contribution numbers (if the work has been published by the author's institution 
%  or organization) should be placed in the acknowledgments rather than as 
%  footnotes to the title or to an author's name.

%%%%%%%%%%%%%%%%%%%%%%%%%%%%%%%%%%%%%%%%%%%%%%%%%%%%%%%%%%%%%%%%%%%%%
% DATA AVAILABILITY STATEMENT
%%%%%%%%%%%%%%%%%%%%%%%%%%%%%%%%%%%%%%%%%%%%%%%%%%%%%%%%%%%%%%%%%%%%%
% 
%
\datastatement
In order to recreate the AIFS predictions historical dataset, recent ECMWF IFS step-0 forecast fields are available via ECMWF Open Data (https://data.ecmwf.int/forecasts/), but historical operational IFS initial conditions are only available through ECMWF archive access subject to ECMWF member access conditions. CombiPrecip data are publicly available (https://opendatadocs.meteoswiss.ch/); access requests can be addressed to MeteoSwiss directly. The SwissTopo DHM25 digital elevation model is openly available from \url{https://www.swisstopo.admin.ch/}. The source code for SwAIther-Precip, including training configurations and evaluation notebooks, is publicly available at \url{https://github.com/danassou/swaither-precip} under the Apache 2.0 license.
%  The data availability statement is where authors should describe how the data underlying 
%  the findings within the article can be accessed and reused. Authors should attempt to 
%  provide unrestricted access to all data and materials underlying reported findings. 
%  If data access is restricted, authors must mention this in the statement. See
%  {http://www.ametsoc.org/PubsDataPolicy} for more info.

%%%%%%%%%%%%%%%%%%%%%%%%%%%%%%%%%%%%%%%%%%%%%%%%%%%%%%%%%%%%%%%%%%%%%
% APPENDIXES
%%%%%%%%%%%%%%%%%%%%%%%%%%%%%%%%%%%%%%%%%%%%%%%%%%%%%%%%%%%%%%%%%%%%%
%

%%%%%%%%%%%%%
\appendix[A] 
%% Appendix title is necessary! For appendix title:
\appendixtitle{Appendix: Model architecture details}
\label{app:archi}

\begin{sidewaysfigure*}
\centering
\includegraphics[width=1\textheight]{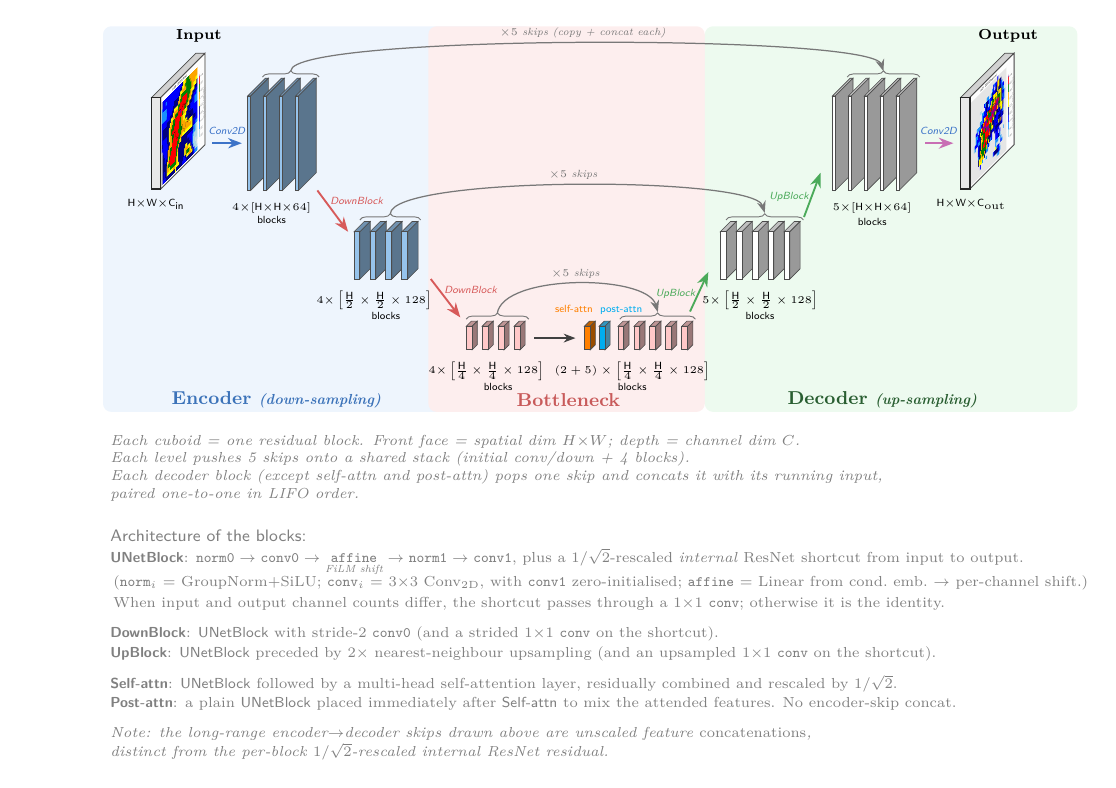}
\caption{\textbf{UNet Architecture, shared by both steps of the SwAIther-Precip pipelines.}}
\label{Fig:UNet_archi}
\end{sidewaysfigure*}

\begin{figure*}
\centering
\includegraphics[width=1\textwidth]{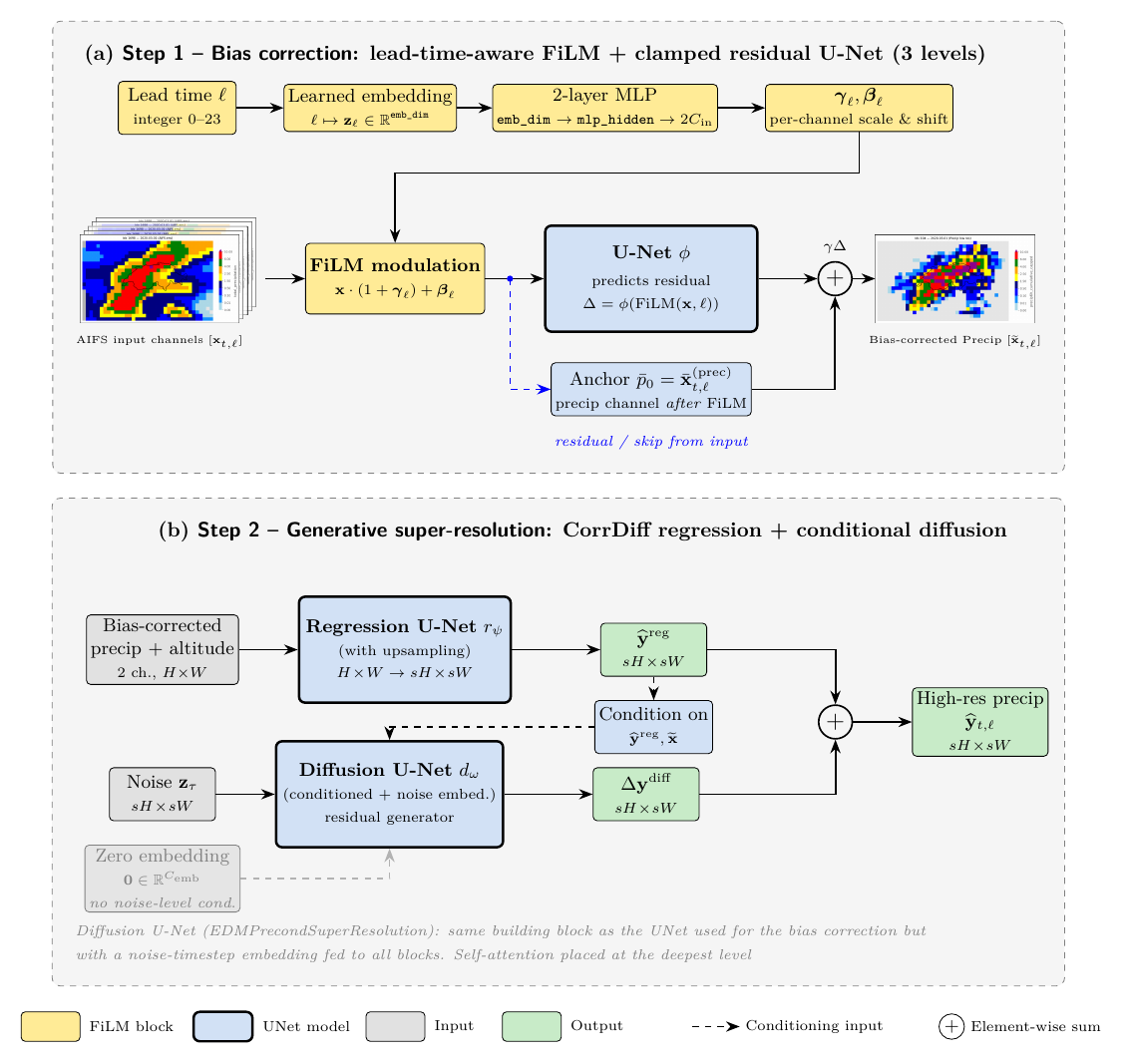}
\caption{\textbf{Detailed architecture for Step 1 (Bias correction) and Step 2 (Super-resolution).} The U-Net architecture used within the different steps is detailed in Figure \ref{Fig:UNet_archi}. Note that the Super-resolution architecture (b) is mostly taken from the CorrDiff implementation.}
\label{Fig:Archi_detailed}
\end{figure*}

We present in Figure \ref{Fig:UNet_archi} the detailed architecture of the UNet model we use in both steps of the SwAIther framework, as well as architectural details for both steps in Figure \ref{Fig:Archi_detailed}. For a detailed list of hyperparameters, and discussions regarding choices the bias correction output clamp value and the noise-level embedding of the residual diffusion sub-network of Step~2, please see the Supplementary Material (Section 3).

\clearpage

%%%%%%%%%%%%%%%%%%%%%%%%%%%%%%%%%%%%%%%%%%%%%%%%%%%%%%%%%%%%%%%%%%%%%
% REFERENCES
%%%%%%%%%%%%%%%%%%%%%%%%%%%%%%%%%%%%%%%%%%%%%%%%%%%%%%%%%%%%%%%%%%%%%
% Make your BibTeX bibliography by using these commands:
\bibliographystyle{ametsocV6}
\bibliography{references_revision}

\clearpage
\onecolumn
\appendix[S]
\appendixtitle{Supplementary Material}
\input{supplementary_with_corrections}

\end{document}

%% file: supplementary_with_corrections.tex
\section{AIFS-Single-1.0: Input/Output Variable List}
\label{sec_SI:aifs_variables}
%%%%%%%%%%%%%%%%%%%%%%%%%%%%%%%%%%%%%%%%%%%%%%%%%%%%%%%%%%%%%%%%%%%%%

\smallskip
\noindent We present in Table \ref{tab:AIFS_variables} and Table \ref{tab:AIFS_inputs_outputs} the list of input and output fields for the \texttt{AIFS-single-1.0} model we use to create a dataset of precipitation forecasts, input to our downscaling framework.

\begin{table}[h]
\centering
\caption{Variables of interest for the \texttt{AIFS-single-1.0} model. The ``Used'' column indicates the variables retained as inputs to the SwAIther-Precip framework (these variables are also listed in bold for better visilibity); for pressure-level variables, the specific levels used are given in parentheses. Note: an additional static altitude channel from the SwissTopo DHM25 digital elevation model is used as an invariant input alongside the marked AIFS variables.}
\label{tab:AIFS_variables}
\resizebox{0.75\textwidth}{!}{%
\begin{tabular}{llll}
\toprule
Variable name & Description & Units & Used \\
\midrule
% Pressure-level variables
\texttt{z} & \textbf{Geopotential} (pressure levels) & m$^2$ s$^{-2}$ & \checkmark (500, 850 hPa) \\
\texttt{t} & \textbf{Air temperature} (pressure levels) & K & \checkmark (500, 850 hPa) \\
\texttt{u} & Zonal wind component (pressure levels) & m s$^{-1}$ & \\
\texttt{v} & Meridional wind component (pressure levels) & m s$^{-1}$ & \\
\texttt{w} & Vertical velocity (pressure coordinates) & Pa s$^{-1}$ & \\
\texttt{q} & \textbf{Specific humidity} (pressure levels) & kg kg$^{-1}$ & \checkmark (500, 850 hPa) \\
% Surface variables
\texttt{sp} & Surface pressure & Pa & \\
\texttt{msl} & Mean sea-level pressure & Pa & \\
\texttt{skt} & Skin temperature & K & \\
\texttt{sst} & Sea surface temperature & K & \\
\texttt{2t} & 2 m air temperature & K & \\
\texttt{2d} & 2 m dewpoint temperature & K & \\
\texttt{10u} & \textbf{10 m zonal wind component} & m s$^{-1}$ & \checkmark \\
\texttt{10v} & \textbf{10 m meridional wind component} & m s$^{-1}$ & \checkmark \\
\texttt{100u} & 100 m zonal wind component & m s$^{-1}$ & \\
\texttt{100v} & 100 m meridional wind component & m s$^{-1}$ & \\
\texttt{tcw} & Total column water & kg m$^{-2}$ & \\
% Soil variables
\texttt{swvl1} & Soil moisture layer 1 (volumetric) & m$^{3}$ m$^{-3}$ & \\
\texttt{swvl2} & Soil moisture layer 2 (volumetric) & m$^{3}$ m$^{-3}$ & \\
\texttt{stl1} & Soil temperature layer 1 & K & \\
\texttt{stl2} & Soil temperature layer 2 & K & \\
% Precipitation & hydrology
\texttt{tp} & \textbf{Total precipitation} & kg m$^{-2}$ & \checkmark \\
\texttt{cp} & \textbf{Convective precipitation} & kg m$^{-2}$ & \checkmark \\
\texttt{sf} & Snowfall water equivalent & kg m$^{-2}$ & \\
\texttt{rowe} & Runoff water equivalent & kg m$^{-2}$ & \\
% Radiation
\texttt{ssrd} & Surface solar radiation downwards & J m$^{-2}$ & \\
\texttt{strd} & Surface thermal radiation downwards & J m$^{-2}$ & \\
% Clouds
\texttt{tcc} & \textbf{Total cloud cover} & 0--1 (fraction) & \checkmark \\
\texttt{hcc} & High cloud cover & 0--1 (fraction) & \\
\texttt{mcc} & Medium cloud cover & 0--1 (fraction) & \\
\texttt{lcc} & Low cloud cover & 0--1 (fraction) & \\
% Static / forcing fields
\texttt{lsm} & Land-sea mask & 0--1 (fraction) & \\
\texttt{sdor} & Standard deviation of sub-grid orography & m & \\
\texttt{slor} & Slope of sub-grid orography & --- & \\
\texttt{lat} & Latitude & degrees & \\
\texttt{lon} & Longitude & degrees & \\
\texttt{insol} & Insolation forcing & --- & \\
\texttt{tod} & Time of day (cyclical encoding) & --- & \\
\texttt{doy} & Day of year (cyclical encoding) & --- & \\
\bottomrule
\end{tabular}%
}
\end{table}

\begin{table}[h]
\centering
\caption{Full list of input and output fields for the \texttt{AIFS-single-1.0} model.}
\label{tab:AIFS_inputs_outputs}
\resizebox{0.65\textwidth}{!}{%
\begin{tabular}{p{6cm} p{3cm} p{2cm}}
\toprule
\textbf{\textsf{Field}} & \textbf{\textsf{Level type}} & \textbf{\textsf{Input/Output}} \\
\midrule
Geopotential, horizontal and vertical wind components, specific humidity, temperature & Pressure level: 50, 100, 150, 200, 250, 300, 400, 500, 600, 700, 850, 925, 1000 & Both \\
\midrule
Surface pressure, mean sea-level pressure, skin temperature, 2 m temperature, 2 m dewpoint temperature, 10 m horizontal wind components, total column water & Surface & Both \\
\midrule
Soil moisture and soil temperature (layers 1 and 2) & Surface & Both \\
\midrule
100m horizontal wind components, solar radiation (Surface short-wave (solar) radiation downwards and Surface long-wave (thermal) radiation downwards), cloud variables (\texttt{tcc}, \texttt{hcc}, \texttt{mcc}, \texttt{lcc}), runoff and snow fall & Surface & Output \\
\midrule
Total precipitation, convective precipitation & Surface & Output \\
\midrule
Land-sea mask, orography, standard deviation of sub-grid orography, slope of sub-scale orography, insolation, latitude/longitude, time of day/day of year & Surface & Input \\
\bottomrule
\end{tabular}%
}
\end{table}

%%%%%%%%%%%%%%%%%%%%%%%%%%%%%%%%%%%%%%%%%%%%%%%%%%%%%%%%%%%%%%%%%%%%%
\newpage
\section{Detailed results by lead time}
\label{sec_SI:metrics_per_lead}
%%%%%%%%%%%%%%%%%%%%%%%%%%%%%%%%%%%%%%%%%%%%%%%%%%%%%%%%%%%%%%%%%%%%%

%%%%%
\subsection{Bias correction metrics by lead time}  
%%%%%

\begin{figure}[h]
\hspace{-1cm}
\includegraphics[width=1.1\textwidth]{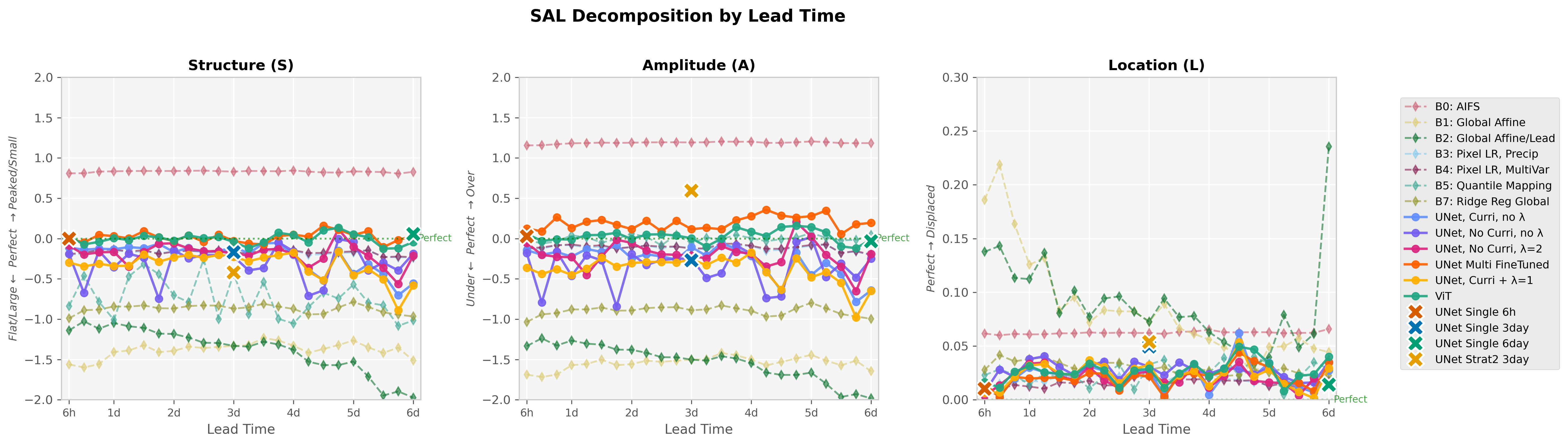}
\caption{\textbf{Lead-time evolution of SAL metrics for the bias correction task (Step 1).} S, A and L metrics are shown across all lead times. Results are shown for multi-lead UNet and ViT variants, all baselines, and single-lead models evaluated at their respective target lead times.}
\label{fig:SAL_BC_perLead}
\end{figure}

\begin{figure}[h]
\hspace{-1cm}
\includegraphics[width=1.1\textwidth]{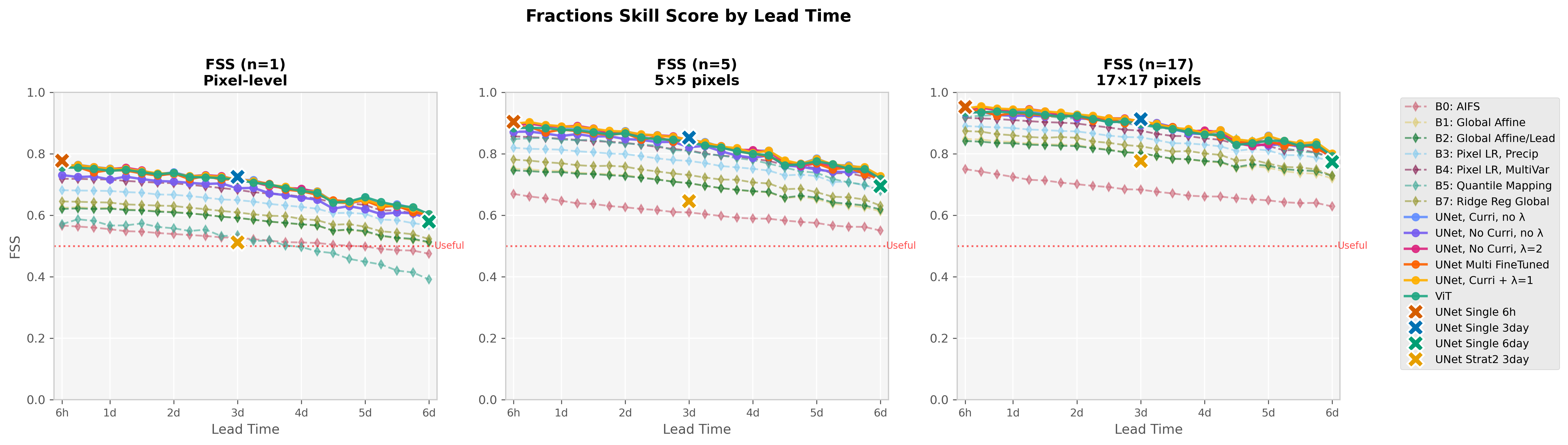}
\caption{\textbf{Lead-time evolution of FSS metrics for the bias correction task (Step 1).} Three different values of neighborhoods are considered here, across all lead times, to complement the results shown in the main text. Results are shown for multi-lead UNet and ViT variants, all baselines, and single-lead models evaluated at their respective target lead times.}
\label{fig:FSS_BC_perLead}
\end{figure}

\noindent The bias correction performance results were presented in the main text for a limited choice of lead times and neighborhood sizes regarding the SAL and FSS metric. To complement these results, Figures \ref{fig:SAL_BC_perLead} and \ref{fig:FSS_BC_perLead} present detailed results for the SAL and FSS metrics, across all lead times. In particular, S, A and L metrics are shown in Figure \ref{fig:SAL_BC_perLead}, and FSS metrics are given for three values of neighborhood size ($n=1$, $n=5$, and $n=17$) in Figure \ref{fig:FSS_BC_perLead}. 

%%%%%
\subsection{Downscaling metrics by lead time}
%%%%%

\begin{figure}[h]
\hspace{-1cm}
\includegraphics[width=1.1\textwidth]{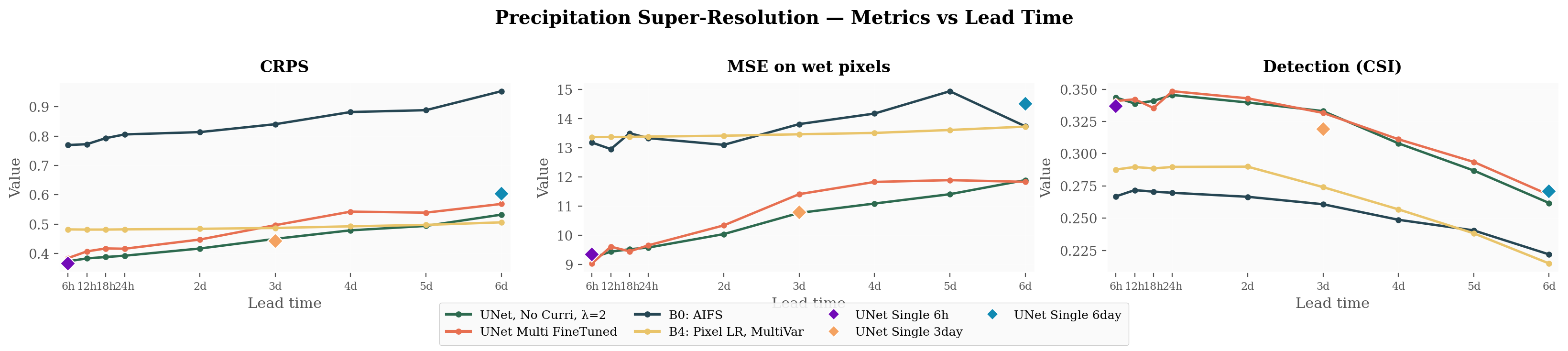}
\caption{\textbf{Lead-time evolution of key downscaling metrics for the full two-step pipeline.} Results are shown for multi-lead UNet variants, baselines (B0, B4), and single-lead models evaluated at their respective target lead times.}
\label{fig:DIFF_plot_metrics_PerLead}
\end{figure}

\begin{figure*}[t!]
\hspace{-1cm}
\includegraphics[width=1.1\textwidth]{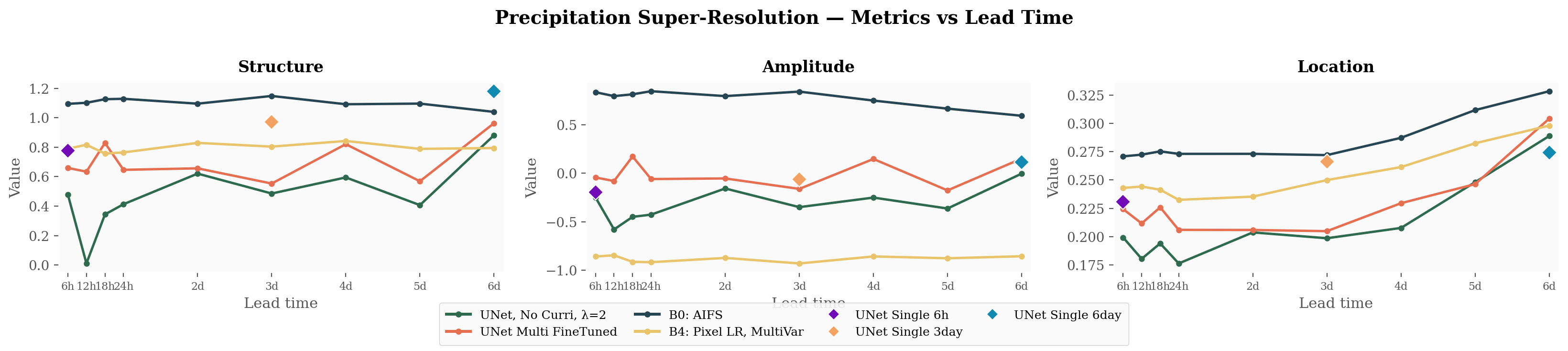}
\caption{Lead-time evolution of SAL metrics for the full two-step pipeline. Results are shown for multi-lead UNet variants, baselines (B0, B4), and single-lead models evaluated at their respective target lead times.}
\label{DIFF_plot_metrics_PerLead_SAL}
\end{figure*}

\begin{figure*}[t!]
\hspace{-1cm}
\includegraphics[width=1.1\textwidth]{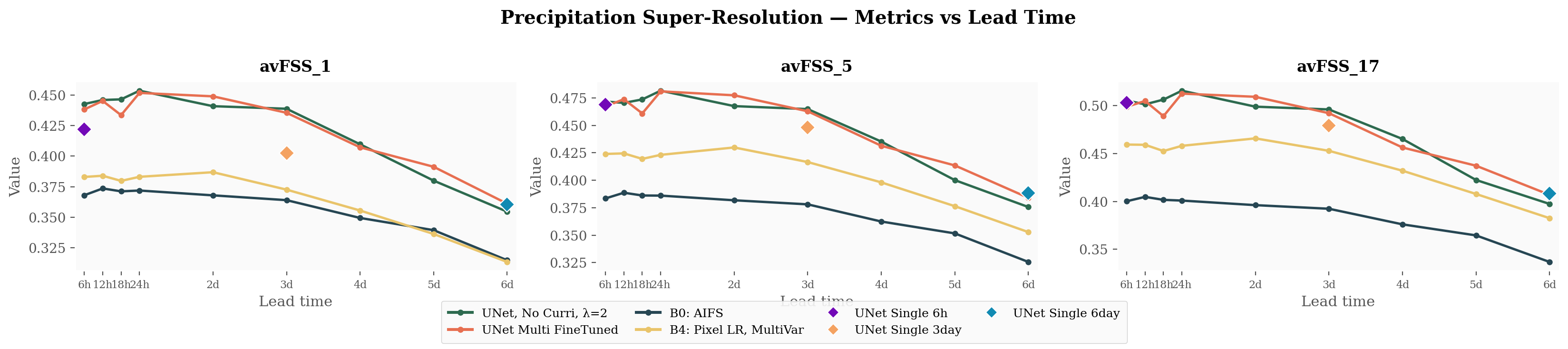}
\caption{Lead-time evolution of avFSS metrics for the full two-step pipeline. Results are shown for multi-lead UNet variants, baselines (B0, B4), and single-lead models evaluated at their respective target lead times.}
\label{DIFF_plot_metrics_PerLead_avFSS}
\end{figure*}

\noindent Table \ref{tab:sr_combined} shows per-lead-time  detailed results for the downscaling pipelines we consider in the main manuscript. In particular, it shows CRPS, CSI, member-averaged FSS$_5$, and MSE (wet pixels) by lead time across our models and baselines (including single-lead models). Figure \ref{fig:DIFF_plot_metrics_PerLead} visualizes the evolution of CRPS, CSI and MSE (wet pixels) with lead time, using line plots. Figures \ref{DIFF_plot_metrics_PerLead_avFSS} and \ref{DIFF_plot_metrics_PerLead_SAL} visualize the evolution of avFSS for different neighborhood choices and eS, eA and eL within the ensemble SAL metric with lead time, using line plots.

\begin{table}[h]
\centering
\caption{Downscaling (Step 2): CRPS, CSI, member-averaged FSS$_5$, and MSE (wet pixels) by lead time across our models and baselines (including single-lead models).}
\label{tab:sr_combined}
\resizebox{0.95\textwidth}{!}{%
\begin{tabular}{lcccccccccc}
\toprule
Model & 6h & 12h & 18h & 1d & 2d & 3d & 4d & 5d & 6d & Avg \\
\midrule
\midrule
\multicolumn{11}{c}{\normalsize{\textit{CRPS $\downarrow$}}} \\
B0: AIFS & 0.770 & 0.772 & 0.793 & 0.806 & 0.814 & 0.841 & 0.882 & 0.888 & 0.953 & 0.835 \\
B4: Pixel LR, MultiVar & 0.482 & 0.481 & 0.481 & 0.482 & 0.484 & 0.487 & 0.493 & 0.497 & \textbf{0.506} & 0.488 \\
\midrule
SwAIther-Precip (UNet$_{\lambda=2}$) & \textbf{0.373} & \textbf{0.383} & \textbf{0.388} & \textbf{0.392} & \textbf{0.417} & \textbf{0.450} & \textbf{0.479} & \textbf{0.494} & 0.532 & \textbf{0.434} \\
SwAIther-Precip (UNet Multi FT) & 0.384 & 0.407 & 0.417 & 0.416 & 0.448 & 0.497 & 0.543 & 0.539 & 0.569 & 0.469 \\
\midrule
UNet Single 6\,h & 0.366 & --- & --- & --- & --- & --- & --- & --- & --- & --- \\
UNet Single 3\,d & --- & --- & --- & --- & --- & 0.442 & --- & --- & --- & --- \\
UNet Single 6\,d & --- & --- & --- & --- & --- & --- & --- & --- & 0.604 & --- \\
\midrule
\midrule
\multicolumn{11}{c}{\normalsize{\textit{CSI $\uparrow$}}} \\
B0: AIFS & 0.267 & 0.272 & 0.271 & 0.270 & 0.267 & 0.261 & 0.249 & 0.241 & 0.222 & 0.258 \\
B4: Pixel LR, MultiVar & 0.288 & 0.290 & 0.289 & 0.290 & 0.290 & 0.274 & 0.257 & 0.238 & 0.215 & 0.270 \\
\midrule
SwAIther-Precip (UNet$_{\lambda=2}$) & \textbf{0.344} & 0.339 & \textbf{0.341} & 0.346 & 0.340 & \textbf{0.333} & 0.308 & 0.287 & 0.262 & 0.322 \\
SwAIther-Precip (UNet Multi FT) & 0.341 & \textbf{0.342} & 0.335 & \textbf{0.348} & \textbf{0.343} & 0.332 & \textbf{0.311} & \textbf{0.294} & \textbf{0.268} & \textbf{0.324} \\
\midrule
UNet Single 6\,h & 0.337 & --- & --- & --- & --- & --- & --- & --- & --- & --- \\
UNet Single 3\,d & --- & --- & --- & --- & --- & 0.319 & --- & --- & --- & --- \\
UNet Single 6\,d & --- & --- & --- & --- & --- & --- & --- & --- & 0.271 & --- \\
\midrule
\midrule
\multicolumn{11}{c}{\normalsize{\textit{FSS$_5$ $\uparrow$}}} \\
B0: AIFS & 0.383 & 0.389 & 0.386 & 0.386 & 0.382 & 0.378 & 0.362 & 0.352 & 0.326 & 0.372 \\
B4: Pixel LR, MultiVar & 0.424 & 0.424 & 0.419 & 0.423 & 0.430 & 0.417 & 0.398 & 0.376 & 0.353 & 0.407 \\
\midrule
SwAIther-Precip (UNet$_{\lambda=2}$) & \textbf{0.472} & 0.470 & \textbf{0.474} & \textbf{0.482} & 0.468 & \textbf{0.465} & \textbf{0.435} & 0.400 & 0.376 & 0.449 \\
SwAIther-Precip (UNet Multi FT) & 0.467 & \textbf{0.474} & 0.461 & 0.481 & \textbf{0.477} & 0.463 & 0.431 & \textbf{0.413} & \textbf{0.384} & \textbf{0.450} \\
\midrule
UNet Single 6\,h & 0.469 & --- & --- & --- & --- & --- & --- & --- & --- & --- \\
UNet Single 3\,d & --- & --- & --- & --- & --- & 0.448 & --- & --- & --- & --- \\
UNet Single 6\,d & --- & --- & --- & --- & --- & --- & --- & --- & 0.388 & --- \\
\midrule
\midrule
\multicolumn{11}{c}{\normalsize{\textit{MSE (wet pixels) $\downarrow$}}} \\
B0: AIFS & 13.179 & 12.955 & 13.500 & 13.331 & 13.101 & 13.812 & 14.174 & 14.940 & 13.738 & 13.637 \\
B4: Pixel LR, MultiVar & 13.369 & 13.370 & 13.371 & 13.384 & 13.413 & 13.463 & 13.509 & 13.610 & 13.726 & 13.469 \\
\midrule
SwAIther-Precip (UNet$_{\lambda=2}$) & 9.224 & \textbf{9.437} & 9.520 & \textbf{9.578} & \textbf{10.040} & \textbf{10.766} & \textbf{11.087} & \textbf{11.408} & 11.889 & \textbf{10.328} \\
SwAIther-Precip (UNet Multi FT) & \textbf{9.032} & 9.608 & \textbf{9.447} & 9.655 & 10.338 & 11.409 & 11.828 & 11.890 & \textbf{11.832} & 10.560 \\
\midrule
UNet Single 6\,h & 9.347 & --- & --- & --- & --- & --- & --- & --- & --- & --- \\
UNet Single 3\,d & --- & --- & --- & --- & --- & 10.786 & --- & --- & --- & --- \\
UNet Single 6\,d & --- & --- & --- & --- & --- & --- & --- & --- & 14.506 & --- \\
\bottomrule
\end{tabular}
}
\end{table}

%%%%%%%%%%%%%%%%%%%%%%%%%%%%%%%%%%%%%%%%%%%%%%%%%%%%%%%%%%%%%%%%%%%%%
\clearpage 
\section{Architectural details - hyperparameters }
\label{sec_SI:archi_details}
%%%%%%%%%%%%%%%%%%%%%%%%%%%%%%%%%%%%%%%%%%%%%%%%%%%%%%%%%%%%%%%%%%%%%
\subsection{Detailed hyperparameter lists}

The choice of hyperparameters for the different models used in the methodology are presented in the following section. Note that we only present the two top performing bias correction models. Regarding Step~2 the same CorrDiff submodels (one regression network and one residual diffusion network) are used on top of all bias correction models tested; the hyperparameters for these two models are also presented here. 

\begin{hpbox}{(Step 1) UNet$_{\lambda=2}$ hyperparameters}
\begin{lstlisting}[style=hpconfig]
# --- Architecture (FiLM bias correction U-Net) ---
model_channels: 64
channel_mult: [1, 2, 2]
attn_resolutions: [4]
N_grid_channels: 4
embedding_type: zero
clamp_min: 0
clamp_max: 20
# --- FiLM lead-time conditioning ---
n_leads: 24
emb_dim: 64
mlp_hidden: 256
# --- Loss ---
huber_delta: 1.0
lambda_time: 2.0
lambda_time_warmup_epochs: 5
# --- Optimization ---
lr: 2e-4
seed: 42
total_batch_size: 1024
fp_optimizations: amp-bf16
\end{lstlisting}
\end{hpbox}

%# --- Curriculum learning ---
%curriculum.enabled: False

\begin{hpbox}{(Step 1) MultiFT hyperparameters}
\begin{lstlisting}[style=hpconfig]
# --- Architecture (FiLM bias correction U-Net) ---
model_channels:        64
channel_mult:          [1, 2, 2]
attn_resolutions:      [4]
N_grid_channels:       4
embedding_type:        zero 
# --- FiLM lead-time conditioning ---
n_leads:               24
emb_dim:               64
mlp_hidden:            256
# --- Output ---
clamp_min:             0
clamp_max:             20
# --- Training (constant across stages) ---
total_batch_size:      1024
optimizer:             Adam
loss:                  Huber (delta=1.0)
lr_decay_rate:         5e5
fp_optimizations:      amp-bf16
seed:                  42
# --- Multi-stage fine-tuning schedule ---
# stage  lead_index   lr      lam_time   lam_warmup
  1:     [0, 7]       2e-4    0.0          --
  2:     [8, 15]      2e-4    0.0          --
  3:     [0, 23]      2e-5    1.0          10
  4:     [12, 23]     1e-5    0.3          3
  5:     [16, 23]     2e-5    0.0          3
\end{lstlisting}
\end{hpbox}

\begin{hpbox}{(Step 2) Regression U-Net hyperparameters}
\begin{lstlisting}[style=hpconfig]
# --- Architecture (CorrDiff regression U-Net) ---
model_type:            SongUNetPosEmbd
model_channels:        64
channel_mult:          [1, 2, 2]
attn_resolutions:      [152]
N_grid_channels:       4
gridtype:              sinusoidal     # default; positional grid embed
hr_mean_conditioning:  False
# --- Training ---
optimizer:             Adam
loss:                  L2 (regression)
lr:                    2e-4
lr_decay_rate:         5e5
total_batch_size:      2
fp_optimizations:      amp-bf16
seed:                  42
\end{lstlisting}
\end{hpbox}

\begin{hpbox}{(Step 2) Diffusion U-Net hyperparameters}
\begin{lstlisting}[style=hpconfig]
# --- Architecture (CorrDiff diffusion U-Net, EDM) ---
model_type:            SongUNetPosEmbd
model_channels:        64
channel_mult:          [1, 2, 2]
attn_resolutions:      [152]
N_grid_channels:       4
gridtype:              sinusoidal
embedding_type:        zero      # noise-level conditioning disabled
                            # (departs from standard EDM; see SI)
hr_mean_conditioning:  True      # conditions on regression output
# --- Training ---
optimizer:             Adam
loss:                  EDM denoising
lr:                    2.e-4
lr_decay_rate:         5e5
total_batch_size:      2
fp_optimizations:      amp-bf16
seed:                  42
# --- Conditioning ---
regression_checkpoint: Step 2 regression U-Net (above)
\end{lstlisting}
\end{hpbox}

\subsection{Sensitivity to the output cap}
\label{app:clipping}

\smallskip
\noindent We retain a $[0, 20]$~mm/6\,h output cap for the Step~1 bias-correction model. The cap was originally introduced as a numerical safeguard during early training. While the climatological $99^{\text{th}}$ percentile of \textsc{CombiPrecip} wet-pixel intensities reaches $\sim 23$~mm/6\,h on our domain, we observed that raising the cap to $[0, 100]$~mm/6\,h gave only marginal differences at Step~1 (CSI: $0.539 \to 0.542$; FSS$_5$: $0.828 \to 0.834$; MSE$_\mathrm{wet}$: $0.066 \to 0.071$) and a small but consistent degradation in end-to-end skill once composed with the Step~2 generative super-resolution model (CRPS: $0.434 \to 0.437$; CSI: $0.322 \to 0.317$; FSS$_5$: $0.449 \to 0.440$; MSE$_\mathrm{wet}$: $10.33 \to 10.51$). We hypothesize that the tighter cap acts as a soft regularization on Step~1 outputs, producing a more spatially coherent and climatologically smoother conditioning signal for the Step~2 generative model, which is trained on bilinearly-downscaled \textsc{CombiPrecip} and therefore expects similarly bounded inputs. A more thorough analysis of this interaction is left for future work.

\subsection{Sensitivity to the noise-level embedding}
\label{app:embedding}

\smallskip
\noindent We tested two configurations for the noise-level embedding of the residual sub-network of Step~2 (the second sub-network of CorrDiff): \texttt{embedding\_type=zero} (constant zero embedding, no explicit noise-level conditioning) and \texttt{embedding\_type=positional} (the standard EDM sinusoidal positional embedding of the diffusion time-step). The \texttt{positional} variant achieved lower training loss and produced deterministically sharper samples (MSE$_\mathrm{wet}$: $10.33 \to 10.29$; CSI: $0.322 \to 0.325$), but generated less-dispersed ensembles, resulting in worse probabilistic skill (CRPS: $0.434 \to 0.447$). This is consistent with a known tension between sharpness and calibration in probabilistic forecasting \citep{gneiting2007strictly}: the simpler \texttt{zero} variant, by averaging implicitly across the diffusion noise schedule, appears to produce ensembles whose spread better matches the actual forecast uncertainty distribution, while the \texttt{positional} variant produces more accurate but underdispersed samples. We retain \texttt{embedding\_type=zero} as the final configuration based on CRPS, which is the primary probabilistic skill metric in our evaluation framework.

%%%%%%%%%%%%%%%%%%%%%%%%%%%%%%%%%%%%%%%%%%%%%%%%%%%%%%%%%%%%%%%%%%%%%

\section{Additional Experiments}
\label{sec_SI:additional_results}
%%%%%%%%%%%%%%%%%%%%%%%%%%%%%%%%%%%%%%%%%%%%%%%%%%%%%%%%%%%%%%%%%%%%%

%%%%%%%%%%%%%%%%%%%%%%%%%%%
\subsection{Summary}
%%%%%%%%%%%%%%%%%%%%%%%%%%%

\smallskip
\noindent This section presents additional diagnostic experiments mentioned in Section 3c (Data Section, Weekly train-test split subsection) of the main text. We address two potential concerns regarding the weekly train--test split: (i)~whether temporal autocorrelation leads to an overestimation of model performance, and (ii)~whether the model trained on the 06--12 and 18--00\,UTC windows can generalize to the held-out windows. Numerical results are reported in Tables~\ref{tab:sr_metrics_compact_addResults},~\ref{tab:sr_mse_wet_addResults} and~\ref{tab:sr_crps_addResults}.

\smallskip
\noindent We summarize the three main findings:

\begin{enumerate}
    \item \textbf{Temporal autocorrelation is not a significant confound.} The field-level autocorrelation function of 6-hour accumulated precipitation over Switzerland drops below 0.2 within 12\,hours (see supplementary material, Figure 4), indicating that consecutive samples are largely independent at the pixel level. A neutral-terrain comparison further confirms this: the weekly-trained model evaluated exclusively on 2023 (CRPS\,=\,0.533) performs comparably to a yearly-trained model on the same period (CRPS\,=\,0.507), showing that the weekly split does not inflate performance through leaked autocorrelation (Tables 4, 5 and 6 in the supplementary material).
    \item \textbf{Performance differences across years are driven by distribution shift, not leakage.} The seasonal decomposition of the yearly model's 2023 evaluation reveals a $\times$13 variation in MSE$_{\mathrm{wet}}$ between winter (2.4) and summer (31.5), explaining the apparent degradation when evaluating on the full year 2023 compared to the weekly validation set.
    \item \textbf{Performance on held-out time windows reflects intrinsic difficulty, not a generalization gap.} A model-independent comparison of raw AIFS forecasts against observations shows that the 12--18\,UTC window (afternoon convection) is nearly twice as difficult as the 00--06\,UTC window ($\times$1.96 MSE$_{\mathrm{wet}}$ ratio, Table 7 in the supplementary material). The training set already includes a window of comparable difficulty (18--00\,UTC, $\times$1.85).
\end{enumerate}

%%%%%%%%%%%%%%%%%%%%%%%%%%%
\subsection{Experiments}
%%%%%%%%%%%%%%%%%%%%%%%%%%%

\smallskip
\noindent We define multiple experiment labels used throughout, based on the same architecture for the bias correction model, the UNet$_{\lambda=2}$ model. For every experiment, we evaluate the full pipeline inference, at the high resolution level. The experiments are as follows:
\begin{itemize}
    \item \textbf{UNet$_{\lambda=2}$}: reference model trained on the weekly split, evaluated on its own test set (06--12 and 18--00\,UTC windows only).
    \item \textbf{UNet$_{\lambda=2}$, weekly, new Hours}: same weekly-trained model, evaluated on the held-out windows (00--06 and 12--18\,UTC combined).
    \item \textbf{UNet$_{\lambda=2}$, weekly new Hours, only 18}: same model, evaluated on the 12--18\,UTC window only.
    \item \textbf{UNet$_{\lambda=2}$, yearly, val 2023}: model trained on a yearly split (2019--2022), evaluated on a representative sample of 2023 spanning all months and seasons.
    \item \textbf{UNet$_{\lambda=2}$, weekly, on 2023}: weekly-trained model evaluated exclusively on the year 2023.
    \item \textbf{UNet$_{\lambda=2}$, yearly, on Nov--Dec}: yearly-trained model evaluated on November--December only.
    \item \textbf{UNet$_{\lambda=2}$, yearly, val 2023, \{season\}}: yearly-trained model evaluated on 2023, stratified by season.
\end{itemize}

\begin{table}[h]
\centering
\caption{Additional downscaling validation experiments: aggregate metrics for the pipeline using the bias correction model (UNet, $\lambda_{\mathrm{time}}=2$) under different train--test splits and evaluation conditions. All metrics are averaged over lead times 6\,h to 6\,days.}
\label{tab:sr_metrics_compact_addResults}
\resizebox{\textwidth}{!}{%
\begin{tabular}{lcccccccc}
\toprule
Model & crps & mse & mse\_wet & csi & avfss\_5 & eS & eA & eL \\
\midrule
UNet$_{\lambda=2}$ & 0.434 & 3.570 & 10.328 & 0.322 & 0.449 & 0.473 & -0.313 & 0.211 \\
UNet$_{\lambda=2}$, weekly, new Hours & 0.374 & 3.011 & 10.039 & 0.306 & 0.423 & 0.596 & -0.276 & 0.205 \\
UNet$_{\lambda=2}$, weekly new Hours, only 18 & 0.500 & 4.836 & 13.733 & 0.325 & 0.443 & 0.602 & -0.392 & 0.201 \\
UNet$_{\lambda=2}$, yearly, on Nov-dec & 0.352 & 1.642 & 3.802 & 0.376 & 0.505 & 0.137 & -0.506 & 0.195 \\
UNet$_{\lambda=2}$, yearly, val 2023 & 0.507 & 3.833 & 12.968 & 0.361 & 0.471 & 0.541 & -0.292 & 0.235 \\
UNet$_{\lambda=2}$, weekly, on 2023 & 0.533 & 4.937 & 12.226 & 0.303 & 0.423 & 0.062 & -0.515 & 0.238 \\
UNet$_{\lambda=2}$, yearly, val 2023, winter & 0.305 & 1.165 & 2.380 & 0.392 & 0.546 & 0.398 & -0.244 & 0.162 \\
UNet$_{\lambda=2}$, yearly, val 2023, spring & 0.422 & 2.065 & 7.924 & 0.409 & 0.544 & 0.499 & -0.445 & 0.158 \\
UNet$_{\lambda=2}$, yearly, val 2023, summer & 0.754 & 8.338 & 31.521 & 0.302 & 0.411 & 0.698 & -0.522 & 0.244 \\
UNet$_{\lambda=2}$, yearly, val 2023, autumn & 0.493 & 3.164 & 6.456 & 0.343 & 0.450 & 0.199 & -0.325 & 0.221 \\
\bottomrule
\end{tabular}%
}
\end{table}

\begin{table}[h]
\centering
\caption{MSE on wet pixels (observed precipitation $>0.1$\,mm) resolved by lead time for the additional validation experiments. The seasonal decomposition of the yearly model on 2023 (bottom four rows) reveals a $\times$13 variation between winter and summer, illustrating the dominant role of convective regimes in driving prediction difficulty.}
\label{tab:sr_mse_wet_addResults}
\resizebox{\textwidth}{!}{%
\begin{tabular}{lcccccccccc}
\toprule
Model & 6h & 12h & 18h & 1d & 2d & 3d & 4d & 5d & 6d & Avg \\
\midrule
UNet$_{\lambda=2}$ & 9.224 & 9.437 & 9.520 & 9.578 & 10.040 & 10.766 & 11.087 & 11.408 & 11.889 & 10.328 \\
UNet$_{\lambda=2}$, weekly, new Hours & 9.228 & 10.225 & 9.620 & 9.730 & 9.679 & 10.118 & 10.395 & 10.640 & 10.718 & 10.039 \\
UNet$_{\lambda=2}$, weekly new Hours, only 18 & 12.768 & 14.704 & 13.348 & 13.591 & 13.168 & 13.331 & 14.023 & 14.004 & 14.660 & 13.733 \\
UNet$_{\lambda=2}$, yearly, on Nov-dec & 2.797 & 2.935 & 3.129 & 3.290 & 3.518 & 3.664 & 4.411 & 5.074 & 5.399 & 3.802 \\
UNet$_{\lambda=2}$, yearly, val 2023 & 12.314 & 12.170 & 12.406 & 12.438 & 12.698 & 12.885 & 13.278 & 13.913 & 14.607 & 12.968 \\
UNet$_{\lambda=2}$, weekly, on 2023 & 10.686 & 10.666 & 10.961 & 11.483 & 11.887 & 13.308 & 13.390 & 13.746 & 13.910 & 12.226 \\
UNet$_{\lambda=2}$, yearly, val 2023, winter & 1.582 & 1.859 & 1.746 & 2.031 & 2.434 & 2.310 & 2.811 & 3.123 & 3.522 & 2.380 \\
UNet$_{\lambda=2}$, yearly, val 2023, spring & 7.209 & 6.888 & 7.124 & 7.012 & 7.897 & 7.846 & 8.456 & 9.115 & 9.765 & 7.924 \\
UNet$_{\lambda=2}$, yearly, val 2023, summer & 31.539 & 31.101 & 31.339 & 30.981 & 31.076 & 31.440 & 31.308 & 32.419 & 32.488 & 31.521 \\
UNet$_{\lambda=2}$, yearly, val 2023, autumn & 5.253 & 5.251 & 5.767 & 6.152 & 5.868 & 6.356 & 7.020 & 7.384 & 9.049 & 6.456 \\
\bottomrule
\end{tabular}
}
\end{table}

\begin{table}[h]
\centering
\caption{CRPS resolved by lead time for the additional validation experiments. The weekly-trained model evaluated on 2023 (\emph{weekly, on 2023}) and the yearly-trained model (\emph{yearly, val 2023}) track each other closely across all lead times, confirming that the weekly split does not produce inflated short-range scores through temporal autocorrelation.}
\label{tab:sr_crps_addResults}
\resizebox{\textwidth}{!}{%
\begin{tabular}{lcccccccccc}
\toprule
Model & 6h & 12h & 18h & 1d & 2d & 3d & 4d & 5d & 6d & Avg \\
\midrule
UNet$_{\lambda=2}$ & 0.373 & 0.383 & 0.388 & 0.392 & 0.417 & 0.450 & 0.479 & 0.494 & 0.532 & 0.434 \\
UNet$_{\lambda=2}$, weekly, new Hours & 0.330 & 0.342 & 0.332 & 0.337 & 0.356 & 0.374 & 0.406 & 0.438 & 0.452 & 0.374 \\
UNet$_{\lambda=2}$, weekly new Hours, only 18 & 0.449 & 0.472 & 0.454 & 0.461 & 0.485 & 0.492 & 0.540 & 0.568 & 0.581 & 0.500 \\
UNet$_{\lambda=2}$, yearly, on Nov-dec & 0.293 & 0.303 & 0.311 & 0.321 & 0.327 & 0.343 & 0.384 & 0.424 & 0.462 & 0.352 \\
UNet$_{\lambda=2}$, yearly, val 2023 & 0.435 & 0.450 & 0.461 & 0.475 & 0.486 & 0.511 & 0.545 & 0.559 & 0.635 & 0.507 \\
UNet$_{\lambda=2}$, weekly, on 2023 & 0.460 & 0.465 & 0.485 & 0.507 & 0.513 & 0.576 & 0.581 & 0.599 & 0.607 & 0.533 \\
UNet$_{\lambda=2}$, yearly, val 2023, winter & 0.242 & 0.259 & 0.257 & 0.276 & 0.290 & 0.287 & 0.329 & 0.375 & 0.425 & 0.305 \\
UNet$_{\lambda=2}$, yearly, val 2023, spring & 0.368 & 0.369 & 0.384 & 0.397 & 0.412 & 0.439 & 0.454 & 0.457 & 0.517 & 0.422 \\
UNet$_{\lambda=2}$, yearly, val 2023, summer & 0.681 & 0.702 & 0.706 & 0.718 & 0.737 & 0.774 & 0.803 & 0.805 & 0.857 & 0.754 \\
UNet$_{\lambda=2}$, yearly, val 2023, autumn & 0.397 & 0.420 & 0.444 & 0.458 & 0.453 & 0.488 & 0.539 & 0.549 & 0.687 & 0.493 \\
\bottomrule
\end{tabular}
}
\end{table}

\subsubsection{Temporal autocorrelation of precipitation fields}
\label{subsec_SI:acf_analysis}

\smallskip
\noindent To quantify the temporal dependence between consecutive samples, we compute the temporal autocorrelation function (ACF) of the 6-hour accumulated precipitation fields from the CombiPrecip observational dataset over Switzerland, at the 12-hour sampling cadence corresponding to the actual spacing between consecutive samples in the dataset.

\smallskip
\noindent We consider two complementary ACF estimates. The \emph{spatial-mean ACF} is computed on the domain-averaged precipitation time series and captures the persistence of large-scale synoptic patterns. The \emph{field-level ACF} is computed independently at each grid point and then averaged spatially: it captures the local decorrelation relevant to the pixel-level predictions made by the model.

\begin{figure}[h]
    \centering
    \includegraphics[width=\textwidth]{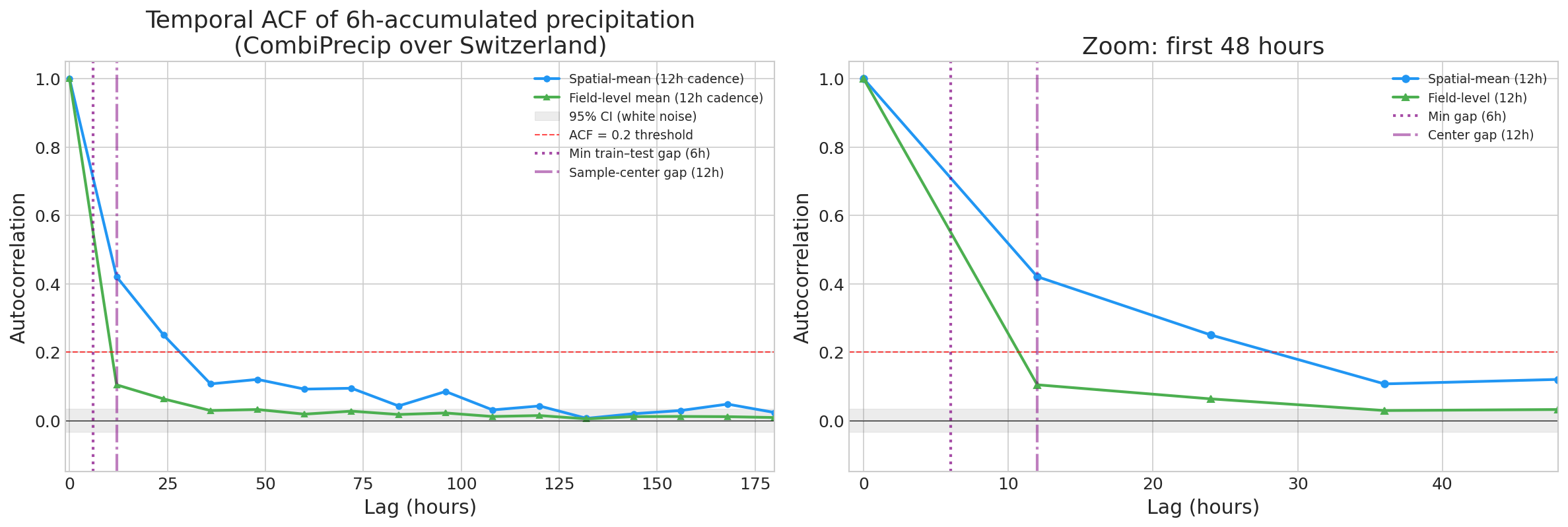}
    \caption{Temporal autocorrelation function (ACF) of 6-hour accumulated precipitation from CombiPrecip over Switzerland. \textbf{Left:} Full ACF up to 15\,days. \textbf{Right:} Zoom on the first 48\,hours. The spatial-mean ACF (blue) captures synoptic-scale persistence and decorrelates at approximately 36\,h. The field-level ACF (green), computed per pixel and then spatially averaged, decorrelates within 12\,h. Vertical lines indicate the minimum train--test boundary gap (6\,h, dotted) and the sample-center gap (12\,h, dash-dotted). The red dashed line marks the ACF\,=\,0.2 threshold.}
    \label{fig:acf}
\end{figure}

\smallskip
\noindent Results are shown in Figure~\ref{fig:acf}. The spatial-mean ACF exhibits a lag-1 (12\,h) autocorrelation of 0.42 and falls below the 0.2 threshold at approximately 36\,hours (1.5\,days), reflecting the persistence of synoptic-scale weather regimes over the domain. In contrast, the field-level ACF decorrelates much faster, dropping below 0.2 within 12\,hours (0.5\,days). This indicates that while the domain-averaged precipitation signal may persist for 1--2 days, the local pixel-level structure---which is what the bias correction and downscaling models must predict---changes substantially between consecutive samples.

\smallskip
\noindent At the weekly split boundary, the minimum gap between the end of the last training accumulation window and the start of the first test accumulation window is 6\,hours. The gap between consecutive sample centers is 12\,hours. Since the field-level ACF is already below 0.2 at a 12-hour lag, the effective independence between train and test samples at the boundary is high at the spatial scale relevant to the model.

\subsubsection{Neutral-terrain comparison: weekly vs.\ yearly split}
\label{subsec_SI:weekly_vs_yearly}

\smallskip
\noindent To empirically verify that the weekly split does not inflate model performance through temporal leakage, we compare two models on a common, previously unseen test period. Both models use the same architecture (UNet, $\lambda_{\mathrm{time}}=2$) and hyperparameters; only the splitting strategy differs.

\smallskip
\noindent The key comparison involves three rows in Table~\ref{tab:sr_metrics_compact_addResults}:
\begin{enumerate}
    \item \emph{UNet$_{\lambda=2}$, weekly}: the reference model on its own weekly validation set (CRPS\,=\,0.434, MSE$_{\mathrm{wet}}$\,=\,10.3).
    \item \emph{UNet$_{\lambda=2}$, weekly, on 2023}: the same weekly-trained model, evaluated exclusively on 2023 (CRPS\,=\,0.533, MSE$_{\mathrm{wet}}$\,=\,12.2).
    \item \emph{UNet$_{\lambda=2}$, yearly, val 2023}: the yearly-trained model, evaluated on the same 2023 period (CRPS\,=\,0.507, MSE$_{\mathrm{wet}}$\,=\,13.0).
\end{enumerate}

\smallskip
\noindent If the weekly split had produced an artificially strong model through temporal leakage, one would expect it to degrade substantially on the neutral 2023 test set, performing significantly worse than a yearly-trained model that has never relied on neighboring training days. Instead, the two models perform comparably on 2023, with the weekly-trained model even slightly better on MSE$_{\mathrm{wet}}$ (12.2 vs.\ 13.0). This indicates that the weekly split does not produce an inflated model.

\smallskip
\noindent The degradation from the weekly validation set (CRPS\,=\,0.434) to the 2023 evaluation (CRPS\,=\,0.533) is therefore attributable to a distribution shift rather than to the disappearance of leaked autocorrelation. This interpretation is strongly supported by the seasonal decomposition of the yearly model's 2023 metrics (Table~\ref{tab:sr_mse_wet_addResults}): MSE$_{\mathrm{wet}}$ ranges from 2.4 in winter to 31.5 in summer---a $\times$13 factor---revealing that performance is dominated by the intrinsic difficulty of convective precipitation regimes rather than by the choice of splitting strategy.

\smallskip
\noindent The result on November--December alone (\emph{UNet$_{\lambda=2}$, yearly, on Nov--Dec}: CRPS\,=\,0.352, MSE$_{\mathrm{wet}}$\,=\,3.8) further illustrates this seasonal effect: evaluating on a period dominated by stratiform precipitation yields substantially better metrics, regardless of the splitting strategy used.

\smallskip
\noindent Lead-time-resolved results (Tables~\ref{tab:sr_mse_wet_addResults} and~\ref{tab:sr_crps_addResults}) confirm that the degradation pattern is consistent across all lead times: the weekly-on-2023 and yearly-on-2023 models track each other closely from 6\,h through 6\,days, with both exhibiting the expected monotonic increase in error with lead time.

\subsubsection{Intrinsic difficulty across accumulation windows}
\label{subsec_SI:intrinsic_difficulty}

\smallskip
\noindent To disentangle the effect of the diurnal cycle from potential generalization gaps, we measure the intrinsic prediction difficulty of each 6-hour accumulation window by comparing raw AIFS forecasts directly against CombiPrecip observations coarsened to the AIFS grid via bilinear interpolation. This comparison involves no trained ML model and thus provides a model-independent baseline of difficulty.

Table~\ref{tab:intrinsic_difficulty} reports the MSE on wet pixels (observed precipitation $>0.1$\,mm) for each window, averaged over all lead times.

\begin{table}[h]
    \centering
    \caption{Intrinsic difficulty of raw AIFS predictions vs.\ observations by accumulation window. MSE$_{\mathrm{wet}}$ is computed on the AIFS grid after coarsening observations via bilinear interpolation. The difficulty ratio is relative to the easiest window (00--06\,UTC).}
    \label{tab:intrinsic_difficulty}
    \begin{tabular}{lccc}
        \toprule
        Window (UTC) & Role & MSE$_{\mathrm{wet}}$ & Difficulty ratio \\
        \midrule
        00--06 & held-out & 13.6 & $\times 1.00$ \\
        06--12 & training & 15.9 & $\times 1.17$ \\
        12--18 & held-out & 26.6 & $\times 1.96$ \\
        18--00 & training & 25.1 & $\times 1.85$ \\
        \bottomrule
    \end{tabular}
\end{table}

\smallskip
\noindent The 12--18\,UTC window (afternoon convection) is nearly twice as difficult as the 00--06\,UTC window (nighttime, predominantly stratiform precipitation). Crucially, the 18--00\,UTC window---which \emph{is} included in the training set---exhibits a comparable level of difficulty ($\times 1.85$), reflecting the persistence of convective activity into the evening hours. The training set therefore already exposes the model to the challenging convective regime.

\smallskip
\noindent This difficulty structure explains the results observed when evaluating the weekly-trained model on the held-out time windows (Table~\ref{tab:sr_metrics_compact_addResults}):
\begin{itemize}
    \item \emph{UNet$_{\lambda=2}$, weekly, new Hours} (both held-out windows combined): CRPS\,=\,0.374, MSE$_{\mathrm{wet}}$\,=\,10.0. This is \emph{better} than the reference model's own validation score (CRPS\,=\,0.434), because the easy 00--06 window dominates the average.
    \item \emph{UNet$_{\lambda=2}$, weekly new Hours, only 18} (12--18\,UTC window alone): CRPS\,=\,0.500, MSE$_{\mathrm{wet}}$\,=\,13.7. The degradation relative to the reference is consistent with the $\times 1.96$ intrinsic difficulty ratio of this window compared to the easiest window.
\end{itemize}

\smallskip
\noindent The model thus does not exhibit a generalization \emph{failure} on the held-out windows; rather, its performance reflects the diurnal cycle of precipitation predictability over Switzerland.

%%%%%%%%%%%%%%%%%%%%%%%%%%%%%%%%%%%%%%%%%%%%%%%%%%%%%%%%%%%%%%%%%%%%%
\newpage
\section{Additional Visualizations: November 2023 Heavy Precipitation Event}
\label{sec_SI:nov_event}
%%%%%%%%%%%%%%%%%%%%%%%%%%%%%%%%%%%%%%%%%%%%%%%%%%%%%%%%%%%%%%%%%%%%%

\smallskip
\noindent We present test diffusion samples from the yearly trained SwAIther-Precip pipeline based on the UNet$_{\lambda=2}$ bias correction model, for a particularly heavy precipitation event from November 12$^{\text{th}}$ to November 15$^{\text{th}}$, 2023. For visualization purposes, we show the first ensemble sample (member 0) out of 12 diffusion samples. Each figure shows the downscaling output across all forecast lead times.

\begin{figure}[h]
\hspace{-1cm}
\includegraphics[width=1.1\textwidth]{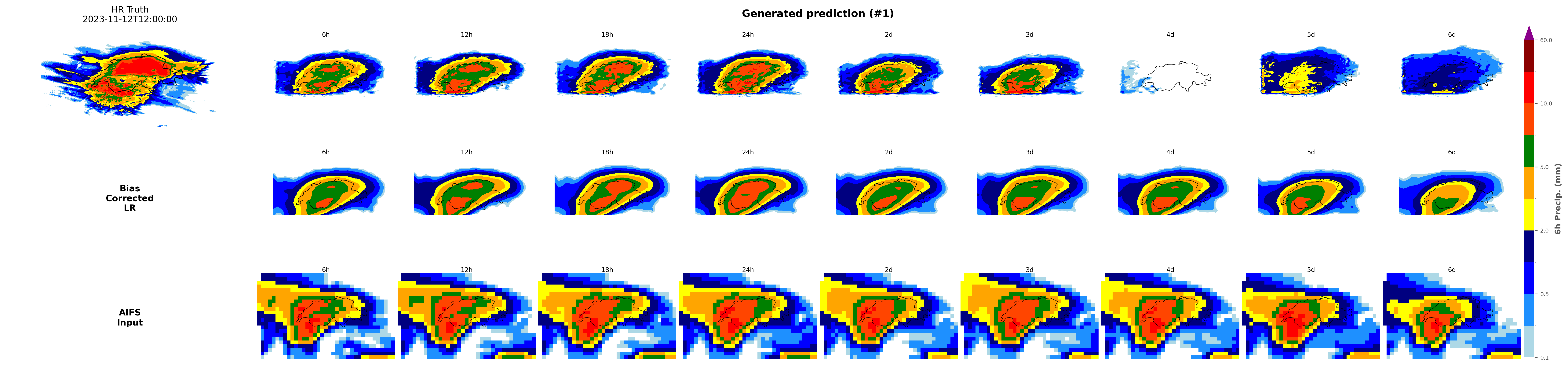}
\caption{Downscaling samples by lead time (member 0) for the November 2023 heavy precipitation event (1/7).}
\label{fig:eventNov_1}
\end{figure}

\begin{figure}[h]
\hspace{-1cm}
\includegraphics[width=1.1\textwidth]{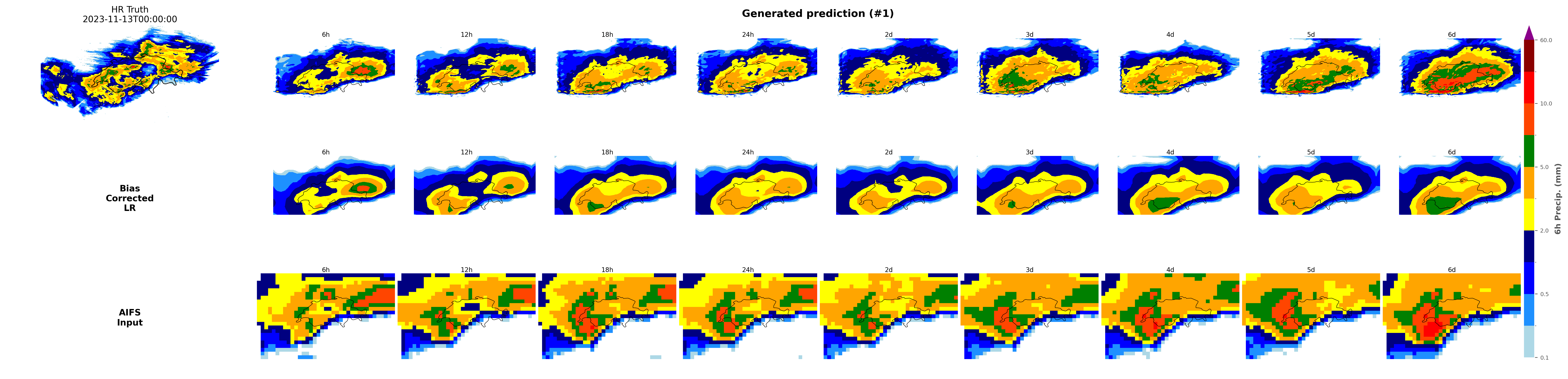}
\caption{Downscaling samples by lead time (member 0) for the November 2023 heavy precipitation event (2/7).}
\label{fig:eventNov_2}
\end{figure}

\begin{figure}[h]
\hspace{-1cm}
\includegraphics[width=1.1\textwidth]{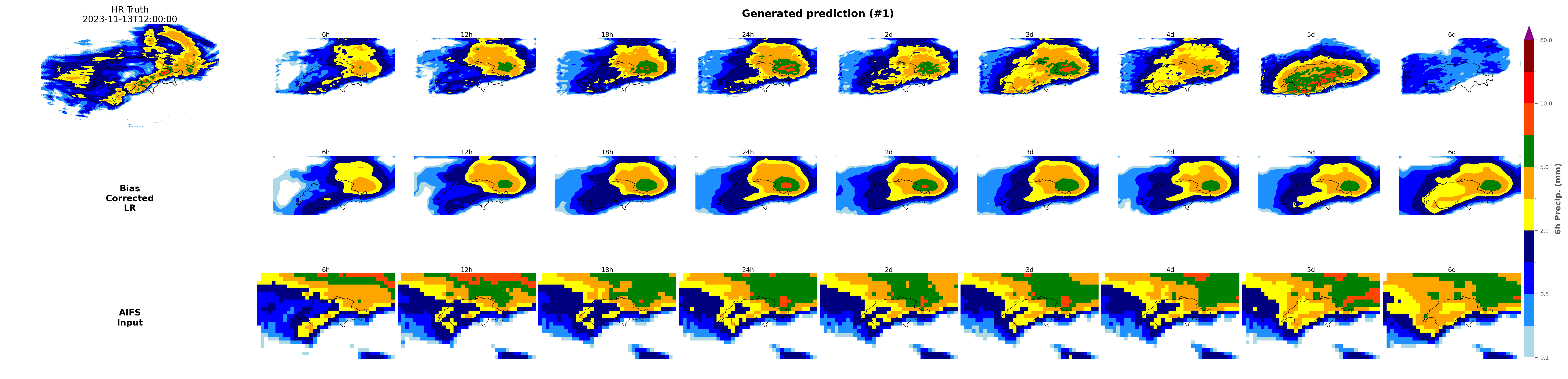}
\caption{Downscaling samples by lead time (member 0) for the November 2023 heavy precipitation event (3/7).}
\label{fig:eventNov_3}
\end{figure}

\begin{figure}[h]
\hspace{-1cm}
\includegraphics[width=1.1\textwidth]{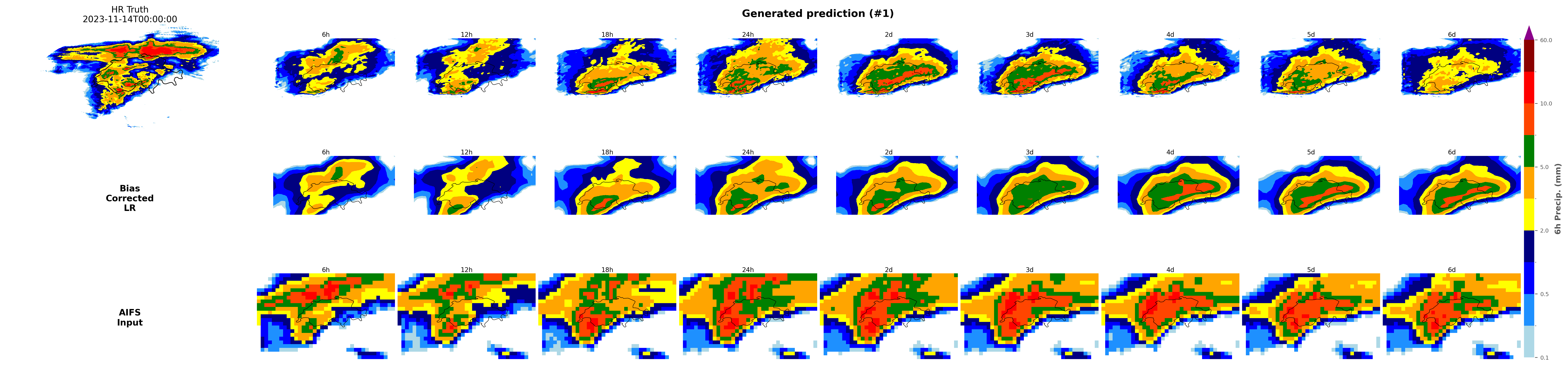}
\caption{Downscaling samples by lead time (member 0) for the November 2023 heavy precipitation event (4/7).}
\label{fig:eventNov_4}
\end{figure}

\begin{figure}[h]
\hspace{-1cm}
\includegraphics[width=1.1\textwidth]{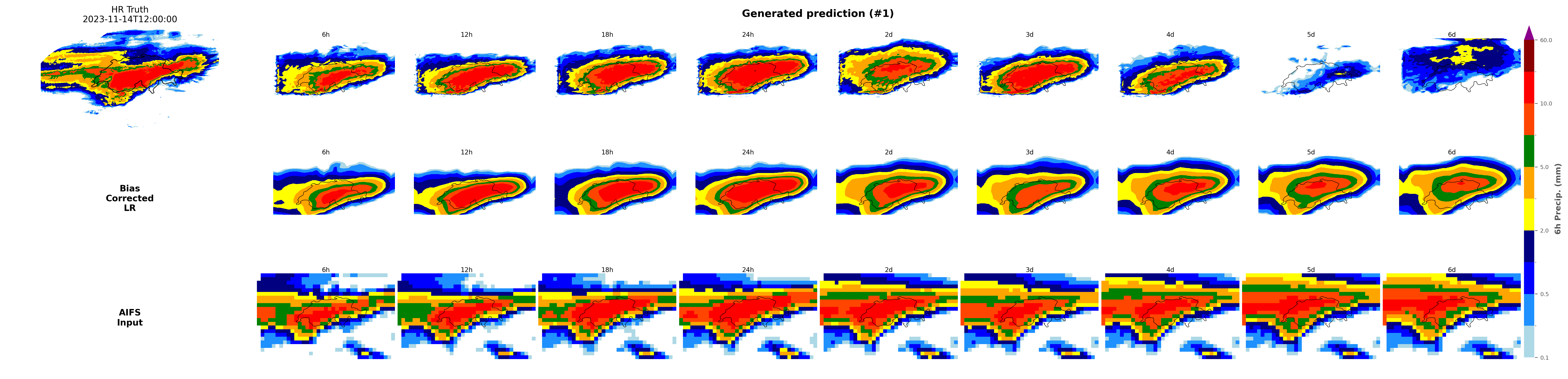}
\caption{Downscaling samples by lead time (member 0) for the November 2023 heavy precipitation event (5/7).}
\label{fig:eventNov_5}
\end{figure}

\begin{figure}[h]
\hspace{-1cm}
\includegraphics[width=1.1\textwidth]{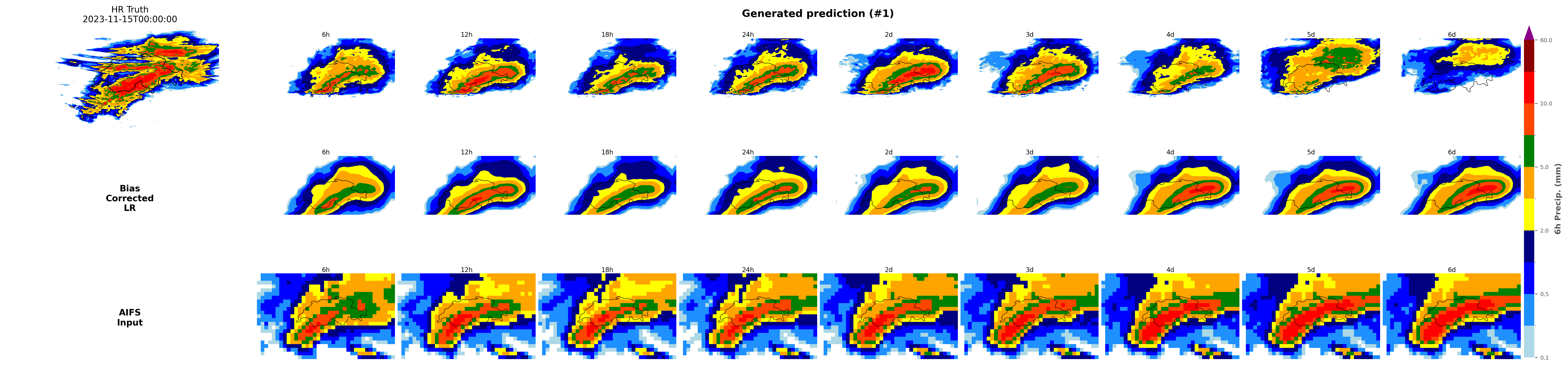}
\caption{Downscaling samples by lead time (member 0) for the November 2023 heavy precipitation event (6/7).}
\label{fig:eventNov_6}
\end{figure}

\begin{figure}[h]
\hspace{-1cm}
\includegraphics[width=1.1\textwidth]{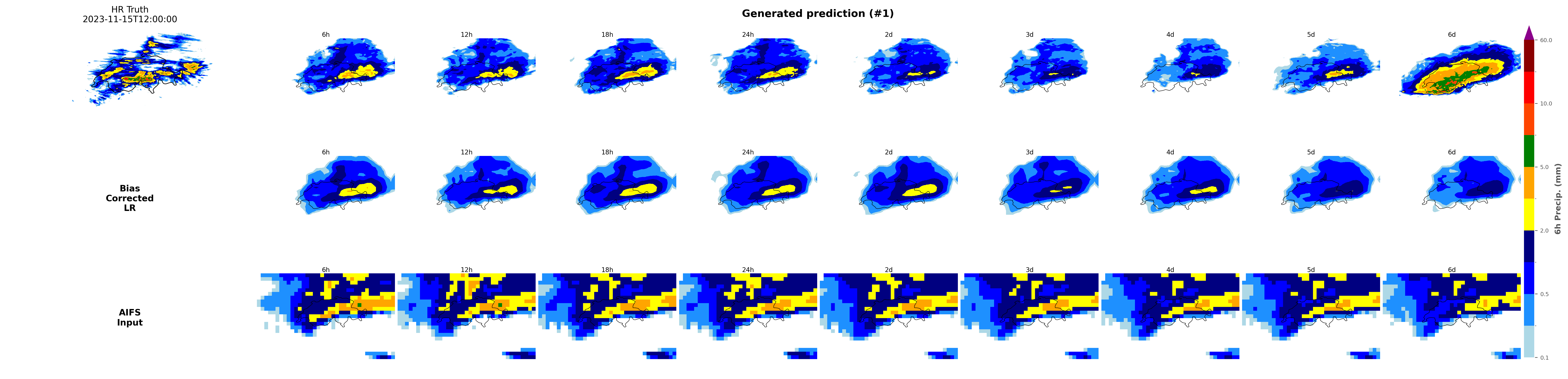}
\caption{Downscaling samples by lead time (member 0) for the November 2023 heavy precipitation event (7/7).}
\label{fig:eventNov_7}
\end{figure}

%%%%%%%%%%%%%%%%%%%%%%%%%%%%%%%%%%%%%%%%%%%%%%%%%%%%%%%%%%%%%%%%%%%%%
\newpage
\section{Forecasting verification metrics: detailed definitions}
\label{sec_SI:metrics_detailed}
%%%%%%%%%%%%%%%%%%%%%%%%%%%%%%%%%%%%%%%%%%%%%%%%%%%%%%%%%%%%%%%%%%%%%

\smallskip
\noindent We evaluate the performance of our bias correction models as well the resulting downscaling models using a comprehensive set of metrics that capture different aspects of precipitation forecast quality. These include point-wise error metrics, categorical event-based metrics, and spatial focused metrics.

\subsubsection{Pointwise and categorical verification metrics}

\noindent Point-wise metrics measure the magnitude of prediction errors at each grid point independently. Since precipitation distributions are typically dominated by dry grid points, we compute an  MSE\textsubscript{wet} restricted to wet pixels, meaning pixels where the precipitation is larger than 0.1 mm/6h. 

\smallskip
\noindent Event-based metrics evaluate the model's ability to correctly detect precipitation occurrence by treating the forecast as a binary classification problem. Wet pixels are defined by a threshold of 0.1 mm/6h, as mentioned in Section 3d of the main text. This yields a contingency table with four categories: true positives (TP, correctly predicted rain), false positives (FP, predicted rain that did not occur), false negatives (FN, missed rain events), and true negatives (TN, correctly predicted dry conditions).

\smallskip
\noindent The Critical Success Index (CSI) is then computed to measure overall detection skills: $\text{CSI} = \text{TP}/(\text{TP} + \text{FP} + \text{FN})$, ranging from 0 to 1. By excluding true negatives, CSI is well suited for precipitation verification where dry conditions dominate.

\subsubsection{Spatial Verification Metrics}

\noindent Traditional point-wise metrics suffer from the ``double penalty'' problem, where spatially displaced but otherwise correct forecasts are penalized twice. Spatial verification metrics address this by evaluating quality within a neighborhood context.

\smallskip
\noindent \textbf{Fractions Skill Score (FSS) and member-averaged FSS (avFSS).} The FSS \citep{roberts2008scale} evaluates spatial agreement between forecast and observed precipitation exceedance fields smoothed over a neighborhood of size $n \times n$. We adopt the reformulation of \cite{necker2024fss} based on Binary Probabilities (BP):
\begin{equation}
    \text{BP}_i = \frac{1}{n^2} \sum_{j \in \mathcal{N}_i} \mathbf{1}[F_j \geq \tau]
\end{equation}
The FSS is defined as $\text{FSS} = 1 - \text{FBS}/\text{FBS}_{\text{ref}}$, where $\text{FBS}$ is the mean squared difference between forecast and observed BP fields, and $\text{FBS}_{\text{ref}}$ is the worst-case reference assuming zero spatial overlap. FSS ranges from 0 to 1, with 1 indicating perfect spatial agreement. For ensemble forecasts, we employ the member-averaged FSS (avFSS) \citep{necker2024fss}, computed as $\text{avFSS} = M^{-1}\sum_{m=1}^{M} \text{FSS}(F_m, O)$, which preserves member sharpness unlike FSS of the ensemble mean.
 
\smallskip
\noindent We compute FSS and avFSS at three neighborhood sizes: $n = 1$ (pixel-level), $n = 5$ ($\sim$25\,km), and $n = 17$ ($\sim$85\,km), as summarized in Table~\ref{tab:fss_interpretation}.

 \begin{table}[t]
\centering
\caption{Interpretation of FSS at different spatial scales}
\label{tab:fss_interpretation}
\begin{tabular}{cll}
\toprule
Neighborhood & Scale & Interpretation \\
\midrule
$n = 1$ & Pixel-level & Exact location skill (strict) \\
$n = 5$ & $\sim$25 km & Small displacement tolerance \\
$n = 17$ & $\sim$85 km & Synoptic-scale skill \\
\bottomrule
\end{tabular}
\end{table}

\smallskip
\noindent \textbf{Structure-Amplitude-Location (SAL) and Ensemble SAL (eSAL).} The SAL diagnostic \citep{wernli2008sal} decomposes forecast error into three independent components. \textbf{Structure (S)} compares the shape and peakedness of precipitation objects ($S = (V_F - V_O) / [0.5(V_F + V_O)]$, where $V = R/R_{\text{max}}$); negative values indicate overly peaked fields, positive values overly diffuse ones. \textbf{Amplitude (A)} compares domain-average precipitation ($A = (R_F - R_O) / [0.5(R_F + R_O)]$); negative indicates underestimation, positive overestimation. \textbf{Location (L)} quantifies spatial displacement based on precipitation-weighted centers of mass. All three components equal zero for a perfect forecast and range within $[-2, 2]$ for S and A, and $[0, 2]$ for L.
 
\smallskip
\noindent For ensemble forecasts, we report the ensemble SAL (eSAL) \citep{radanovics2018verification}, defined as the median of each SAL component computed across ensemble members: $\text{eS} = \text{median}(\{S_m\})$, $\text{eA} = \text{median}(\{A_m\})$, $\text{eL} = \text{median}(\{L_m\})$.

\subsubsection{Probabilistic Verification Metrics}

\noindent Step~2 of \textit{SwAIther-Precip} employs a residual diffusion sub-network to generate an ensemble of high-resolution precipitation fields, providing a probabilistic estimate of the downscaled precipitation. Evaluating probabilistic forecasts requires metrics that assess both the reliability (statistical consistency) and sharpness (concentration) of the predicted distributions. We employ the CRPS to to evaluate the quality of our probabilistic predictions.

\smallskip
\noindent \textbf{Continuous Ranked Probability Score (CRPS).} The CRPS \citep{matheson1976scoring, hersbach2000decomposition} is a strictly proper scoring rule that generalizes MAE to probabilistic forecasts. For an ensemble of $M$ members $\{x_1, \ldots, x_M\}$ and observation $y$, it is computed as \citep{gneiting2007strictly}:
\begin{equation}
    \text{CRPS}(F, y) = \frac{1}{M} \sum_{m=1}^{M} |x_m - y| - \frac{1}{2M^2} \sum_{m=1}^{M} \sum_{m'=1}^{M} |x_m - x_{m'}|
\end{equation}
where the first term measures ensemble--observation distance and the second rewards spread. CRPS has the same unit as the predicted variable and reduces to MAE for deterministic forecasts. We report the spatial average over all grid points. 

\smallskip
\noindent \textbf{Probability Integral Transform (PIT).} The PIT \citep{gneiting2007strictly, dawid1984present} assesses ensemble calibration by evaluating the quantile at which the observation falls within the predicted distribution. For a calibrated forecast, PIT values are uniformly distributed on $[0, 1]$; deviations reveal underdispersion (U-shaped histogram), overdispersion (dome-shaped), or systematic bias (skewed). Because precipitation has a point mass at zero, we use the randomized PIT \citep{gneiting2007strictly}:
\begin{equation} 
    \text{PIT} \sim \mathcal{U}\!\left[F(y^{-}),\; F(y)\right]
\end{equation}
where $F(y)$ and $F(y^{-})$ are the empirical CDF evaluated at and just below $y$, respectively. We aggregate PIT values into histograms with $K = 20$ bins and quantify departure from uniformity using the Kullback--Leibler divergence \citep{leinonen2023latent}:
\begin{equation}
    D_{\mathrm{KL}} = \sum_{k=1}^{K} h_k \, \log\!\left(\frac{h_k}{1/K}\right)
\end{equation}
where $D_{\mathrm{KL}} = 0$ corresponds to perfect calibration. We additionally compute PIT histograms stratified by precipitation intensity (dry, light, heavy) to diagnose regime-dependent miscalibration.

% For ensemble forecasts represented by $M$ members $\{x_1, \ldots, x_M\}$, the empirical CDF yields:
% \begin{equation}
%     F(y) = \frac{1}{M} \sum_{m=1}^{M} \mathbf{1}(x_m \leq y), \qquad
%     F(y^{-}) = \frac{1}{M} \sum_{m=1}^{M} \mathbf{1}(x_m < y)
% \end{equation}
% Precipitation fields contain a point mass at zero: many ensemble members and observations share the exact value $y = 0$, causing $F(y)$ and $F(y^{-})$ to differ substantially. To avoid artificial spikes in the PIT histogram, we use the randomized PIT \citep{gneiting2007strictly}:
% \begin{equation} 
%     \text{PIT} \sim \mathcal{U}\!\left[F(y^{-}),\; F(y)\right]
% \end{equation}
% where $\mathcal{U}[a, b]$ denotes a uniform draw on $[a, b]$. When no ensemble members share the observed value, $F(y) - F(y^{-}) = 1/M$ and the randomization has negligible effect; when many members equal the observation (typically at zero), the randomization spreads the PIT values across the appropriate interval, preserving the uniformity property under correct calibration.

\subsubsection{Spectral Evaluation}
 
\noindent Point-wise and categorical metrics do not characterize the spatial scales at which the downscaling pipeline produces reliable precipitation structures. A model may achieve low MSE while systematically under-representing variability at certain scales---a deficiency consequential for hydrological applications. Spectral analysis addresses this by decomposing forecast fields into contributions at each spatial wavelength.
 
\smallskip
\noindent \textbf{Radially-averaged power spectral density.} Following \cite{sinclair2005empirical, pulkkinen2019pysteps}, we compute the radially-averaged power spectral density (RAPSD) of log-transformed precipitation fields ($\tilde{P} = \log(P + \varepsilon)$, $\varepsilon = 0.1$~mm/6h). After removing the spatial mean and applying a Tukey tapering window ($\alpha = 0.1$), we compute the 2D PSD and radially average it to obtain a 1D spectrum as a function of wavelength $\lambda = 1/k$ (km).
 
\smallskip
\noindent \textbf{Co-masked spectral comparison.} Since predicted and observed fields generally have different wet-area extents, a direct spectral comparison would conflate coverage differences with structural differences. We therefore adopt a co-masking strategy: for each time step and ensemble member, both fields are restricted to the intersection of their wet areas ($> 0.1$~mm/6h) before computing spectra. Time steps where the common wet area covers less than 3\% of the domain are excluded.

\smallskip
\noindent \textbf{Scale-dependent spectral ratio and effective resolution.} The primary diagnostic is the spectral ratio $\text{PSD}_{\text{pred}}(\lambda) / \text{PSD}_{\text{obs}}(\lambda)$: a ratio of 1 indicates perfect spectral fidelity, values below 1 indicate under-dispersion, and values above 1 indicate over-dispersion. We compute band-averaged ratios over three wavelength bands, as reported in Table~\ref{tab:spectral_bands}.
% ): large scale (100--600\,km, synoptic precipitation), mesoscale (20--100\,km, frontal bands), and small scale (2--20\,km, convective cells).
 
\begin{table}[t]
\centering
\caption{Wavelength bands for scale-dependent spectral analysis}
\label{tab:spectral_bands}
\begin{tabular}{lcl}
\toprule
Band & Wavelength range & Interpretation \\
\midrule
Large scale & 100--600 km & Synoptic precipitation  \\
Mesoscale & 20--100 km & Frontal bands \\
Small scale & 2--20 km & Convective cells, fine-scale \\
\bottomrule
\end{tabular}
\end{table}
 
\smallskip
\noindent Following \cite{klaver2020effective}, we define the \emph{effective resolution} as the wavelength at which the spectral ratio first drops below 0.5, scanning from large to small scales. This provides a single-number characterization of the pipeline's resolving capability; for reference, NWP models typically achieve effective resolutions of 4--8 times their grid spacing \citep{skamarock2004evaluating, klaver2020effective}. In the context of our pipeline, band-averaged ratios at large and mesoscales primarily reflect the bias correction quality (Step~1), while the small-scale band characterizes the ability of the residual diffusion sub-network to generate realistic fine-scale texture (Step~2).

%%%%%%%%%%%%%%%%%%%%%%%%%%%%%%%%%%%%%%%%%%%%%%%%%%%%%%%%%%%%%%%%%%%%%
\newpage
\section{Uncertainty of the bias-correction metrics}
\label{sec_SI:Uncertainty}
%%%%%%%%%%%%%%%%%%%%%%%%%%%%%%%%%%%%%%%%%%%%%%%%%%%%%%%%%%%%%%%%%%%%%

\begin{figure}[ht]
\centering
\includegraphics[width=\textwidth]{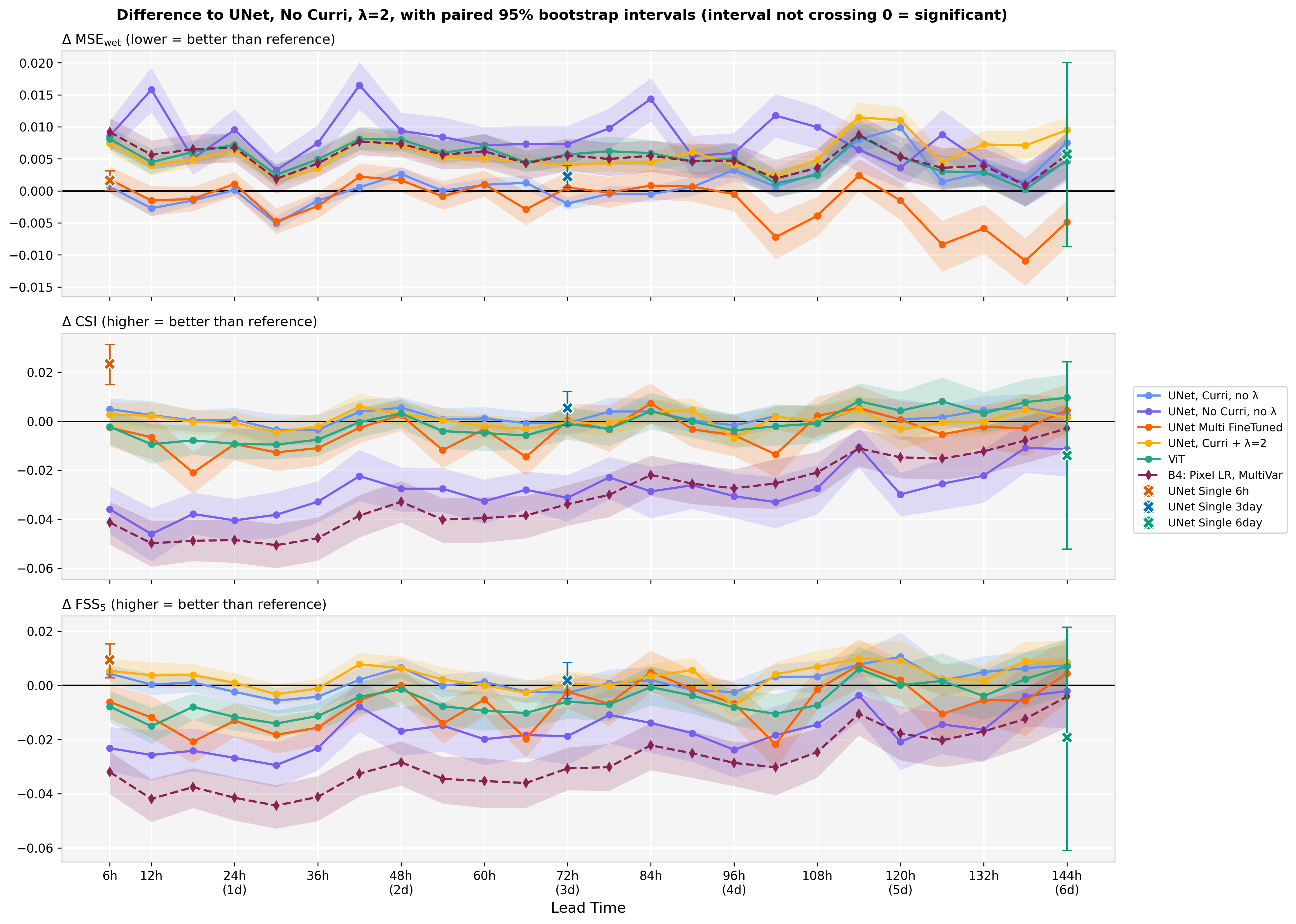}
\caption{Differences between each bias-correction model and UNet$_{\lambda=2}$ by lead time,
for MSE$_\mathrm{wet}$, CSI and FSS$_5$, with paired 95\% confidence intervals from the
moving-block bootstrap. A difference is significant where its interval does not cross zero.
Crosses: single-lead models at their lead time.}
\label{fig:supp_paired}
\end{figure}

We estimate the sampling uncertainty of the bias-correction metrics with a moving-block bootstrap over the validation times (1000 replicates). Each metric is recomputed on every
resample exactly as on the full evaluation set, i.e.\ pooled over the samples of each lead time and then averaged over lead times. All lead times of a valid time are resampled together, as they share the same observation, and blocks of four consecutive validation times (about two weeks) account for the persistence of weather regimes; blocks of one or eight validation times change the interval widths by less than 25\%. Confidence intervals are given by the
2.5th and 97.5th percentiles of the bootstrap distribution. Since the same resamples are applied to all models, differences between models are evaluated with paired intervals, which are much narrower than the intervals of the individual scores because the variability due to
the choice of weather situations is shared by all models (Tables~\ref{tab:supp_ci} and
\ref{tab:supp_paired}, Fig.~\ref{fig:supp_paired}). The SAL components have wide intervals: all deep-learning models are significantly closer to zero than B0, B1, B2 and B7, but the
differences among the deep-learning models are not significant.

\begin{table}[ht]
\centering
\caption{Bias-correction metrics averaged over lead times, with 95\% confidence intervals
(moving-block bootstrap over validation times).}
\label{tab:supp_ci}
\small
\begin{tabular}{lcccc}
\toprule
 & MSE $\downarrow$ & MSE$_\mathrm{wet}$ $\downarrow$ & CSI $\uparrow$ & FSS$_5$ $\uparrow$ \\
\midrule
B0: AIFS (raw)            & 0.050 [0.046, 0.055] & 0.088 [0.083, 0.094] & 0.355 [0.331, 0.377] & 0.607 [0.581, 0.632] \\
B1: Global affine         & 0.024 [0.021, 0.027] & 0.096 [0.085, 0.107] & 0.412 [0.388, 0.432] & 0.695 [0.671, 0.715] \\
B2: Affine / lead         & 0.024 [0.021, 0.027] & 0.096 [0.085, 0.107] & 0.411 [0.387, 0.432] & 0.695 [0.671, 0.716] \\
B3: Pixel LR (precip)     & 0.020 [0.017, 0.023] & 0.084 [0.072, 0.094] & 0.470 [0.443, 0.493] & 0.766 [0.742, 0.785] \\
B4: Pixel LR (multivar)   & 0.018 [0.016, 0.021] & 0.073 [0.064, 0.081] & 0.508 [0.482, 0.532] & 0.800 [0.780, 0.817] \\
B5: Quantile mapping      & 0.040 [0.036, 0.044] & 0.135 [0.125, 0.146] & 0.347 [0.325, 0.367] & 0.792 [0.775, 0.808] \\
B7: Ridge regression      & 0.023 [0.020, 0.026] & 0.095 [0.086, 0.105] & 0.430 [0.406, 0.451] & 0.722 [0.700, 0.740] \\
\midrule
UNet$_{\lambda=0}$, curriculum & 0.016 [0.014, 0.019] & 0.069 [0.061, 0.077] & 0.540 [0.518, 0.563] & 0.830 [0.814, 0.845] \\
UNet$_{\lambda=0}$             & 0.019 [0.016, 0.022] & 0.075 [0.066, 0.085] & 0.510 [0.489, 0.530] & 0.811 [0.795, 0.824] \\
UNet$_{\lambda=2}$             & 0.016 [0.014, 0.019] & 0.067 [0.060, 0.075] & 0.538 [0.516, 0.561] & 0.828 [0.812, 0.844] \\
UNet Multi FT                  & 0.017 [0.015, 0.020] & 0.066 [0.058, 0.073] & 0.534 [0.513, 0.556] & 0.821 [0.805, 0.836] \\
UNet$_{\lambda=2}$, curriculum & 0.017 [0.014, 0.019] & 0.073 [0.065, 0.081] & 0.539 [0.517, 0.561] & 0.831 [0.816, 0.846] \\
ViT                            & 0.017 [0.014, 0.019] & 0.073 [0.064, 0.081] & 0.537 [0.516, 0.559] & 0.823 [0.807, 0.838] \\
\bottomrule
\end{tabular}
\end{table}

\begin{table}[ht]
\centering
\caption{Paired differences to UNet$_{\lambda=2}$ (model minus UNet$_{\lambda=2}$, all lead
times pooled), with 95\% confidence intervals. Negative MSE and positive CSI or FSS$_5$
differences indicate a better model. *: significant (interval excludes zero).}
\label{tab:supp_paired}
\small
\resizebox{\columnwidth}{!}{%
\begin{tabular}{lcccc}
\toprule
 & $\Delta$MSE & $\Delta$MSE$_\mathrm{wet}$ & $\Delta$CSI & $\Delta$FSS$_5$ \\
\midrule
B0: AIFS (raw)            & $+0.0338$ [0.0303, 0.0380]* & $+0.0208$ [0.0143, 0.0280]* & $-0.181$ [$-0.194$, $-0.168$]* & $-0.220$ [$-0.239$, $-0.202$]* \\
B1: Global affine         & $+0.0074$ [0.0063, 0.0087]* & $+0.0288$ [0.0240, 0.0337]* & $-0.125$ [$-0.136$, $-0.114$]* & $-0.132$ [$-0.146$, $-0.119$]* \\
B2: Affine / lead         & $+0.0074$ [0.0063, 0.0086]* & $+0.0287$ [0.0239, 0.0336]* & $-0.125$ [$-0.136$, $-0.114$]* & $-0.131$ [$-0.146$, $-0.118$]* \\
B3: Pixel LR (precip)     & $+0.0037$ [0.0030, 0.0045]* & $+0.0161$ [0.0119, 0.0203]* & $-0.066$ [$-0.075$, $-0.056$]* & $-0.060$ [$-0.072$, $-0.049$]* \\
B4: Pixel LR (multivar)   & $+0.0017$ [0.0013, 0.0022]* & $+0.0052$ [0.0029, 0.0074]* & $-0.029$ [$-0.035$, $-0.022$]* & $-0.027$ [$-0.035$, $-0.021$]* \\
B5: Quantile mapping      & $+0.0237$ [0.0220, 0.0256]* & $+0.0679$ [0.0636, 0.0719]* & $-0.190$ [$-0.201$, $-0.179$]* & $-0.032$ [$-0.045$, $-0.020$]* \\
B7: Ridge regression      & $+0.0064$ [0.0053, 0.0075]* & $+0.0280$ [0.0238, 0.0322]* & $-0.107$ [$-0.116$, $-0.098$]* & $-0.106$ [$-0.116$, $-0.096$]* \\
\midrule
UNet$_{\lambda=0}$, curriculum & $-0.0000$ [$-0.0002$, 0.0002] & $+0.0013$ [0.0005, 0.0022]* & $+0.003$ [0.000, 0.006]* & $+0.003$ [0.000, 0.006]* \\
UNet$_{\lambda=0}$             & $+0.0023$ [0.0019, 0.0028]* & $+0.0080$ [0.0053, 0.0106]* & $-0.027$ [$-0.033$, $-0.021$]* & $-0.016$ [$-0.023$, $-0.010$]* \\
UNet Multi FT                  & $+0.0012$ [0.0008, 0.0017]* & $-0.0019$ [$-0.0037$, 0.0000] & $-0.004$ [$-0.009$, 0.002] & $-0.007$ [$-0.012$, $-0.001$]* \\
UNet$_{\lambda=2}$, curriculum & $+0.0003$ [0.0001, 0.0006]* & $+0.0058$ [0.0046, 0.0070]* & $+0.002$ [$-0.002$, 0.005] & $+0.004$ [0.001, 0.007]* \\
ViT                            & $+0.0005$ [0.0002, 0.0009]* & $+0.0051$ [0.0036, 0.0068]* & $-0.000$ [$-0.005$, 0.005] & $-0.005$ [$-0.010$, $-0.000$]* \\
\bottomrule
\end{tabular}
}
\end{table}

% %%%%%%%%%%%%%%%%%%%%%%%%%%%%%%%%%%%%%%%%%%%%%%%%%%%%%%%%%%%%%%%%%%%%%
% % REFERENCES
% %%%%%%%%%%%%%%%%%%%%%%%%%%%%%%%%%%%%%%%%%%%%%%%%%%%%%%%%%%%%%%%%%%%%%
% \newpage
% % Make your BibTeX bibliography by using these commands:
% \bibliographystyle{ametsocV6}
% \bibliography{references}

% \end{document}